\documentclass[a4,11pt]{article}
\usepackage{graphicx} 
\usepackage{float}
\usepackage{amsmath,amsthm}
\usepackage{amsfonts}
\usepackage{latexsym}
\usepackage{amssymb}
\usepackage{setspace}
\usepackage{comment}

\makeatletter
 
  \@addtoreset{equation}{section}
 \makeatother

\begin{document}
\title{Solving Backward Stochastic Differential Equations \\
with quadratic-growth drivers \\
 by Connecting the Short-term Expansions~\footnote{
 Accepted for publication in Stochastic Processes and their Applications.
All the contents expressed in this research are solely those of the author and do not represent any views or 
opinions of any institutions. 
}
}

\author{Masaaki Fujii\footnote{Quantitative Finance Course, Graduate School of Economics, The University of Tokyo. }
~~\&~~Akihiko Takahashi\footnote{Quantitative Finance Course, Graduate School of Economics, The University of Tokyo.} 
}
\date{ \small 
May 23, 2018\\
}
\maketitle



\newtheorem{definition}{Definition}[section]
\newtheorem{assumption}{Assumption}[section]
\newtheorem{condition}{$[$ C}
\newtheorem{lemma}{Lemma}[section]
\newtheorem{proposition}{Proposition}[section]
\newtheorem{theorem}{Theorem}[section]
\newtheorem{remark}{Remark}[section]
\newtheorem{example}{Example}[section]
\newtheorem{corollary}{Corollary}[section]
\def\n{{\bf n}}
\def\A{{\bf A}}
\def\B{{\bf B}}
\def\C{{\bf C}}
\def\D{{\bf D}}
\def\E{{\bf E}}
\def\F{{\bf F}}
\def\G{{\bf G}}
\def\H{{\bf H}}
\def\I{{\bf I}}
\def\J{{\bf J}}
\def\K{{\bf K}}
\def\L{{\bf L}}
\def\M{{\bf M}}
\def\N{{\bf N}}
\def\O{{\bf O}}
\def\P{{\bf P}}
\def\Q{{\bf Q}}
\def\R{{\bf R}}
\def\S{{\bf S}}
\def\T{{\bf T}}
\def\U{{\bf U}}
\def\V{{\bf V}}
\def\W{{\bf W}}
\def\X{{\bf X}}
\def\Y{{\bf Y}}
\def\Z{{\bf Z}}
\def\cala{{\cal A}}
\def\calb{{\cal B}}
\def\calc{{\cal C}}
\def\cald{{\cal D}}
\def\cale{{\cal E}}
\def\calf{{\cal F}}
\def\calg{{\cal G}}
\def\calh{{\cal H}}
\def\cali{{\cal I}}
\def\calj{{\cal J}}
\def\calk{{\cal K}}
\def\call{{\cal L}}
\def\calm{{\cal M}}
\def\caln{{\cal N}}
\def\calo{{\cal O}}
\def\calp{{\cal P}}
\def\calq{{\cal Q}}
\def\calr{{\cal R}}
\def\cals{{\cal S}}
\def\calt{{\cal T}}
\def\calu{{\cal U}}
\def\calv{{\cal V}}
\def\calw{{\cal W}}
\def\calx{{\cal X}}
\def\caly{{\cal Y}}
\def\calz{{\cal Z}}
%
\def\sskip{\hspace{0.5cm}}
\def\simleq{ \raisebox{-.7ex}{\em $\stackrel{{\textstyle <}}{\sim}$} }
\def\leqsim{ \raisebox{-.7ex}{\em $\stackrel{{\textstyle <}}{\sim}$} }
\def\ep{\epsilon}
\def\half{\frac{1}{2}}
\def\iku{\rightarrow}
\def\Iku{\Rightarrow}
\def\ikup{\rightarrow^{p}}
\def\inclusion{\hookrightarrow}
\def\cadlag{c\`adl\`ag\ }
\def\up{\uparrow}
\def\down{\downarrow}
\def\doti{\Leftrightarrow}
\def\douti{\Leftrightarrow}
\def\dochi{\Leftrightarrow}
\def\douchi{\Leftrightarrow}%
\def\yy{\\ && \nonumber \\}
\def\y{\vspace*{3mm}\\}
\def\nn{\nonumber}
\def\be{\begin{equation}}
\def\ee{\end{equation}}
\def\bea{\begin{eqnarray}}
\def\eea{\end{eqnarray}}
\def\beas{\begin{eqnarray*}}
\def\eeas{\end{eqnarray*}}
%
\def\hd{\hat{D}}
\def\hv{\hat{V}}
\def\hsd{{\hat{d}}}
\def\hx{\hat{X}}
\def\hsx{\hat{x}}
\def\bsx{\bar{x}}
\def\bsd{{\bar{d}}}
\def\bx{\bar{X}}
\def\ba{\bar{A}}
\def\bb{\bar{B}}
\def\bc{\bar{C}}
\def\bv{\bar{V}}
\def\balpha{\bar{\alpha}}
\def\bbalpha{\bar{\bar{\alpha}}}
\def\combi{\l(\begin{array}{c}\alpha\\ \beta \end{array}\r)}
\def\f{^{(1)}}
\def\s{^{(2)}}
\def\ss{^{(2)*}}
\def\l{\left}
\def\r{\right}
\def\a{\alpha}
\def\b{\beta}
\def\L{\Lambda}


\def\E{{\bf E}}
\def\P{{\bf P}}
\def\Q{{\bf Q}}
\def\R{{\bf R}}

\def\cadlag{{c\`adl\`ag~}}

\def\calf{{\cal F}}
\def\calp{{\cal P}}
\def\calq{{\cal Q}}
\def\wt{\widetilde}
\def\wh{\widehat}

\def\nab{\nabla}
\def\ol{\overline}
\def\mbb{\mathbb}
\newcommand{\bvec}[1]{\mbox{\boldmath $#1$}}
\def\bpi{\bvec{\pi}}

\newcommand{\bbf}[1]{{\bold{#1}}}

\def\bchi{\bar{\chi}}
\def\bby{\bar{\bold{y}}}
\def\bbg{\bar{\bold{g}}}

\def\LDis{\frac{\bigl.}{\bigr.}}
\def\ep{\epsilon}
\def\varep{\varepsilon}
\def\al{\alpha}

\def\bep{{\bar{\epsilon}}}
\def\br{{\bar{r}}}
\def\bq{{\bar{q}}}

\def\del{\delta}
\def\Del{\Delta}
\def\part{\partial}
\def\bsigma{\bar{\sigma}}
\def\yy{\\ && \nonumber \\}
\def\y{\vspace*{3mm}\\}
\def\nn{\nonumber}
\def\be{\begin{equation}}
\def\ee{\end{equation}}
\def\bea{\begin{eqnarray}}
\def\eea{\end{eqnarray}}
\def\beas{\begin{eqnarray*}}
\def\eeas{\end{eqnarray*}}

\def\bull{$\bullet~$}

\newcommand{\Slash}[1]{{\ooalign{\hfil/\hfil\crcr$#1$}}}
\vspace{-11mm}

\begin{abstract}
This article proposes a new approximation scheme for 
quadratic-growth BSDEs in a Markovian setting
by connecting a series of
semi-analytic asymptotic expansions applied to short-time intervals.
Although there remains a condition which needs to be checked a posteriori, one can avoid altogether time-consuming Monte Carlo
simulation and other numerical integrations for estimating conditional expectations at each space-time node.
Numerical examples of quadratic-growth  as well as Lipschitz BSDEs 
suggest that the scheme works well even for large quadratic coefficients, and a fortiori for large Lipschitz constants.

\end{abstract}
\vspace{0mm}
\vspace{-4mm}
\section{Introduction}
The research on backward stochastic differential equations (BSDEs) was initiated by 
Bismut (1973)~\cite{Bismut} for a linear case and followed by Pardoux \& Peng (1990)~\cite{Pardoux-Peng} for  general 
non-linear setups. Since then, BSDEs have attracted strong interests among researchers and now 
large amount of literature is available.
See for example, El Karoui et al. (1997)~\cite{ElKaroui-Quenez},
El Karoui \& Mazliak (eds.) (1997)~\cite{ElKaroui-Mazliak}, 
Ma \& Yong (2000)~\cite{Ma-Yong}, Yong \& Zhou (1999)~\cite{Yong-Zhou}, Cvitani\'c \& Zhang (2013)~\cite{Cvitanic}
and Delong (2013)~\cite{Delong} for excellent reviews and various applications,
and also Pardoux \& Rascanu (2014)~\cite{Pardoux-Rascanu} for a recent thorough textbook for 
BSDEs in the diffusion setup.
In particular, since the financial crisis in 2009, the importance of BSDEs in the financial industry
has grown significantly. This is because BSDEs have been found to be indispensable to understand
various valuation adjustments collectively referred to XVAs as well as the optimal risk control
under the new regulations. For market developments and related issues, see
Brigo, Morini \& Pallavicini (2013)~\cite{Brigo-book}, 
Bianchetti \& Morini (eds.) (2013)~\cite{Bianchetti-Riskbook},
Cr\'epey \& Bielecki (2014)~\cite{Crepey-book} and references therein.

In the past decade, there has been also significant progress of numerical computation methods
for BSDEs. In particular, based on the so-called $\mbb{L}^2$-regularity 
of the control variables established by Zhang (2001,~2004)~\cite{JZhangPHD,JZhang}, 
now standard backward Monte Carlo schemes for Lipschitz BSDEs have been developed by
Bouchard \& Touzi (2004)~\cite{Bouchard-Touzi},
Gobet, Lemor \& Warin (2005)~\cite{Gobet-Warin}.  One can find many variants and extensions such as
Bouchard \& Elie (2008)~\cite{Bouchard-Elie} for BSDEs with jumps,
Bouchard \& Chassagneux (2008)~\cite{Bouchard-Chassagneux} for reflected BSDEs,
and Chassagneux \& Richou (2016)~\cite{Chassagneux-reflected} for a system of reflected BSDEs
arising from optimal switching problems.
Bender \& Denk (2007)~\cite{Bender-Denk} proposed a forward scheme free from 
the  linearly growing regression errors existing in the backward schemes;
Bender \& Steiner (2012)~\cite{Bender-Steiner} suggested a possible improvement of the scheme~\cite{Gobet-Warin}
by using martingale basis functions for regressions;
Crisan \& Monolarakis (2014)~\cite{Crisan-Monolarakis} developed a second-order discretization
using the cubature method. A different scheme based on the optimal quantization was proposed by
Bally \& Pag\`es (2003)~\cite{Bally-Pages}. See
Pag\`es \& Sagna (2017)~\cite{Pages-Sagna} and references therein for its recent developments.
Delarue \& Menozzi (2006, 2008)~\cite{DM06,DM08} studied a class of quasi-linear PDEs via a coupled forward-backward SDE 
with Lipschitz functions,
which, in particular,  becomes equivalent to a special type of quadratic-growth (qg) BSDE (so called deterministic KPZ equation) under a certain setting.
Recently, Chassagneux \& Richou (2016)~\cite{Richou}
extended the standard backward scheme to qg-BSDEs 
with bounded terminal conditions in a Markovian setting.

As another approach,
a semi-analytic  approximation scheme was proposed 
by Fujii \& Takahashi (2012)~\cite{FT-analytic}
and justified in the Lipschitz case by Takahashi \& Yamada (2015)~\cite{Takahashi-Yamada}.
An efficient implementation algorithm based on an interacting particle method by Fujii \& Takahashi (2015)~\cite{FT-particle}
has been successfully applied to a large scale credit portfolio by 
Cr\'epey \& Song (2016)~\cite{Crepey-Song}. 
This is an asymptotic expansion
around a linear driver motivated by the observation that, for many financial applications,
the non-linear part of the driver is proportional
to an interest rate spread and/or a default intensity which is, at most, of the order of a few percentage points.
Although it cuts the cost of numerical computation drastically under many interesting situations, 
the non-linear effects may grow and cease to be perturbative
when one deals with longer maturities,  higher volatilities,
or general risk-sensitive control problems for highly concave utility functions.

In this paper, we propose a new approximation scheme for Markovian qg-BSDEs with bounded terminal functions.
The qg-BSDEs have many applications, in particular, they appear 
in exponential and power utility optimizations as well as mean-variance hedging problem. They are
 also relevant for a class of recursive utilities introduced by Epstein \& Zin (1989)~\cite{Epstein-Zin},
which are widely used in economic theory.
In addition, a successful scheme for qg-BSDEs possibly provides an unified computation method for a wide range of applications,
since it is expected to be applicable also to the standard Lipschitz BSDEs. 
We try to achieve the advantages of both the standard Monte Carlo scheme,
in terms of generality,  and also the semi-analytic approximation scheme, 
in terms of the lesser numerical cost. 
The main idea is to decompose the original qg-BSDE into a sequence of qg-BSDEs 
each of which is defined in a short-time interval.
We then employ an asymptotic expansion method to solve each of them approximately.\footnote{ Similar ideas have been applied to stochastic filtering by Fujii (2014)~\cite{F-Filtering}
and to European option pricing by Takahashi \& Yamada (2016)~\cite{TY-AAP}.}
In order to obtain the total error estimate, we first investigate the propagation of error
for a sequence of qg-BSDEs with terminal conditions
perturbed by bounded functions, say $\{\del^i\}_{1\leq i\leq n}$.
The first main result is thus obtained as Theorem~\ref{th-delY-delZ}.
We then substitute the error function associated to the
asymptotic expansion in each period for the function $\del^i$, 
which then leads to our second main result of Theorem~\ref{th-main} providing 
the total error estimate for the proposed scheme.

Although there remains assumptions which cannot be confirmed a priori, which is a drawback of the 
current scheme, they can still be checked for each model a posteriori by numerical calculation. 
Once it is confirmed, the convergence with the rate of $n^{-1/2}$ in the strong sense is
obtained for a finite range of discretization. 
In the case of the standard scheme with Monte Carlo simulation, although the convergence 
is guaranteed a priori for {\it sufficiently small discretization and many paths},
it is not completely free from a posteriori checks, either.
One still needs to run heavy tests with varying
discretization sizes and number of paths to confirm whether a given result provides a reasonable approximation or not.
The advantage of the proposed scheme is that one can avoid time-consuming simulation 
for estimating conditional expectations at each space-time node
by using simple semi-analytic results and thus finer discretization becomes easier to implement.
We give numerical examples for both qg- and Lipschitz BSDEs
to illustrate the empirical performance. They suggest that the scheme works well
even for very large quadratic coefficients, and all the more so for large Lipschitz constants.
Note also that, the short-term asymptotic expansion of qg-BSDEs in the strong sense
is provided for the first time, which may be useful for other applications.

The organization of the paper is as follows:
Section~2 explains the general setting and notations,
Section~3 gives the time-discretization and investigates a sequence of qg-BSDEs
perturbed in the terminal values; 
Section~4 applies the short-term expansion to the result of Section 3
which yields the total error estimate of the proposed scheme.
Section~5 explains a concrete implementation using a discretized space-time grid
and the corresponding error estimates. Section 6 provides numerical examples
and also explain the relevant modifications when the scheme is applied to the Lipschitz BSDEs.
We finally concludes in Section 7. Appendix A summarizes the properties of BMO-martingales,
Appendices B and C derive the formula of the short-term asymptotic expansion
and obtain the error estimates relevant for the analysis in the main text.

\section{Preliminaries}
\subsection{General setting and notations}
Throughout the paper, we fix the terminal time $T>0$.
We work on the filtered probability space $(\Omega,\calf,\mbb{F}, \mbb{P})$
carrying a $d$-dimensional independent standard Brownian motion $W$.
$\mbb{F}=(\calf_t)_{t\in[0,T]}$ is the Brownian filtration satisfying the usual conditions
augmented by the $\mbb{P}$-null sets.
We denote a generic positive constant by $C$, which may change line by line
and it is sometimes associated with several subscripts (such as $C_{p,K}$)
when there is a need to emphasize its dependency on those parameters.
$\calt^T_0$ denotes the set of all $\mbb{F}$-stopping times $\tau:\Omega\rightarrow [0,T]$.
We denote the sup-norm of $\mbb{R}^k$-valued function $x:[0,T]\rightarrow \mbb{R}^k$, $k\in\mbb{N}$
by the symbol $||x||_{[a,b]}:=\sup\bigl\{|x_t|, t\in[a,b]\bigr\}$
and write $||x||_t:=||x||_{[0,t]}$.

Let us introduce the following spaces for stochastic processes with $p\geq 2$ and $k\in\mbb{N}$.
For the convenience of the reader,  we separately summarize the relevant properties of the BMO martingales
and the associated function spaces in Appendix~\ref{app-BMO}.\\
\bull $\cals^p_{[s,t]}(\mbb{R}^k)$ is the set of $\mbb{R}^k$-valued
adapted  processes $X$ satisfying
\begin{center}
$||X||_{\cals^p_{[s,t]}}:=\mbb{E}\bigl[||X||_{[s,t]}^p\bigr]^{1/p}<\infty~$.
\end{center}
\bull $\cals^\infty_{[s,t]}(\mbb{R}^k)$ is the set of $\mbb{R}^k$-valued
essentially bounded adapted processes $X$ satisfying
\begin{center}
$||X||_{\cals^\infty_{[s,t]}}:=\bigl|\bigl|\sup_{r\in[s,t]}|X_r|\bigr|\bigr|_{\infty}<\infty~$.
\end{center}
\bull $\calh^p_{[s,t]}(\mbb{R}^k)$ is the set of $\mbb{R}^k$-valued progressively measurable processes $Z$
satisfying
\be
||Z||_{\calh^p_{[s,t]}}:=\mbb{E}\Bigl[\Bigl(\int_s^t |Z_r|^2 dr\Bigr)^\frac{p}{2}\Bigr]^\frac{1}{p}<\infty. \nn
\ee
\bull $\calk^p{[s,t]}$ is the set of functions $(Y,Z)$ in the space $\cals^p_{[s,t]}(\mbb{R})\times 
\calh^p_{[s,t]}(\mbb{R}^{1\times d})$ with the norm defined by
\be
||(Y,Z)||_{\calk^p[s,t]}:=\bigl(||Y||^p_{\cals^p_{[s,t]}}+||Z||^p_{\calh^p_{[s,t]}}\bigr)^{1/p}~.\nn
\ee
\bull $\mbb{L}^\infty(\mbb{R}^d;\mbb{R}^k)$ is the set of measurable bounded functions $f:\mbb{R}^d\rightarrow \mbb{R}^k$. \\
\bull $C^m(\mbb{R}^d; \mbb{R}^k)$  is the set of $m$-time continuously 
differentiable functions $f:\mbb{R}^d\rightarrow \mbb{R}^k$. \\
\bull $C^m_b(\mbb{R}^d;\mbb{R}^k)$ is the subset of $C^m(\mbb{R}^d;\mbb{R}^k)$ with bounded derivatives. \\ 
\bull $C^\infty_b(\mbb{R}^d;\mbb{R}^k):=\bigcap_{m\geq 1}C^m_b(\mbb{R}^d;\mbb{R}^k)$.

We frequently omit the arguments such as  $\mbb{R}^d,~\mbb{R}^k$ and subscript $[s,t]$ if they are 
obvious from the context.
We use $\bigl(\Theta_s, s\in[0,T]\bigr)$ as a collective argument $
\Theta_s:=(Y_s,Z_s) $
to lighten the notation.  We use the following notation for partial derivatives with respect to $x\in\mbb{R}^d$
such that
\be
\part_x:=(\part_{x^1},\cdots,\part_{x^d}):=\bigl(\frac{\part}{\part x^1},\cdots, \frac{\part}{\part x^d}\bigr)\nn
\ee
and  $\part_\theta:=(\part_y, \part_z)$ for the collective argument $\Theta$.
We sometimes use the abbreviation $Z*W:=\int_0^\cdot Z_s dW_s$.

\subsection{Setup}
Firstly, we introduce the underlying forward process $X_t,t\in[0,T]$:
\bea
X_t=x_0+\int_0^t b(r,X_r)dr+\int_0^t \sigma(r,X_r)dW_r~,
\label{eq-SDE-org}
\eea
where $x_0\in \mbb{R}^d$ and $b:[0,T]\times \mbb{R}^d\rightarrow \mbb{R}^d$, 
$\sigma:[0,T]\times \mbb{R}^d\rightarrow \mbb{R}^{d\times d}$ are measurable functions.\footnote{Useful standard estimates on the Lipschitz SDEs can be found, for example, in Appendix A of \cite{FT-AE}.}
\begin{assumption}
\label{assumption-X}
(i) For all $t,t^\prime \in[0,T]$ and $x,x^\prime \in\mbb{R}^d$, there exists a positive constant $K$ such that
$~|b(t,x)-b(t^\prime,x^\prime)|+|\sigma(t,x)-\sigma(t^\prime,x^\prime)|\leq K\bigl(|t-t^\prime|^\frac{1}{2}+|x-x^\prime|\bigr)$.\\
(ii) $||b(\cdot,0)||_T+||\sigma(\cdot,0)||_T\leq K$. \\
(iii) $b$ and $\sigma$ are 3-time continuously differentiable with respect to $x$ 
and satisfy 
\bea
&&|\part_x^m b(t,x)|+|\part_x^m \sigma(t,x)|\leq K \nn~, \\
&&|\part_x^m b(t,x)-\part_x^m b(t^\prime,x)|+|\part_x^m \sigma(t,x)-\part_x^m \sigma(t^\prime,x)|\leq K|t-t^\prime|^{1/2}~,
\label{eq-H-conti}
\eea
for all $1\leq m\leq 3$, $t,t^\prime \in [0,T]$ and $x \in \mbb{R}^d$.
\end{assumption}

Let us now introduce a qg-BSDE which is a target of our investigation:
\bea
Y_t=\xi(X_T)+\int_t^T f(r,X_r,Y_r,Z_r)dr-\int_t^T Z_r dW_r, \quad t\in[0,T]
\label{eq-BSDE-org}
\eea
where $\xi:\mbb{R}^d\rightarrow \mbb{R}$, $f:[0,T]\times \mbb{R}^d\times \mbb{R}\times \mbb{R}^{1\times d}\rightarrow \mbb{R}$
are measurable functions.
\begin{assumption}
\label{assumption-Y}
(i) $f$ satisfies the quadratic structure condition~\cite{Barrieu-ElKaroui}:
\be
|f(t,x,y,z)|\leq l_t+\beta|y|+\frac{\gamma}{2}|z|^2 \nn
\ee
for all $(t,x,y,z)\in[0,T]\times \mbb{R}^d\times \mbb{R}\times \mbb{R}^{1\times d}$,
where $\beta\geq 0, \gamma>0$ are constants, $l:[0,T]\rightarrow  \mbb{R}_+$ is a positive function bounded by 
a constant $K$, i.e. $||l||_T\leq K$. \\
(ii) $f$ satisfies the continuity conditions such that, for all $t,t^\prime \in[0,T]$, $y,y^\prime \in \mbb{R}$, $x,x^\prime \in\mbb{R}^d$, $z,z^\prime \in\mbb{R}^{1\times d}$, 
\bea
|f(t,x,y,z)-f(t^\prime,x,y,z)|&\leq &K|t-t^\prime|^{1/2}~,\nn \\
|f(t,x,y,z)-f(t,x,y^\prime,z)|&\leq&  K|y-y^\prime|~,\nn \\
|f(t,x,y,z)-f(t,x,y,z^\prime)|&\leq & K\bigl(1+|z|+|z^\prime|\bigr)|z-z^\prime|~, \nn \\
|f(t,x,y,z)-f(t,x^\prime,y,z)|&\leq & K\bigl(1+|y|+|z|^2\bigr)|x-x^\prime|. \nn 
\eea
(iii) the driver $f$ is 1-time continuously differentiable with respect to the spatial variables
with continuous derivatives.
In particular, we assume that
\bea
|\part_y f(t,x,y,z)|\leq K,~\quad |\part_z f(t,x,y,z)|\leq K(1+|z|),\quad |\part_x f(t,x,y,z)|\leq K(1+|y|+|z|^2) \nn
\eea
for all $(t,x,y,z)\in[0,T]\times \mbb{R}^d\times \mbb{R}\times \mbb{R}^{1\times d}$.\\
(iv) $\xi$ is a 3-time continuously differentiable function satisfying 
$||\xi(\cdot)||_{\mbb{L}^\infty} \leq K$ and $||\part_x^m \xi(\cdot)||_{\mbb{L}^\infty}\leq K$
for every $1\leq m\leq 3$.\\

\end{assumption}

\begin{remark}
\label{remark-assumption}
The 2nd- and 3rd-order differentiability in Assumptions~\ref{assumption-X} and \ref{assumption-Y} is relevant only for the later 
part of the discussions (Section~\ref{sec-connection}$\sim$), where the error estimate of the 
short-term expansions are required.
\end{remark}

It has been well-known since the work of Kobylanski (2000)~\cite{Kobylanski}
that there exists a unique solution $(Y,Z)$ to (\ref{eq-BSDE-org})
in the space $(Y,Z)\in \cals^\infty\times \calh^2_{BMO}$.
\begin{lemma}(universal bound)
\label{lemma-universal}
Under Assumptions \ref{assumption-X} and \ref{assumption-Y}, the solution $(Y,Z)\in \cals^\infty\times \calh^2_{BMO}$ of (\ref{eq-BSDE-org}) 
satisfies
\bea
&&||Y||_{\cals^\infty}\leq e^{\beta T}\Bigl(||\xi(\cdot)||_{\mbb{L}^\infty}+T||l||_T\Bigr)~,\nn \\
&&||Z||_{\calh^2_{BMO}}^2\leq \frac{e^{4\gamma ||Y||_{\cals^\infty}}}{\gamma^2}
\Bigl(3+6\gamma T(\beta||Y||_{\cals^\infty}+||l||_T)\Bigr)~.\nn
\eea
\begin{proof}
This follows straightforwardly from the quadratic structure condition \cite{Barrieu-ElKaroui}
which is given by Assumption \ref{assumption-Y} (i).
See, for example, Lemma 3.1 and 3.2 in \cite{FT-QJ}.
\end{proof}
\end{lemma}

\section{A sequence of qg-BSDEs perturbed in terminals}
\label{sec-sequence}
In this section, we investigate a sequence of qg-BSDEs.
For each connecting point, we introduce a bounded measurable function $\del^i:\mbb{R}^d\rightarrow \mbb{R}$ as perturbation. 
We then investigate the propagation of its effects to the total error.

\subsection{Setup}
Let us introduce a time partition $\pi:0=t_0<t_1<\cdots<t_n=T$.
We put $h_i:=t_i-t_{i-1}$, $|\pi|:=\max_{1\leq i\leq n}h_i$. 
We denote each interval by $I_i:=[t_{i-1},t_i]$, $i\in\{1,\cdots,n\}$
and assume that there exists some positive constant $C$ such that $|\pi| n\leq C$ as well as $|\pi|/h_i\leq C$
for every $i\in\{1,\cdots,n\}$. 
In order to approximate the BSDE (\ref{eq-BSDE-org}), we 
introduce a sequence of qg-BSDEs perturbed in the terminal values
for each interval $t\in I_i$, $i\in\{1,\cdots,n\}$ in the following way:
\bea
\label{eq-BSDE-barY}
\ol{Y}^i_t=\wh{u}^{i+1}(X_{t_i})+\int_t^{t_i}f(r,X_r,\ol{Y}^i_r,\ol{Z}_r^i)dr-\int_t^{t_i}\ol{Z}_r^i dW_r~,
\eea
where $\wh{u}^{i+1}:\mbb{R}^d\rightarrow \mbb{R}$
with $\wh{u}^{n+1}(x):=\xi(x)$.

Each terminal function $\wh{u}^{i+1}(x)$, $x\in\mbb{R}^d$ of the period $I_i$ is defined by
$\bigl(\ol{Y}^{i+1,t_i,x}_t,t\in [t_i,t_{i+1}]\bigr)$, which is the solution of (\ref{eq-BSDE-barY}) for the period $I_{i+1}$ 
corresponding to the underlying process $X$ with 
the initial data $(t_i,X_{t_i}=x)$~\footnote{
In other words, the underlying forward process is given by $X_s^{t_i,x}=x+\int_{t_i}^s b(r,X_r^{t_i,x}) dr+\int_{t_i}^s \sigma(r,X_r^{t_i,x})
dW_r$, $s\in I_{i+1}$, and hence $\ol{Y}^{i+1,t_i,x}_{t_i}$ is a deterministic function of $x\in \mbb{R}^d$.},
and the additional perturbation term $\del^{i+1}$
\bea
\wh{u}^{i+1}(x)&:=&\ol{Y}^{i+1,t_i,x}_{t_i}-\del^{i+1}(x), \quad i\in\{1,\cdots,n-1\}.
\label{def-delta}
\eea
In the next section, $\{\del^i\}_{i\leq n} $ will be related to the errors from the short-term expansion.
\begin{assumption}\label{assumption-barY}
(i)The perturbation terms $\del^{i+1}:\mbb{R}^d\rightarrow \mbb{R}$, $i\in\{1,\cdots,n\}$ are absolutely bounded 
measurable functions that keep $(\wh{u}^{i+1})_{1\leq i\leq n}$ satisfying the conditions
\bea
(a)&&\max_{1\leq i\leq n}||\wh{u}^{i+1}(\cdot)||_{\mbb{L}^{\infty}}\leq K^\prime,\nn \\
(b)&&\max_{1\leq i\leq n}||\part_x \wh{u}^{i+1}(\cdot)||_{\mbb{L}^\infty}\leq K^\prime, \nn \\
(c)&&\max_{1\leq i \leq n} ||\part_x^m \wh{u}^{i+1}(\cdot)||_{\mbb{L}^{\infty}}\leq K^\prime,~(m\in\{2,3\}) \nn 
\eea 
with some $n$-independent positive constant 
$K^\prime$.\footnote{The exact size of $K^\prime$ is somewhat arbitrary if it is big enough not to 
contradict the true solution of (\ref{eq-BSDE-org}). 
}\\
(ii) There exists an $n$-independent positive constant $C$ such that
$\sum_{i=1}^{n-1}||\del^{i+1}(\cdot)||_{\mbb{L}^\infty}\leq C$.
\end{assumption}
We use the convention $\del^{n+1}\equiv 0$ and $\ol{Y}^{n+1}_{t_n}=\xi(X_{t_n})$ in the following.

\begin{remark}
The condition (i)(c) becomes relevant only for the short-term expansion in the next section.
\end{remark}
\begin{remark}
The classical (as well as variational) differentiability of qg-BSDEs is well-known
by the works of Ankirchner et al. (2007)~\cite{Imkeller-Reis}, Briand \& Confortola (2008)~\cite{Briand-Confortola}
and Imkeller \& Reis (2010)~\cite{Imkeller}. See Fujii \& Takahashi (2017)~\cite{FT-QJ} 
for the extension of these results to qg-BSDEs with Poisson random measures. 
\end{remark}

\subsection{Properties of the solution}
Applying the known results of qg-BSDE for each period, one sees that 
there exists a unique solution $(\ol{Y}^i,\ol{Z}^i)\in \cals^{\infty}_{[t_{i-1},t_i]}\times \calh^2_{BMO[t_{i-1},t_i]}$.
Applying Lemma~\ref{lemma-universal} for each period $I_i$, one also sees
\bea
||\ol{Y}^i||_{\cals^\infty[t_{i-1},t_i]}\leq ||\ol{Y}||_{\cals^\infty}:=e^{\beta |\pi|}\Bigl(K^\prime+|\pi| ||l||_T\Bigr)~,
\label{eq-barY-bound}
\eea
which is bounded uniformly in $i\in\{1,\cdots,n\}$, and so is $||\ol{Z}^i||_{\calh^2_{BMO}[t_{i-1},t_i]}$.

\begin{proposition}\label{prop-barZ-sup}
Under Assumptions~\ref{assumption-X}, \ref{assumption-Y} and Assumption~\ref{assumption-barY}~[i(a,b)], 
there exists some positive $(i,n)$-independent constant $C$  such that
the process $\ol{Z}^i_t, t\in I_i$ of the solution to the BSDE (\ref{eq-BSDE-barY}) satisfies
\be |\ol{Z}^i_t|\leq C(1+|X_t|), ~ t\in I_i, ~\text{$\mbb{P}$-a.s.} \nn \ee
uniformly in $i\in\{1,\cdots,n\}$.
\begin{proof}
We use the representation theorem for the control variable (Theorem 8.5 in \cite{Imkeller-Reis})
and follow the arguments of Theorem 3.1 in Ma \& Zhang (2002)~\cite{Ma-Zhang}.
Let us introduce the parameterized solution $(X^{t,x},\ol{Y}^{i,t,x}, \ol{Z}^{i,t,x})$
with the initial data $(t,x)\in[t_{i-1},t_i]\times \mbb{R}^d$:
\bea
\label{eq-X-tx}
&&X_s^{t,x}=x+\int_t^s b(r,X_r^{t,x})dr+\int_t^s \sigma(r,X_r^{t,x})dW_r, \\
&&\ol{Y}^{i,t,x}_s=\wh{u}^{i+1}(X_{t_i}^{t,x})+\int_s^{t_i}f(r,X_r^{t,x},\ol{Y}^{i,t,x}_r,\ol{Z}_r^{i,t,x})dr
-\int_s^{t_i}\ol{Z}_r^{i,t,x}dW_r,
\label{eq-barY-tx}
\eea
$s\in[t,t_i]$ where the classical differentiability of (\ref{eq-X-tx}) and (\ref{eq-barY-tx}) with respect to the 
position $x$ is known~\cite{Imkeller-Reis}.
The differential processes $(\part_x X^{t,x}, \part_x \ol{Y}^{i,t,x}, \part_x \ol{Z}^{i,t,x})$ are given by
the solutions to the following forward- and backward-SDE:
\bea
&&\part_x X_s^{t,x}=\mbb{I}+\int_t^s \part_x b(r,X_r^{t,x})\part_x X_r^{t,x}dr+\int_t^s \part_x \sigma(r,X_r^{t,x})\part_x X_r^{t,x}
dW_r, \nn \\
&&\part_x \ol{Y}_s^{i,t,x}=\part_x \wh{u}^{i+1}(X_{t_i}^{t,x})\part_x X_{t_i}^{t,x}+
\int_t^{t_i} \Bigl\{\part_x f(r,X_r^{t,x},\ol{\Theta}^{i,t,x}_r)\part_x X_r^{t,x}\nn \\
&&\qquad+\part_\theta f (r,X_r^{t,x},\ol{\Theta}_r^{i,t,x})\part_x \ol{\Theta}^{i,t,x}_r\Bigr\}dr-\int_t^{t_i}
\part_x \ol{Z}_r^{i,t,x}dW_r, 
\label{eq-bar-nabY}
\eea
where $\mbb{I}$ is the $d\times d$ identity matrix and $\part_x \ol{\Theta}^{i,t,x}=(\part_x\ol{Y}^{i,t,x},\part_x \ol{Z}^{i,t,x})$.
Note that $|\part_y f|$ is bounded and $
|\part_z f(r,X_r^{t,x},\ol{\Theta}_r^{i,t,x})|\leq K\bigl(1+|\ol{Z}_r^{i,t,x}|\bigr)$
by Assumption~\ref{assumption-Y} (iii). 
By the facts given in (\ref{eq-barY-bound}) and the remark that follows, one sees
$||\part_z f(\cdot, X^{t,x}_\cdot, \ol{\Theta}^{i,t,x}_\cdot)||_{\calh^2_{BMO[t,t_i]}}\leq C $
with some constant $C$. Thus Corollary 9 in \cite{Briand-Confortola} or Theorem A.1 in \cite{FT-QJ} implies that
the BSDE (\ref{eq-bar-nabY}) has a unique solution satisfying, for any $p\geq 2$,
\bea
&&\hspace{-5mm}\bigl|\bigl|\part_x \ol{\Theta}^{i,t,x}\bigr|\bigr|^p_{\calk^p[t,t_i]} \leq C_{p,\bq}\mbb{E}\Bigl[
|\part_x \wh{u}^{i+1}(X_{t_i}^{t,x})\part_x X_{t_i}^{t,x}|^{p\bq^2}+\Bigl(\int_t^{t_i}|\part_x f(r,X_r^{i,t,x},\ol{\Theta}_r^{i,t,x})\part_x X_r^{i,t,x}|dr\Bigr)^{p\bq^2}\Bigr]^{\frac{1}{\bq^2}}\nn \\
&&\quad \leq C_{p,\bq}\mbb{E}\Bigl[||\part_x X^{t,x}||^{2p\bq^2}_{[t,t_i]}\Bigr]^\frac{1}{2\bq^2}
\Bigl(1+h_i^p\mbb{E}\Bigl[||\ol{Y}^i||_{[t,t_i]}^{2p\bq^2}\Bigr]^\frac{1}{2\bq^2}+\mbb{E}\Bigl[\Bigl(\int_t^{t_i}|\ol{Z}_r^{i,t,x}|^2dr\Bigr)^{2p\bq^2}\Bigr]^\frac{1}{2\bq^2}
\Bigr)
\label{eq-nab-Theta}
\eea
where $\bq$ is a positive  constant satisfying $q_*\leq \bq<\infty$. Here, $q_*=\frac{r^*}{r^*-1}>1$
is the conjugate exponent of $r^*$ the upper bound 
of power with which the Reverse H\"older inequality holds for $\cale(\part_z f*W)$.
We have used  Assumption~\ref{assumption-Y} (iv) and H\"older inequality 
in the last line.

By the standard estimate of SDE~\footnote{See, for example, Appendix A in \cite{FT-AE}.}, 
one can show that $||\part_x X^{t,x}||_{\cals^{2p\bq^2}}\leq C$ with some positive  constant $C$
that is independent of the initial data $(t,x)$.  The boundedness of $\ol{Y}^i$ in (\ref{eq-barY-bound}) and the following remark on $\ol{Z}^i$ together with Lemma~\ref{lemma-energy}
show that the right-hand side of (\ref{eq-nab-Theta}) is bounded by some positive constant.
In particular, one can choose a common constant $C$ for every $i\in\{1,\cdots,n\}$ such that
$|\part_x \ol{Y}^{i,t,x}_t|\leq ||\part_x \ol{Y}^{i,t,x}||_{\cals^p[t,t_i]}\leq C $
uniformly in $(t,x)\in[t_{i-1},t_i]\times \mbb{R}^d$.
By the representation theorem~\cite{Imkeller-Reis, Ma-Zhang}, we have
$\ol{Z}_t^i=\part_x \ol{u}^i(t,X_t)\sigma(t,X_t), ~\forall t\in[t_{i-1},t_i]$, $\mbb{P}$-a.s. 
where the function $\part_x \ol{u}^i:[t_{i-1},t_i]\times \mbb{R}^d\rightarrow \mbb{R}^{1\times d}$ 
is defined by $\part_x \ol{u}^i(t,x):=\part_x \ol{Y}^{i,t,x}_t$.
Now the Lipschitz property of $\sigma$ gives the desired result.
\end{proof}
\end{proposition}

Let us now define a progressively measurable process $\bigl(\ol{Z}_t,t\in[0,T]\bigr)$ by
\be 
\ol{Z}_t:=\sum_{i=1}^n \ol{Z}_t^i \bold{1}_{\{t_{i-1}\leq t<t_i\}}, ~t\in[0,T]~
\label{eq-bar-Z}
\ee
so that 
$\int_0^T |\ol{Z}_t|^2 dt=\sum_{i=1}^n \int_{t_{i-1}}^{t_i}|\ol{Z}_t^i|^2 dt$.

\begin{proposition}\label{prop-barZ-bmo}
Under Assumptions~\ref{assumption-X}, \ref{assumption-Y} and \ref{assumption-barY} [i(a), ii], 
the process $\bigl(\ol{Z}_t,~t\in[0,T]\bigr)$ defined by (\ref{eq-bar-Z}) belongs to $\calh^2_{BMO[0,T]}$
satisfying $||\ol{Z}||_{\calh^2_{BMO}[0,T]}\leq C$ with some $n$-independent positive constant $C$.
\begin{proof}
Applying It\^o formula to $e^{2\gamma \ol{Y}^i}$, one obtains for $t\in I_i$ that
\bea
\int_t^{t_i}e^{2\gamma \ol{Y}^i_r}2\gamma^2 |\ol{Z}_r^i|^2 dr=e^{2\gamma \ol{Y}^i_{t_i}}
-e^{2\gamma \ol{Y}^i_t}+\int_t^{t_i}e^{2\gamma \ol{Y}^i_r} 2\gamma f(r,X_r,\ol{Y}_r^i,\ol{Z}_r^i)dr
-\int_t^{t_i}e^{2\gamma \ol{Y}^i_r}2\gamma \ol{Z}_r^i dW_r~.\nn
\eea
The quadratic structure condition in Assumption~\ref{assumption-Y} (i) gives
\bea
\int_t^{t_i}e^{2\gamma\ol{Y}^i_r}\gamma^2 |\ol{Z}_r^i|^2 dr\leq e^{2\gamma \ol{Y}^i_{t_i}}-e^{2\gamma \ol{Y}^i_t}
+\int_t^{t_i}e^{2\gamma \ol{Y}^i_r}2\gamma\bigl(l_r+\beta|\ol{Y}_r^i|\bigr)dr-
\int_t^{t_i}e^{2\gamma \ol{Y}^i_r}2\gamma \ol{Z}_r^i dW_r\nn ~.
\eea
Since $\ol{Y}^i_{t_i}=\wh{u}^{i+1}(X_{t_i})=\ol{Y}^{i+1}_{t_i}-\del^{i+1}(X_{t_i})$ 
and $||\del^{i+1}(\cdot)||_{\mbb{L}^\infty}\leq C$ uniformly in $i\in\{1,\cdots,n\}$, 
one has
$ e^{2\gamma{\ol{Y}^i_{t_i}}}\leq e^{2\gamma\ol{Y}^{i+1}_{t_i}}+C e^{2\gamma \ol{Y}^{i+1}_{t_i}}
|\del^{i+1}(X_{t_i})|$  
with some positive constant $C$.
 It follows that, with the choice $t=t_{i-1}$,
\bea
&&\int_{t_{i-1}}^{t_i}e^{2\gamma \ol{Y}^i_r}\gamma^2 |\ol{Z}_r^i|^2 dr\leq 
\bigl(e^{2\gamma \ol{Y}^{i+1}_{t_i}}-e^{2\gamma\ol{Y}^i_{t_{i-1}}}\bigr)+Ce^{2\gamma \ol{Y}^{i+1}_{t_i}}|\del^{i+1}(X_{t_i})|\nn \\
&&\qquad\quad+\int_{t_{i-1}}^{t_i} e^{2\gamma \ol{Y}^i_r}2\gamma \bigl(l_r+\beta|\ol{Y}_r^i|\bigr)dr-
\int_{t_{i-1}}^{t_i}e^{2\gamma \ol{Y}^i_r}2\gamma \ol{Z}_r^i dW_r.\nn
\eea
Thus for any $\tau\in\calt^T_0$ and $j:=\min\bigl(j\in\{1,\cdots,n\}: \tau\leq t_j\bigr)$,
\bea
&&\hspace{-10mm}\int_\tau^{t_j}e^{2\gamma \ol{Y}^j_r}\gamma^2 |\ol{Z}^j_r|^2 dr+\sum_{i=j+1}^n\int_{t_{i-1}}^{t_i}
e^{2\gamma \ol{Y}^i_r}\gamma^2 |\ol{Z}_r^i|^2 dr \leq e^{2\gamma \ol{Y}^{n+1}_{t_n}}-e^{2\gamma \ol{Y}^j_\tau}+C\sum_{i=j}^ne^{2\gamma\ol{Y}_{t_i}^{i+1}}
|\del^{i+1}(X_{t_i})|\nn \\
&&\hspace{35mm}+\int_\tau^{t_j}e^{2\gamma \ol{Y}^j_r}2\gamma \bigl(l_r+\beta|\ol{Y}^j_r|\bigr)dr+
\sum_{i=j+1}^n \int_{t_{i-1}}^{t_i}e^{2\gamma \ol{Y}^i_r}2\gamma \bigl(l_r+\beta|\ol{Y}^i_r|\bigr)dr\nn \\
&&\hspace{35mm}-\int_\tau^{t_j}e^{2\gamma \ol{Y}^j_r}2\gamma \ol{Z}_r^j dW_r
-\sum_{i=j+1}^n \int_{t_{i-1}}^{t_i} e^{2\gamma \ol{Y}^i_r}2\gamma \ol{Z}^i_r dW_r\nn~.
\eea
Since $e^{2\gamma \ol{Y}^j_\tau}>0$ and $\del^{n+1}\equiv 0$, one obtains
\bea
&&\mbb{E}\Bigl[\int_\tau^{t_j}e^{2\gamma \ol{Y}^j_r}\gamma^2 |\ol{Z}_r^j|^2 dr+
\sum_{i=j+1}^n \int_{t_{i-1}}^{t_i}e^{2\gamma \ol{Y}^i_r}\gamma^2 |\ol{Z}_r^i|^2dr \Bigr|\calf_\tau \Bigr]\nn \\
&&\qquad\leq
\mbb{E}\Bigl[e^{2\gamma \xi(X_T)}+C\sum_{i=j}^{n-1}e^{2\gamma \ol{Y}^{i+1}_{t_i}}|\del^{i+1}(X_{t_i})|
+\sum_{i=j}^n \int_{t_{i-1}\vee\tau}^{t_i} e^{2\gamma \ol{Y}^i_r} 2\gamma \bigl(l_r+\beta|\ol{Y}^i_r|\bigr)dr
\Bigr|\calf_\tau \Bigr]~.\nn
\eea
By Assumption~\ref{assumption-barY} $[i(a),ii]$ and (\ref{eq-barY-bound}), the above inequality implies that
there exists some $n$-independent constant $C$ such that
\bea
\mbb{E}\Bigl[\int_\tau^T |\ol{Z}_r|^2 dr\Bigr|\calf_\tau\Bigr]
\leq \frac{e^{4\gamma ||\ol{Y}||_{\cals^\infty}}}{\gamma^2}
\Bigl( 1+2\gamma T\bigl(||l||_T+\beta||\ol{Y}||_{\cals^\infty}\bigr)+C\sum_{i=1}^{n-1}||\del^{i+1}(\cdot)||_{\mbb{L}^{\infty}}\Bigr)\leq C~, \nn
\eea
and thus the claim is proved.
\end{proof}
\end{proposition}

\subsection{Error estimates for the perturbed BSDEs in the terminals}
Let $(Y,Z)$ be the solution to the BSDE (\ref{eq-BSDE-org})
and $(\ol{Y}^i,\ol{Z}^i)$ to (\ref{eq-BSDE-barY}).
Let us put
\bea
\del Y^i_t:=Y_t-\ol{Y}^i_t, \quad \del Z_t^i:=Z_t-\ol{Z}_t^i,\quad \del f^i(t):=f(t,X_t,Y_t,Z_t)-f(t,X_t,\ol{Y}_t^i,\ol{Z}_t^i)~, \nn
\eea
for $t\in I_i$, $i\in\{1,\cdots,n\}$. Then $(\del Y^i,\del Z^i)$ follows
the BSDE
\bea
\del Y^i_t=\del Y^{i+1}_{t_i}+\del^{i+1}(X_{t_i})+\int_t^{t_i} \del f^i(r)dr-\int_t^{t_i}\del Z_r^i dW_r~,
\label{eq-BSDE-delY}
\eea
for $t\in I_i$. Our first main result is given as follows.

\begin{theorem}
\label{th-delY-delZ}
Under Assumptions~\ref{assumption-X}, \ref{assumption-Y} and \ref{assumption-barY}~[i(a,b),ii],  there exist some $n$-independent positive constants $\bq>1$ and $C_{p,\bq}$ such that, for any $p>1$,
\bea
\max_{1\leq i\leq n}\mbb{E}\left[\sup_{r\in I_i}|\del Y_r^i|^{2p}\right]^\frac{1}{p}
+\sum_{i=1}^n \int_{t_{i-1}}^{t_i}\mbb{E}|\del Z_r^i|^2 dr
\leq \frac{C_{p,\bq}}{|\pi|}\mbb{E}\Bigl[\Bigl(\sum_{i=1}^{n-1}|\del^{i+1}(X_{t_i})|^2\Bigr)^{p\bq}\Bigr]^{\frac{1}{p\bq}}~\nn.
\eea
\begin{proof}
It directly follows from the next Propositions~\ref{prop-delZ} and \ref{prop-delY}.
\end{proof}
\end{theorem}
\begin{remark}
\label{remark-del-order}
Theorem~\ref{th-delY-delZ} implies that we need $\mbb{E}|\del|^2\propto n^{-k}$ with $k>2$ for the right-hand side to converge.
We shall see in fact that $k=3$ is achieved by the short-term expansion. 
\end{remark}
\begin{proposition}
\label{prop-delZ}
Under Assumptions~\ref{assumption-X}, \ref{assumption-Y} and Assumption \ref{assumption-barY} [i(a,b)], the inequality 
\bea
\sum_{i=1}^n \int_{t_{i-1}}^{t_i}\mbb{E}|\del Z_r^i|^2 dr\leq C_p \max_{1\leq i\leq n}
\mbb{E}\Bigl[\sup_{r\in I_i}|\del Y^i_r|^{2p}\Bigr]^{1/p}+\frac{C}{|\pi|}
\sum_{i=1}^{n-1} \mbb{E}|\del^{i+1}(X_{t_i})|^2\nn~,
\eea
holds for any $p>1$ with some $n$-independent positive constants $C$ and $C_p$.
\begin{proof}
For each interval $I_i$, let us define new progressively measurable processes
$\bigl(\beta^i_r, r\in I_i\bigr)$ and $\bigl(\gamma^i_r, r\in I_i\bigr)$ as follows:
\bea
\hspace{-5mm} \beta_r^i:=\frac{f(r,X_r,Y_r,Z_r)-f(r,X_r,\ol{Y}_r^i,Z_r)}{\del Y^i_r}\bold{1}_{\del Y^i_r\neq 0},~~\gamma^i_r:=\frac{f(r,X_r,\ol{Y}^i_r,Z_r)-f(r,X_r,\ol{Y}_r^i,\ol{Z}_r^i)}{|\del Z_r^i|^2}
\bold{1}_{\del Z_r^i\neq 0}(\del Z_r^i)^\top \nn~.
\eea
Then, $|\beta^i|\leq K$ is a bounded process by the Lipschitz property,  and by Proposition~\ref{prop-barZ-sup}, 
there exists some $(i,n)$-independent positive constant $C$ such that
\be
|\gamma_r^i|\leq K(1+|Z_r|+|\ol{Z}_r^i|)\leq C(1+|X_r|), ~\forall r\in I_i,~\mbb{P}\text{-a.s.} 
\label{eq-gamma-sup}
\ee
with $i\in \{1,\cdots,n\}$.
The BSDE (\ref{eq-BSDE-delY}) can now be written as
\bea
\del Y^i_t=\del Y^{i+1}_{t_i}+\del^{i+1}(X_{t_i})+\int_t^{t_i}\Bigl(\beta_r^i \del Y_r^i+\del Z_r^i \gamma_r^i\Bigr)dr
-\int_t^{t_i} \del Z_r^i dW_r~.
\label{eq-BSDE-delY-decomp}
\eea
A simple application of It\^o formula gives
\bea
&&\mbb{E}|\del Y^i_{t_{i-1}}|^2+\int_{t_{i-1}}^{t_i}\mbb{E}|\del Z_r^i|^2 dr=\mbb{E}\bigl| \del Y^{i+1}_{t_i}+\del^{i+1}(X_{t_i})\bigr|^2+\int_{t_{i-1}}^{t_i}\mbb{E}
\Bigl[2\del Y^i_r\bigl(\beta_r^i \del Y_r^i+\del Z_r^i\gamma_r^i\bigr)\Bigr]dr~. \nn
\eea
By H\"older inequality and (\ref{eq-gamma-sup}), one obtains with some positive constants $C, C_p$ that
\bea
&&\frac{1}{2}\int_{t_{i-1}}^{t_i}\mbb{E}|\del Z_r^i|^2 dr\leq
\Bigl(\mbb{E}|\del Y^{i+1}_{t_i}|^2-\mbb{E}|\del Y^i_{t_{i-1}}|^2\Bigr)+C|\pi|\mbb{E}|\del Y^{i+1}_{t_i}|^2+\frac{C}{|\pi|}
\mbb{E}|\del^{i+1}(X_{t_i})|^2  \nn \\
&&\hspace{30mm}+C\int_{t_{i-1}}^{t_i}\mbb{E}\Bigl[|\del Y^i_r|^2 (1+|\gamma_r^i|^2)\Bigr]dr\nn \\
&&\leq \Bigl(\mbb{E}|\del Y^{i+1}_{t_i}|^2-\mbb{E}|\del Y^i_{t_{i-1}}|^2\Bigr)+C|\pi|\mbb{E}|\del Y^{i+1}_{t_i}|^2+\frac{C}{|\pi|}
\mbb{E}|\del^{i+1}(X_{t_i})|^2+C_p|\pi|\mbb{E}\Bigl[\sup_{r\in I_i}|\del Y^i_r|^{2p}\Bigr]^{1/p} \nn~,
\eea
where $p$ is an arbitrary constant satisfying $p>1$.
Summing up for $i\in\{1,\cdots,n\}$, one obtains
\bea
\frac{1}{2}\sum_{i=1}^n \int_{t_{i-1}}^{t_i}\mbb{E}|\del Z_r^i|^2 dr &\leq& \mbb{E}|\del Y^{n+1}_{t_n}|^2-\mbb{E}|\del Y^1_{t_0}|^2+C|\pi|\sum_{i=1}^n \mbb{E}|\del Y^{i+1}_{t_i}|^2\nn \\
&+&\frac{C}{|\pi|}\sum_{i=1}^n \mbb{E}|\del^{i+1}(X_{t_i})|^2+C_p|\pi|\sum_{i=1}^n 
\mbb{E}\Bigl[\sup_{r\in I_i} |\del Y_r^i|^{2p}\Bigr]^{1/p}\nn~.
\eea
Since $\del Y^{n+1}_{t_n}=\del^{n+1}=0$, one gets the desired result.
\end{proof}
\end{proposition}

\begin{proposition}
\label{prop-delY}
Under Assumptions~\ref{assumption-X}, \ref{assumption-Y} and \ref{assumption-barY} [i(a),ii], there exists some $n$-independent positive constants $\bq>1$ and $C_{p,\bq}$ such that, for any $p>1$,
\bea
\mbb{E}\Bigl[\max_{1\leq i\leq n}\sup_{r\in I_i}|\del Y^i_r|^p\Bigr]
\leq C_{p,\bq} \mbb{E}\Bigl[\Bigl(\sum_{i=1}^{n-1}|\del^{i+1}(X_{t_i})|\Bigr)^{p\bq}\Bigr]^{1/\bq}~.\nn
\eea
\begin{proof}
Let us use the same notations $\beta^i_r, \gamma^i_r$ defined in Proposition~\ref{prop-delZ}.
We also introduce the process $(\ol{\gamma}_r, r\in[0,T])$ by
$\ol{\gamma}_r:=\sum_{i=1}^n \gamma_r^i \bold{1}_{\{t_{i-1}\leq r<t_i\}}$. 
With $\ol{Z}$ defined by (\ref{eq-bar-Z}), it satisfies
$|\ol{\gamma}_r|\leq K(1+|Z_r|+|\ol{Z}_r|)$.
By Lemma~\ref{lemma-universal} and Proposition~\ref{prop-barZ-bmo}, both  $Z$ and $\ol{Z}$ are in $\calh^2_{BMO}$,
and so is $\ol{\gamma}$. In particular, $||\ol{\gamma}||_{\calh^2_{BMO}}\leq C$
by some $n$-independent constant.

From the remark following Definition~\ref{def-h2bmo}, one can show that $\ol{\gamma}*W \in BMO(\mbb{P})$.
Thus the new probability measure $\mbb{Q}$ can be defined by $d\mbb{Q}/d\mbb{P}=\cale_T$,
where $\cale$ is a Dol\'eans-Dade exponential $\cale_t:=\cale\bigl(\int_0^t \ol{\gamma}_r^\top dW_r\bigr)$.
The Brownian motion $W^{\mbb{Q}}$ under the measure $\mbb{Q}$ is
given by $W^{\mbb{Q}}_t=W_t-\int_0^t\ol{\gamma}_r dr$
for $t\in [0,T]$. Furthermore, it follows from Lemma~\ref{lemma-reverse} that there exists a constant $r^*$ satisfying $1<r^*<\infty$ such that, for every
$1<\br\leq r^*$, the reverse H\"older inequality of power $\br$ holds:
$\bigl(1/\cale_\tau\bigr)\mbb{E}\bigl[\cale_T^{\br} |\calf_\tau\bigr]^{1/\br} \leq C_{\br}$.
Here, $\tau\in\calt^T_0$ is an arbitrary $\mbb{F}$-stopping time,  
$C_{\br}$ is some positive constant depending only on $\br$ and $||\ol{\gamma}||_{\calh^2_{BMO}}$.
We put $\bq>1$ as the conjugate exponent of this $\br$ in the following.
By the last observation, all of these constants can be chosen independently of $n$.

Under the new measure $\mbb{Q}$,  the BSDE (\ref{eq-BSDE-delY-decomp}) is given by
\bea
\del Y^i_t=\del Y^{i+1}_{t_i}+\del^{i+1}(X_{t_i})+\int_t^{t_i}\beta_r^i \del Y^i_r dr-\int_t^{t_i}\del Z_r^i dW_r^{\mbb{Q}}~, \nn
\eea
which can be solved as
$\del Y^i_t=\mbb{E}^{\mbb{Q}}\bigl[e^{\int_t^{t_i}\beta_r^i dr}\bigl(\del Y^{i+1}_{t_i}+\del^{i+1}(X_{t_i})\bigr)\bigr|\calf_t\bigr]$
for all $t\in I_i$.
Since $|\beta^i|\leq K$, one obtains
$|\del Y^i_t|\leq e^{K h_i}\mbb{E}^{\mbb{Q}}\bigl[|\del Y^{i+1}_{t_i}|+|\del^{i+1}(X_{t_i})|\bigr|\calf_t\bigr]$.
It then follows by iteration
\bea
|\del Y^i_t|\leq \mbb{E}^{\mbb{Q}}\Bigl[ e^{K\sum_{j=i}^n h_j}|\del Y^{n+1}_{t_n}|
+\sum_{j=i}^n e^{K\sum_{k=i}^j h_k}|\del^{j+1}(X_{t_j})|\Bigr|\calf_t\Bigr]\nn~.
\eea
Since $\del Y^{n+1}_{t_n}=\del^{n+1}(X_{t_n})=0$, one concludes
$|\del Y^i_t|\leq \mbb{E}^{\mbb{Q}}\bigl[\sum_{j=i}^{n-1}e^{K\sum_{k=i}^jh_k}|\del^{j+1}(X_{t_j})| \bigr|\calf_t\bigr]$
for $t\in I_i$, $i\in\{1,\cdots,n\}$.  
The reverse H\"older inequality gives
\bea
|\del Y^i_t|&\leq& e^{KT}\mbb{E}^{\mbb{Q}}\Bigl[\sum_{j=i}^{n-1}|\del^{j+1}(X_{t_j})|\Bigr|\calf_t\Bigr]
=\frac{e^{KT}}{\cale_t}\mbb{E}\Bigl[\cale_T \sum_{j=i}^{n-1}|\del^{j+1}(X_{t_j})|\Bigr|\calf_t\Bigr]\nn \\
&\leq & C_{\bq}e^{KT}\mbb{E}\Bigl[\Bigl(\sum_{j=i}^{n-1}|\del^{j+1}(X_{t_j})|\Bigr)^{\bq}\Bigr|\calf_t \Bigr]^{1/\bq}\nn,
\eea
which then yields
\bea
\max_{1\leq i\leq n}\sup_{t\in I_i}|\del Y^i_t|\leq C_\bq \sup_{t\in[0,T]}\mbb{E}\Bigl[
\Bigl(\sum_{i=1}^{n-1}|\del^{i+1}(X_{t_i})|\Bigr)^{\bq}\Bigr|\calf_t\Bigr]^{1/\bq}\nn.
\eea
Using Jensen and Doob's maximal inequalities, one finally obtains
\bea
&&\mbb{E}\Bigl[\max_{1\leq i\leq n}\sup_{t\in I_i}|\del Y^i_t|^p\Bigr]\leq
C_{p,\bq}\mbb{E}\Bigl[\sup_{t\in[0,T]}\mbb{E}\Bigl[
\Bigl(\sum_{i=1}^{n-1}|\del^{i+1}(X_{t_i})|\Bigr)^{\bq}\Bigr|\calf_t\Bigr]^{p/\bq}\Bigr]\nn \\
&&\qquad\leq C_{p,\bq}\mbb{E}\Bigl[\sup_{t\in[0,T]}\mbb{E}\Bigl[
\Bigl(\sum_{i=1}^{n-1}|\del^{i+1}(X_{t_i})|\Bigr)^{\bq}\Bigr|\calf_t\Bigr]^{p}\Bigr]^{1/\bq} \leq C_{p,\bq}\mbb{E}\Bigl[\Bigl(\sum_{i=1}^{n-1}|\del^{i+1}(X_{t_i})|\Bigr)^{p\bq} \Bigr]^{1/\bq} \nn
\eea
which proves the claim.
\end{proof}
\end{proposition}

\section{Connecting the sequence of qg-BSDEs}
\label{sec-connection}
\subsection{Short-term expansion of a qg-BSDE}
We now give an analytic approximate solution  $(\ol{Y}^i,\ol{Z}^i)$ 
of the BSDE (\ref{eq-BSDE-barY}) as a short-term expansion $(\wh{Y}^i, \wh{Z}^i)$.
We need two steps involving the linearization method as well as the small-variance
expansion method for BSDEs proposed in Fujii \& Takahashi (2012)~\cite{FT-analytic} and (2015)~\cite{FT-AE},
respectively. 
We set aside technical details  until  Appendices~\ref{sec-short1} and \ref{sec-short2} so that we can focus on the main story.

We obtain the approximated solution $(\wh{Y}^i,\wh{Z}^i)$ as
\bea
\label{eq-YZhat-i}
\wh{Y}^i_t:=\wh{Y}^{i,[0]}_t+\wh{Y}^{i,[1]}_t,\quad \wh{Z}^i_t:=\wh{Z}^{i,[0]}_t, 
\eea
for every interval $t\in I_i,i\in\{1,\cdots,n\}$, for which the exact expressions can be read from 
(\ref{eq-Yhat-i0}), (\ref{eq-Zhat-i0}), (\ref{eq-Yhat-i1}) and (\ref{eq-Zhat-i1}).
For numerical implementation, the values at each connecting point $\{t_i\}$ are the most relevant;
Under the condition $X_{t_{i-1}}=x, x\in \mbb{R}^d$, the 
approximate solutions 
$\wh{Y}^{i,t_{i-1},x}_{t_{i-1}}:=\wh{Y}^{i}_{t_{i-1}}\bigr|_{X_{t_{i-1}}=x}, \quad \wh{Z}^{i,t_{i-1},x}_{t_{i-1}}:=\wh{Z}^{i}_{t_{i-1}}\bigr|_{X_{t_{i-1}}=x} $
are given by the following simple explicit formulas;
\bea
\label{eq-Yhat-tim1}
\wh{Y}^{i,t_{i-1},x}_{t_{i-1}}&=&\ol{y}(t_{i-1},x)+\frac{1}{2}\ol{y}^{[2]}_0(t_{i-1},x) \nn\\
&&+h_i f\Bigl(t_{i-1},x,\ol{y}(t_{i-1},x)+\frac{1}{2}\ol{y}^{[2]}_0(t_{i-1},x), \ol{\bold{y}}^{[1]\top}(t_{i-1},x)\sigma(t_{i-1},x)
\Bigr)~,  \\
\label{eq-Zhat-tim1}
\wh{Z}^{i,t_{i-1},x}_{t_{i-1}}&=&\ol{\bold{y}}^{[1]\top}(t_{i-1},x)\sigma(t_{i-1},x)~, \\
\text{where}&&
\begin{cases}
\ol{\chi}(t_i,x)=x+h_i b\bigl(t_{i-1},x\bigr)~, \nn \\ 
\ol{y}(t_{i-1},x)=\wh{u}^{i+1}\bigl(\ol{\chi}(t_i,x)\bigr)~,\nn \\
\ol{\bold{y}}^{[1]}(t_{i-1},x)=\Bigl( \mbb{I}+h_i\bigl[\part_x b(t_i,\ol{\chi}(t_i,x))\bigr]\Bigr)\part_x \wh{u}^{i+1}
(\ol{\chi}(t_i,x))~,\nn \\
\ol{y}^{[2]}_0(t_{i-1},x)=h_i{\rm Tr}\Bigl(\part^2_{x,x}\wh{u}^{i+1}(\ol{\chi}(t_i,x))\bigl[\sigma\sigma^\top\bigr]
(t_i,\ol{\chi}(t_i,x))\Bigr)~, \nn \\
\end{cases}
\eea
where $\mbb{I}$  denotes $d\times d$-identity matrix.
The main result regarding the error estimate for the short-term approximation is 
given by Theorem~\ref{th-barY-hatY}.
We emphasize that the theorem is interesting in its own sake. It provides the short-term asymptotic expansion of a $qg$-BSDE
explicitly in the strong sense.

\subsection{Connecting procedures}
We now connect these approximate solutions by the following scheme.
\begin{definition}(Connecting Scheme)\label{def-connection}\\
(i) Setting $\wh{u}^{n+1}(x)=\xi(x)$, $x\in \mbb{R}^d$. \\
(ii) Repeating from $i=n$ to $i=1$ that \\
(a) Calculate the short-term approximation of the BSDE (\ref{eq-BSDE-barY})
by using (\ref{eq-Yhat-tim1}) \\ 
and store the 
values $\bigl\{\wh{Y}^{i,t_{i-1},x^\prime}_{t_{i-1}}\bigr\}_{x^\prime \in B_i}$ for a finite subset $B_i$ of  $\mbb{R}^d$.\\
(b) Define the terminal function $\wh{u}^i(x),x\in \mbb{R}^d$ for the next period $I_{i-1}$ by
\bea
\wh{u}^{i}(x):={\rm Interpolation}\bigl(\bigl\{\wh{Y}^{i,t_{i-1},x^\prime}_{t_{i-1}}\bigr\}_{x^\prime\in B_i} \bigr)(x)\nn
\eea
where ``Interpolation" stands for some smooth interpolating function satisfying
the bounds in Assumption~\ref{assumption-barY} (i)~.
\end{definition}

From the definition of $\del^{i}$ in (\ref{def-delta}), we have
\bea
&&\del^{i}(x)=\ol{Y}^{i,t_{i-1},x}_{t_{i-1}}-\wh{u}^{i}(x)=\del^i_{SE}(x)+\calr^i(x), \nn \\
\label{eq-delSE-i}
{\text where} &&\qquad \del^i_{SE}(x):=\Bigl(\ol{Y}^{i,t_{i-1},x}_{t_{i-1}}-\wh{Y}^{i,t_{i-1},x}_{t_{i-1}}\Bigr)~, \\
&&\qquad \calr^i(x):=\Bigl(\wh{Y}^{i,t_{i-1},x}_{t_{i-1}}-\wh{u}^i(x)\Bigr)~.
\label{eq-calr-i}
\eea
Here, $\del_{SE}^i$ denotes the error of the short-term approximation (see Theorem~\ref{th-barY-hatY}), and
$\calr^i$ the interpolation error as well as the regularization effects rendering the 
approximated function $\wh{Y}^{i,t_{i-1},x}_{t_{i-1}}$ into the bounds satisfying Assumption~\ref{assumption-barY} (i).

\begin{remark}
Despite the similarity in its appearance, it differs from four-step-scheme (Ma et.al.(1994)~\cite{Ma-Protter-Yong}),
its extensions such as \cite{DM06, DM08}, and other PDE discretization approaches.
They typically require differentiability in time $t$, 
the uniform ellipticity of $\sigma\sigma^\top$ and the global Lipschitz continuity of the driver.
In particular, we do not know any literature
treating Markovian BSDEs with drivers of quadratic growth in the control variables
with this type of methods.
\end{remark}

\subsection{Total error estimate}
\begin{lemma}
\label{lemma-qgY-conti}
Under Assumptions~\ref{assumption-X} and \ref{assumption-Y}, the solution $Y_t,t\in[0,T]$ of the BSDE (\ref{eq-BSDE-org}) satisfies the continuity property
$ \mbb{E}\bigl[\sup_{s\leq u\leq t} \bigl|Y_u-Y_s\bigr|^p\bigr]\leq C_p \bigl|t-s\bigr|^{p/2} $
for any $0\leq s\leq t\leq T$ and $p\geq 2$ with some positive constant $C_p$.
\begin{proof}
Using the Burkholder-Davis-Gundy inequality and Assumption~\ref{assumption-Y} (i), one obtains
\bea
&&\mbb{E}\Bigl[\sup_{s\leq u\leq t}|Y_u-Y_s|^p\Bigr]\leq C_p \mbb{E}\Bigl[
\Bigl(\int_s^t |f(r,X_r,Y_r,Z_r)|dr\Bigr)^p+\Bigl(\int_s^t |Z_r|^2 dr\Bigr)^{p/2}\Bigr]\nn \\
&&\qquad \leq C_p \mbb{E}\Bigl[\Bigl(\int_s^t\bigl[l_r+\beta |Y_r|+\frac{\gamma}{2}|Z_r|^2\bigr]dr\Bigr)^p+
\Bigl(\int_s^t |Z_r|^2 dr\Bigr)^{p/2}\Bigr]\nn~.
\eea
Since $l, |Y|$ are bounded and $|Z_t|\leq C(1+|X_t|)$ with a constant $C$, the claim is proved.
\end{proof}
\end{lemma}
Let us now give the main result of the paper:
\begin{theorem}
\label{th-main}
Define the piecewise constant process $(Y^\pi_t,Z^\pi_t),t\in[0,T]$ by
\bea
Y_t^\pi:=\wh{u}^i(X_{t_{i-1}}), \quad Z_t^\pi:=\ol{\bold{y}}^{[1]\top}(t_{i-1},X_{t_{i-1}})\sigma(t_{i-1},X_{t_{i-1}})~, \nn
\eea
for $t_{i-1}\leq t <t_i$, $i\in\{1,\cdots,n\}$ and $Y^\pi_{t_n}=\xi(X_{t_n}),~Z_{t_n}^\pi=0$,
where the $\wh{u}^i$ and $\ol{\bold{y}}^{[1]}$ are those determined by the 
connecting scheme in Definition~\ref{def-connection}.
Then, under Assumptions~\ref{assumption-X}, \ref{assumption-Y} and \ref{assumption-barY}, 
there exist some $n$-independent positive constants $\bq>1$ and $C_{p,\bq}$ such that
\bea
&&\max_{1\leq i\leq n}\mbb{E}\left[\bigl|\bigl|Y-Y^\pi\bigr|\bigr|_{[t_{i-1},t_i]}^{2p}\right]^\frac{1}{2p}
+\Bigl(\sum_{i=1}^n \int_{t_{i-1}}^{t_i}\mbb{E}\Bigl[\bigl|Z_t-Z_t^\pi\bigr|^2\Bigr]dt\Bigr)^{1/2} \nn \\
&&\quad \leq C_{p,\bq}\sqrt{|\pi|}+C_{p,\bq}\sqrt{n}\mbb{E}\Bigl[
\Bigl(\sum_{i=1}^n \bigl|\calr^{i}(X_{t_{i-1}})\bigr|^2\Bigr)^{p\bq}\Bigr]^\frac{1}{2p\bq}~\nn
\eea
holds for any $p>1$.
\begin{proof}
One obtains, by simple manipulation, that
\bea
&&\max_{1\leq i\leq n}\mbb{E}\left[\bigl|\bigl|Y-Y^\pi\bigr|\bigr|_{[t_{i-1},t_i]}^{2p}\right]^{1/p}+\sum_{i=1}^n \int_{t_{i-1}}^{t_i}
\mbb{E}\bigl|Z_t-Z_t^\pi\bigr|^2 dt\nn \\
&&\qquad \leq C_p\Bigl(
\max_{1\leq i\leq n}\mbb{E}\Bigl[\bigl|Y_{t_{i-1}}-\ol{Y}^i_{t_{i-1}}\bigr|^{2p}\Bigr]^{1/p}+
\sum_{i=1}^n \int_{t_{i-1}}^{t_i}\mbb{E}\bigl|Z_t-\ol{Z}_t^i\bigr|^2 dt\Bigr)\nn \\
&&\qquad+C_p\Bigl(\max_{1\leq i\leq n}\mbb{E}\Bigl[\bigl|\ol{Y}^{i}_{t_{i-1}}-\wh{u}^i(X_{t_{i-1}})\bigr|^{2p}\Bigr]^{1/p}
+\sum_{i=1}^n \int_{t_{i-1}}^{t_i}\mbb{E}\bigl|\ol{Z}_t^i-\wh{Z}_t^i\bigr|^2 dt\Bigr)\nn \\
&&\qquad+C_p\Bigl(\max_{1\leq i\leq n}\mbb{E}\Bigl[\sup_{t\in I_i}\bigl|Y_t-Y_{t_{i-1}}\bigr|^{2p}\Bigr]^{1/p}+
\sum_{i=1}^n \int_{t_{i-1}}^{t_i}\mbb{E}\bigl|\wh{Z}_t^i-\wh{Z}_{t_{i-1}}^i\bigr|^2 dt
\Bigr)\nn~.
\eea
It follows that, by applying Theorems~\ref{th-delY-delZ}, \ref{th-barY-hatY}, 
Lemmas~\ref{lemma-qgY-conti}, \ref{lemma-continuity-Yhat},  and expressions (\ref{eq-delSE-i}), (\ref{eq-calr-i}),
\bea
&&\max_{1\leq i\leq n}\mbb{E}\Bigl[\bigl|\bigl|Y-Y^\pi\bigr|\bigr|_{[t_{i-1},t_i]}^{2p}\Bigr]^{1/p}+\sum_{i=1}^n \int_{t_{i-1}}^{t_i}
\mbb{E}|Z_t-Z_t^\pi|^2 dt \nn \\
&&\quad \leq \frac{C_{p,\bq}}{|\pi|}\mbb{E}\Bigl[\Bigl(\sum_{i=1}^{n-1}|\del^{i+1}(X_{t_i})|^2\Bigr)^{p\bq}\Bigr]^\frac{1}{p\bq}
+C_p\Bigl(\max_{1\leq i\leq n}\mbb{E}\Bigl[|\del^i(X_{t_{i-1}})|^{2p}\Bigr]^{1/p}+\sum_{i=1}^n h_i^3\Bigr)
+C_p|\pi| \nn \\
&&\quad\leq C_p|\pi|+\frac{C_{p,\bq}}{|\pi|}
\Bigl\{ n^{p\bq-1}\sum_{i=1}^n \mbb{E}\Bigl[\bigl|\del^i_{SE}(X_{t_{i-1}})\bigr|^{2p\bq}\Bigr]+
\mbb{E}\Bigl[\Bigl(\sum_{i=1}^n|\calr^i(X_{t_{i-1}})|^2\Bigr)^{p\bq}\Bigr]
\Bigr\}^\frac{1}{p\bq}\nn \\
&&\quad \leq C_{p,\bq}|\pi|+C_{p,\bq}n 
\mbb{E}\Bigl[\Bigl(\sum_{i=1}^n|\calr^i(X_{t_{i-1}})|^2\Bigr)^{p\bq}\Bigr]^\frac{1}{p\bq}\nn~,
\eea
which proves the desired result.
\end{proof}
\end{theorem}

\section{An example of implementation}
\label{sec-FD}
\subsection{Finite-difference scheme}
The remaining problem for us is to find a concrete method of constructing the smooth bounded functions $(\wh{u}^{i})_{1\leq i\leq n+1}$
used in the step (ii) of Definition~\ref{def-connection}. It is important to notice that there is no need to specify $\wh{u}^i(x)$
for the whole space $x\in \mbb{R}^d$ but only for those used in the interpolation.
Based on this observation, we consider a finite-difference scheme as a method of {\it non-parametric} coarse graining.
Let us suppose every $B_i$ a grid-cube in $\mbb{R}^d$ centered at the origin, equally spaced by the size $\Del$.
For a function $v:\mbb{R}^d\rightarrow \mbb{R}$, we denote the 1st-order central difference as 
$\part_\Del v:\mbb{R}^d\rightarrow \mbb{R}^d$;
$(\part_\Del v(x))_j:=(v(x+\Del~e_j)-v(x-\Del~e_j))/(2\Del),~j=\{1,\cdots, d\}$
where $e_j$ is the unit vector of direction $j$. Note that,  when $v\in C^3$, 
$|\part_x v-\part_\Del v|\propto \Del^2$.
For higher-order differences, we use a non-central scheme
denoted similarly as $\part_\Del^m v$, $m={2,\cdots}$. 
Let us arrange a sequence of grids $B_i\subseteq B_{i+1}$ satisfying 
\be
\label{eq-B-size}
M/2 \leq |B_1|\leq |B_{n+1}| \leq M, 
\ee
where $M$ is some sufficiently large constant and $|B_i|$ the length of the cube's edge.
Let us write $x\in \ol{B_i}$ when $x$ is inside the cube but not necessary on the grid point
and denote the boundary of the cube by $\part \ol{B_i}$. 
\begin{lemma}
\label{lemma-uhat-approx}
Suppose $\wh{u}^{i}\in C_b^{\infty}\cap \mbb{L}^\infty$ and $dist(y,\part \ol{B_i})\leq C^\prime \Del$ with some positive constant $C^\prime$.
Then, denoting $x\in B_i$ as the nearest grid point to $y$, there exists a constant $C$ such that
\bea
&&\bigl|\wh{u}^{i}(y)-[\wh{u}^{i}(y)]\bigl|\leq C\Del^3,~
\bigl|\part_x \wh{u}^{i}(y)-[\part_x \wh{u}^{i}(y)]\bigr|\leq C\Del^2,\nn \\
&&\bigl|\part_x^2 \wh{u}^{i}(y)-[\part_x^2 \wh{u}^{i}(y)]\bigr|\leq C\Del~,\nn 
\eea
\bea
\text{where}&&\begin{cases}
[\wh{u}^{i}(y)]:=\wh{u}^{i}(x)+\part_x \wh{u}^{i}(x)\cdot (y-x)+\frac{(y-x)^\top}{2}\part_{x}^2\wh{u}^i(x)\cdot (y-x) \\
[\part_x \wh{u}^{i}(y)]:=\part_x \wh{u}^i(x)+\part_{x}^2 \wh{u}^i(x)\cdot (y-x) \\ 
[\part_{x}^2 \wh{u}^i(y)]:=\part_x^2 \wh{u}^i(x) 
\end{cases}.\nn
\eea
\begin{proof}
It is obvious from the Taylor formula.
\end{proof}
\end{lemma}

Let us scale the size of the grid-cube according to the number of discretization as
\be
\label{eq-M-scaling}
M=Cn^{\del/2}
\ee
with a given constant $C$ and an arbitrary $\del\in (0,1/\rho)$,
where $\rho\geq 2$ is some integer to be specified later. (See (\ref{AE-conditional}) and the following discussion.)
In this case, the quadratic-growth term $h_i|x|^2$ existing in $\wh{Y}^{i,t_{i-1},x}_{t_{i-1}}$ 
of (\ref{eq-Yhat-tim1}) is bounded by $Cn^{-1+\del}$ within the cube.
We now modify the connecting scheme of Definition~\ref{def-connection} as follows.
\begin{definition}\label{def-connection-FD}(Connecting scheme with finite-difference)\\
(0) Set a scaling rule $\Del=\zeta |\pi|^{\nu}$ with constants $\zeta,\nu>0$, 
and then construct a sequence of grid-cubes $B_1\subseteq B_{2}\subseteq\cdots\subseteq B_{n+1}\subseteq \mbb{R}^d$
satisfying (\ref{eq-B-size}) and (\ref{eq-M-scaling}).\\
(i) Suppose $\wh{u}^{i+1}$ is in the class $C_b^{\infty}\cap\mbb{L}^\infty$ and the values of 
$\{\wh{u}^{i+1}(x),\part_x \wh{u}^{i+1}(x), \part_x^2 \wh{u}^{i+1}(x) ~| x\in B_{i+1}\}$ are known.~\footnote{In this scheme, $\wh{u}^{n+1}\in C_b^{\infty}$ is a bounded function constructed as in step (iv) using $\wh{Y}^{n+1,t_n,x}_{t_n}
=\xi(x)$.} \\
(ii) Store the values of  $\{[\wh{Y}^{i,t_{i-1},x}_{t_{i-1}}],x\in B_i\}$
where $[\wh{Y}^{i,t_{i-1},x}_{t_{i-1}}]$ is equal to $\wh{Y}^{i,t_{i-1},x}_{t_{i-1}}$ calculated by (\ref{eq-Yhat-tim1}) 
with $\wh{u}^{i+1}(\ol{\chi}(t_i,x))$ and its derivatives  replaced by their approximations as in 
Lemma~\ref{lemma-uhat-approx}~\footnote{If $b$ is $x$-independent, the adjustment of Lemma~\ref{lemma-uhat-approx}
is unnecessary by shifting the center of the grids. If $b$ is proportional to $x$, $log$-transformation may be 
used to make the drift $x$-independent once again.}.\\
(iii) Store the values of 
$\{\part_\Del[\wh{Y}^{i,t_{i-1},x}_{t_{i-1}}], \part_\Del^2[\wh{Y}^{i,t_{i-1},x}_{t_{i-1}}]~|x\in B_i\}$.\footnote{
If $x$ is at the edge of the grid $B_i$, apply a non-central difference scheme.}\\
(iv) Consider $\wh{u}^i$ as an {\it appropriate} 
$C_b^\infty\cap \mbb{L}^\infty$-class function (see Remark~\ref{remark-candidate}) satisfying  
\be
\Bigl(\wh{u}^{i}(x), \part_x \wh{u}^{i}(x), \part_x^2 \wh{u}^{i}(x)\Bigr)\simeq \Bigl([\wh{Y}^{i,t_{i-1},x}_{t_{i-1}}],
\part_\Del[\wh{Y}^{i,t_{i-1},x}_{t_{i-1}}], \part_\Del^2[\wh{Y}^{i,t_{i-1},x}_{t_{i-1}}]\Bigr), \quad x\in B_i~. \nn
\ee
where $\simeq$ is the approximate equality with the size of error bounded respectively by
$(C \Del ^3, C \Del ^2, C \Del )$ with some constant $C$, and smoothly tracks $\ol{Y}^{i,t_{i-1},x}_{t_{i-1}}$ outside the grid-cube
while keeping the same order of accuracy as inside i.e., $\sup_{x\notin \ol{B_i}}|\ol{Y}^{i,t_{i-1},x}_{t_{i-1}}-\wh{u}^i(x)|
\leq C\sup_{x\in \ol{B_i}}|\ol{Y}^{i,t_{i-1},x}_{t_{i-1}}-\wh{u}^i(x)|$. 
\end{definition}

\begin{remark}
\label{remark-candidate}
The existence of functions $\wh{u}^{i}\in C_b^\infty\cap \mbb{L}^\infty$ satisfying (iv) of the above scheme
can be easily seen. Suppose $v:\mbb{R}^d\rightarrow \mbb{R}$ is an arbitrary smooth and bounded function 
satisfying $\max_{x\in B_i}|v(x)-[\wh{Y}^{i,t_{i-1},x}_{t_{i-1}}]|\leq C\Del^3$.
Then, by construction, $(\part_\Del v, \part_\Del^2 v)$ is equal to $(\part_\Del[\wh{Y}^{i,t_{i-1},x}_{t_{i-1}}],
\part_\Del^2[\wh{Y}^{i,t_{i-1},x}_{t_{i-1}}])$ with the desired accuracy.
Since $(\part_\Del v, \part_\Del^2 v)$ is equal to the true derivatives with the accuracy $(C\Del^2, C\Del)$
with $C=||\part_x^3 v||_\infty$, $v$ is in fact a valid candidate for $\wh{u}^i$.
Although there is no need to single out $\wh{u}^i$, one can just suppose that a function
with the smallest total variation is chosen among the candidates in order to avoid unnecessary oscillations
between the neighboring grid points. Adjusting $v$ outside the cube so that it smoothly 
tracks $\ol{Y}^{i,t_{i-1},x}_{t_{i-1}}$ is always possible. 
\end{remark}

\begin{remark}
One can see from (\ref{eq-Yhat-tim1}) and Definition~\ref{def-connection-FD}
that if $\Del\downarrow 0$ faster than $|\pi|$, the higher-order derivatives diverge due to the fact that $f\in C^1$.
On the other hand, we need at least $\nu>1/3$ for the sum of $C\Del^3$ to converge.
Interestingly, for the numerical example given in the next section, we find that the stability (i) is achieved 
with $\nu=1/2$, i.e., the scaling $\Del=\zeta |\pi|^{1/2}$ with the coefficient $\zeta$ of the order of $X$'s volatility.
The scaling rule also suggests a connection to the stability problem of parabolic PDEs with explicit finite-difference scheme,
but we need further research to understand whether proving  Assumption~\ref{assumption-barY} is possible or not
for the current scheme in the limit $n\rightarrow \infty$ with some $\nu>1/3$.
\end{remark}

\subsection{Error estimate}
Unfortunately, we cannot prove Assumption~\ref{assumption-barY} for the above scheme  in the limit $n\rightarrow \infty$
because of the non-linearity from the driver. 
Therefore, in the following analysis, we are forced to restrict our attention to the 
approximation within a finite range of $n$.
In this case, Assumption~\ref{assumption-barY} becomes trivial by simply taking the maximum for a given range
and all the estimates including Theorem~\ref{th-main} can be used.
However, this does not tell if the approximation is improved by using a finer discretization.
We need to confirm that the bounds of Assumption~\ref{assumption-barY}
remain stable within the relevant range of $n$.
\begin{assumption}
\label{assumption-posteriori}(A posteriori check)\\
The values
\bea
\max_{1\leq i\leq n}\max_{x\in B_i^\prime }\Bigl\{|\wh{u}^i(x)|, |\part_x \wh{u}^i(x)|, |\part_x^2 \wh{u}^i(x)|, |\part_x^3 \wh{u}^i(x)|\Bigr\}\nn
\eea 
are confirmed, a posteriori,  to be stable for a given range of discretization, say,  $n\in[n_0,n_1]$.
Here $B_i^\prime ~(\supset B_i)$ denotes a slightly bigger grid-cube $|B_i^\prime|\geq |B_i|+\lambda ||\sigma_i|| \sqrt{h_i}$
with the same spacing $\Delta$, where $\lambda>1$ is 
some constant and $||\sigma_i||$ the maximum size of the volatility of $X$ 
$\bigl(i.e., ||\sigma(\cdot,x)||_{[t_{i-1},t_i]}\bigr)$ for $x$  
near the boundary $\part\ol{B_i}$.\footnote{ This means that the information from $B_i^\prime-B_i$ is only used 
to guarantee the error estimate. }
\end{assumption}

\begin{remark}
\label{remark-outside}
Note that Assumption~\ref{assumption-posteriori} ensures the uniform boundedness of $|\part_x^m \wh{u}^i(\cdot)|_{0\leq m\leq 3}$
within $(\lambda)$-sigma range of the paths $(X_t, t\in I_i)$ provided $X_{t_{i-1}}\in \ol{B}_i$. 
Now consider the conditional version of Theorem~\ref{th-barY-hatY} for the period $I_i=[t_{i-i},t_i]$ given $X_{t_{i-1}}\in \ol{B_i}$. 
Since the contribution from the paths outside the $(\lambda)$-sigma range
is suppressed by an arbitrary power of $\lambda$, one can choose its size
so that the outside paths do not alter the error estimate for a give range of $n$.
In practice, $\lambda\simeq 5$ would be large enough. 
\end{remark}

From the above Remark~\ref{remark-outside} and Theorem~\ref{th-barY-hatY}, there exists a constant $C$ such that
\bea
|\ol{Y}^{i,t_{i-1},x}_{t_{i-1}}-\wh{Y}^{i,t_{i-1},x}_{t_{i-1}}|&\leq& \mbb{E}\Bigl[
\sup_{t\in I_i}|\ol{Y}^{i,t_{i-1},x}_{t}-\wh{Y}^{i,t_{i-1},x}_{t}|^p\Bigr|X_{t_{i-1}}=x\Bigr]^\frac{1}{p} \nn \\
&\leq& C (1+|x|^{\rho}) h_i^{3/2}\leq C M^\rho h_i^{3/2}, \quad \forall x\in \ol{B_i} 
\label{AE-conditional}
\eea
for a given range of $n\in[n_0,n_1]$, where $\rho$ is some positive integer.
Checking carefully the initial-value dependence in the short term expansion, 
one can actually find that $\rho=3$ in the current setup. However, the exact value of $\rho$
does not affect the following estimates and leaving $\rho$ as a general integer $\rho\geq 2$
is useful for later discussions.

The following result shows that Assumption~\ref{assumption-barY}(ii) is guaranteed for a finite range of $n$ 
by a posteriori check in Assumption~\ref{assumption-posteriori}.
\begin{proposition}
\label{prop-sumdel-bound}
Suppose Assumptions~\ref{assumption-X}, \ref{assumption-Y} and also Assumption~\ref{assumption-posteriori}
with the scaling rule $\Del=\zeta|\pi|^{1/2}$, then there exists a  constant $C$  such that
$\sum_{i=1}^n||\del^{i+1}||_{\mbb{L}^\infty}\leq C$  uniformly for $n\in[n_0,n_1]$.  
In particular $(||\wh{u}^i||_{\mbb{L}^\infty})_{1\leq i\leq n}$  are also uniformly bounded.
\begin{proof}
By definition, we have
\bea
\sum_{i=1}^n||\del^{i+1}||_{\mbb{L}^\infty}\leq
\sum_{i=2}^{n+1}\Bigl\{\sup_{x\in \ol{B_i}}\bigl(|\ol{Y}^{i,t_{i-1},x}_{t_{i-1}}-\wh{Y}^{i,t_{i-1},x}_{t_{i-1}}|
+|\wh{Y}^{i,t_{i-1},x}_{t_{i-1}}-\wh{u}^i(x)|\bigr)+\sup_{x\notin \ol{B_i}}|\ol{Y}^{i,t_{i-1},x}_{t_{i-1}}-\wh{u}^i(x)|\Bigr\}.\nn
\eea
From (\ref{AE-conditional}), one has
\be
\sup_{x\in\ol{B}_i}|\ol{Y}^{i,t_{i-1},x}_{t_{i-1}}-\wh{Y}^{i,t_{i-1},x}_{t_{i-1}}|\leq 
C M^\rho |\pi|^{3/2}\leq Cn^{(\rho\del -3)/2} \nn
\ee
with some constant $C$ uniformly. Next, by construction,  $|\wh{Y}^{i,t_{i-1},x}_{t_{i-1}}-\wh{u}^i(x)|\leq C\Del^3$, $|\part_\Del \wh{Y}^{i,t_{i-1},x}_{t_{i-1}}-\part_x\wh{u}^{i}(x)|\leq C\Del^2$ and
$|\part_\Del^2 \wh{Y}^{i,t_{i-1},x}_{t_{i-1}}-\part_x^2\wh{u}^{i}(x)|\leq C\Del$ at each grid point $x\in B_i$.
The function $\wh{u}^i$ within the grid-cube $\ol{B_i}$ can be separated into two parts; one is the regularization of the first term of 
(\ref{eq-Yhat-tim1}), and the other is the remaining two terms proportional to $h_i$.
Since the first term of (\ref{eq-Yhat-tim1}) is confirmed to have bounded derivatives up to the third order, the difference of the second (first) order derivatives
is bounded by $C\Del$ ($C \Del^2)$ between any interval of the grid points.
For the latter, since they are in class-$C^1$,  proportional to $h$ and with at most quadratic growth in $x$, 
the difference of the first order derivatives 
is bounded by $C(1+|x|^2)h\leq C M^\rho h$. Combining these two, one sees that
\be
|\wh{Y}^{i,t_{i-1},x}_{t_{i-1}}-\wh{u}^i(x)|\leq C\Del^3+C(\Del^2+M^\rho h)\Del \leq Cn^{(\rho\del-3)/2} \nn
\ee
for the whole interval between the two 
neighboring grid points.\footnote{Although one may only have $|\part_\Del \wh{Y}^{i,t_{i-1},x}_{t_{i-1}}-\part_x \wh{u}^i(x)|\leq C\Del$
when $x\in \part \ol{B_i}$ and the direction of the derivative is orthogonal to the boundary
(since the central difference cannot be taken for $\part_\Del[\wh{Y}^{i,t_{i-1},x}_{t_{i-1}}]$
in the step (iii) of Definition~\ref{def-connection-FD}),  one obtains the same conclusion by 
estimating from the neighboring internal point.}
By repeating the same arguments, one sees
$\max_{x\in \ol{B_i}}|\wh{Y}^{i,t_{i-1},x}_{t_{i-1}}-\wh{u}^i(x)|\leq C n^{(\rho\del-3)/2}$.
Combining these two results, one has
\be
\sup_{x\in \ol{B}_i}\Bigl(|\ol{Y}^{i,t_{i-1},x}_{t_{i-1}}-\wh{Y}^{i,t_{i-1},x}_{t_{i-1}}|+|\wh{Y}^{i,t_{i-1},x}_{t_{i-1}}-\wh{u}^i(x)|
\Bigr)\leq C n^{(\rho \del -3)/2}.\nn
\ee

Since $\wh{u}^i(x)$ outside the cube is constructed to follow $\ol{Y}^{i,t_{i-1},x}_{t_{i-1}}\in C_b^1\cap \mbb{L}^\infty$ 
so that it keeps the same order of accuracy, one also has 
$\sup_{x\notin \ol{B_i}}|\ol{Y}^{i,t_{i-1},x}_{t_{i-1}}-\wh{u}^i(x)|\leq C n^{(\rho\del-3)/2}$.
Thus $\sum_{i=1}^n||\del^{i+1}||_{\mbb{L}^\infty}\leq C n^{(\rho\del-1)/2}$ and the first claim is proved.
Using the universal bound iteratively, one also has
\bea
||\wh{u}^i||_{\mbb{L}^\infty}\leq e^{\beta |\pi|}\Bigl(||\wh{u}^{i+1}||_{\mbb{L}^\infty}+|\pi|||l||_T\Bigr)+Cn^{(\rho\del-3)/2}
\leq e^{\beta T}\Bigl(||\xi||_{\mbb{L}^\infty}+T \bigl(||l||_T+Cn^{(\rho \del-1)/2}\bigr)\Bigr)\nn
\eea
which yields the second.
\end{proof}
\end{proposition}

\begin{proposition}
\label{prop-R-fd}
Suppose Assumptions~\ref{assumption-X}, \ref{assumption-Y} and also Assumption~\ref{assumption-posteriori}
with the scaling rule $\Del=\zeta|\pi|^{1/2}$.
Then there exist some constant $C$ satisfying 
\be
\mbb{E}\Bigl[\Bigl(\sum_{i=1}^{n+1}|\calr^i(X_{t_{i-1}})|^2\Bigr)^p\Bigr]^{\frac{1}{2p}}
\leq C n^{-1}
\ee
uniformly for $n\in[n_0,n_1]$, for every $p>1$.
\begin{proof}
Using $|\wh{Y}^{i,t_{i-1},x}_{t_{i-1}}-\wh{u}^i(x)|\leq C (1+|x|^\rho)n^{-3/2},~\forall x\in\ol{B}_i$ given in (\ref{AE-conditional}),
$||\wh{u}^i||_{\mbb{L}^\infty}\leq C$ shown in Proposition~\ref{prop-sumdel-bound}
and the quadratic-growth property of $\wh{Y}^{i,t_{i-1},x}_{t_{i-1}}$ in $x$, one obtains that
\bea
&&\mbb{E}\Bigl[\Bigl(\sum_{i=1}^{n+1}|\calr^i(X_{t_{i-1}})|^2\Bigr)^p\Bigr]^{\frac{1}{2p}} \leq |n+1|^\frac{1}{2}\max_{1\leq i\leq n+1} \mbb{E}\Bigl[\Bigl|\wh{Y}^{i,t_{i-1},X_{t_{i-1}}}_{t_{i-1}}-\wh{u}^i(X_{t_{i-1}})
\Bigr|^{2p}\Bigr]^\frac{1}{2p}\nn \\
&&\leq Cn^{-1}\max_{1\leq i\leq n+1}\mbb{E}\Bigl[(1+||X||_{I_{i}}^\rho)^{2p}\Bigr]^\frac{1}{2p}
+C\sqrt{n}\max_{1\leq i\leq n+1}\mbb{E}\Bigl[(1+||X||_{I_{i}}^\rho)^{2p}\Bigl(\frac{||X||_{I_i}}{M/2}\Bigr)^{4pk}\Bigr]^\frac{1}{2p}\nn \\
&&\leq Cn^{-1}+C_{p,k}n^{-k\del+1/2}. \nn
\eea
Since $k>1$ is arbitrary, one obtains the desired result.
\end{proof}
\end{proposition}

Let us define the approximate piecewise constant processes $(Y_t^{FD},Z_t^{FD}), t\in[0,T]$ by
using the bounded function $(\wh{u}^i)_{1\leq i\leq n+1}$ constructed as in Definition~\ref{def-connection-FD} as
\bea
\label{eq-yz-fd}
&&Y_t^{FD}:=\wh{u}^i(x), \quad Z_t^{FD}:=\ol{\bold{y}}^{[1]\top}(t_{i-1},x)\sigma(t_{i-1},x) \nn
\eea
for $t\in[t_{i-1},t_i), i\in\{1,\cdots,n+1\}$. $x$ is a $B_i$-valued $\calf_{t_{i-1}}$-measurable r.v.
nearest to $X_{t_{i-1}}$.
\begin{corollary}
\label{coro-fd-final}
Suppose Assumptions~\ref{assumption-X}, \ref{assumption-Y} and also Assumption \ref{assumption-posteriori}
with the scaling rule $\Del=\zeta|\pi|^{1/2}$.
Then  there exist positive  constants $\bq>1$ and $C_{p,\bq}$ such that
\bea
&&\max_{1\leq i\leq n}\mbb{E}\Bigl[||Y-Y^{FD}||_{[t_{i-1},t_i]}^{2p}\Bigr]^\frac{1}{2p}
+\Bigl(\sum_{i=1}^n \int_{t_{i-1}}^{t_i}\mbb{E}\Bigl[|Z_t-Z_t^{FD}|^2\Bigr]dt\Bigr)^{1/2} \leq C_{p,\bar{q}} n^{-1/2}\nn
\eea
for $\forall p>1$ uniformly for $n\in[n_0,n_1]$.
\begin{proof}
For $t\in I_i$ and $x\in B_i$, the nearest grid point to $X_{t_{i-1}}$, one has
\bea
|Y_t^\pi-Y_t^{FD}|&\leq& |\wh{u}^i(X_{t_{i-1}})-\wh{u}^i(x)|\bigl(\bold{1}_{\{X_{t_{i-1}}\in \ol{B}_i\}}
+\bold{1}_{\{X_{t_{i-1}}\notin \ol{B}_i\}})\nn \\
 &\leq& C\Del +2||\wh{u}^i||_{\mbb{L}^\infty}\bold{1}_{\{X_{t_{i-1}}\notin \ol{B}_i\}} \nn 
\eea
and similarly 
\bea
|Z_t^\pi-Z_t^{FD}|&\leq & |\ol{\bold{y}}^{[1]}(t_{i-1},X_{t_{i-1}})
-\ol{\bold{y}}^{[1]}(t_{i-1},x)||\sigma(t_{i-1},X_{t_{i-1}})|\nn \\
&&+|\ol{\bold{y}}^{[1]}(t_{i-1},x)||\sigma(t_{i-1},x)-\sigma(t_{i-1},X_{t_{i-1}})|\nn \\
&\leq &C|X_{t_{i-1}}-x|(1+|X_{t_{i-1}}|)\nn \\
&\leq &C\Del(1+|X_{t_{i-1}}|)\bold{1}_{\{X_{t_{i-1}}\in \ol{B}_i\}}
+C(1+|X_{t_{i-1}}|^2)\bold{1}_{\{X_{t_{i-1}}\notin \ol{B}_i\}}\nn~.
\eea
Since Chebyshev's inequality applied for the set $X_{t_{i-1}}\notin \ol{B}_i$ gives a multiplicative factor of $M^{-k}\propto n^{-k\del/2}$
for any $k\geq 1$, one gets
\bea
\mbb{E}\Bigl[||Y^\pi-Y^{FD}||^{2p}_{I_i}\Bigr]^\frac{1}{2p}+
\mbb{E}\Bigl[||Z^\pi-Z^{FD}||^{2p}_{I_i}\Bigr]^\frac{1}{2p}\leq C n^{-1/2}~.
\label{eq-pi-FD}
\eea
Thus the  conclusion follows from Propositions~\ref{prop-sumdel-bound}, \ref{prop-R-fd} and Theorem~\ref{th-main}.
\end{proof}
\end{corollary}

\subsection{Application of sparse grids}
In order to mitigate the so called {\it curse of dimensionality},
there exists a very interesting result on high dimensional polynomial interpolation using
sparse grids.
By Theorem 8 (as well as Remark 9) of Barthelmann et al.~\cite{Novak-Ritter}, 
it is known that there exists an
interpolating function satisfying the following
uniform estimates on the compact set for a function  $f:\mbb{R}^d\rightarrow \mbb{R}$ in the class $C^k$;
\bea
\sup_{|x|\leq M}\Bigl|f(x)-\cala^{q,d}\bigl(f(x)\bigr)\Bigl|\leq C_{q,d} N^{-k}_{(q,d)}(\log(N_{(q,d)}))^{(k+1)(d-1)}~.
\label{eq-sparse}
\eea
Here, $\cala^{q,d}(f):\mbb{R}^d\rightarrow\mbb{R}$ is an 
interpolating  polynomial function  of degree $q ~(\geq d)$  based on 
the Smolyak algorithm. 
The interpolating function is uniquely determined
by the values of $f(x_i), x_i\in H(q,d)$ where $H(q,d)$ is the sparse grid 
whose number of nodes is give by $N_{(q,d)}$.  $C_{q,d}$ is some positive constant depending only on $(q,d)$
and $\sup_{|x|\leq M}|\part_x^m f|$ of $m=\{0,\cdots,k\}$.
The sparse grid $H(q,d)$ is the set of points on which the Chebyshev polynomials take the extrema.
For details, see \cite{Novak-Ritter, Sauer} and references therein.
The sparse grid method looks very attractive since (\ref{eq-sparse})  has only weak $\log$ dependency 
on the dimension $d$. 

For our purpose, we want to interpolate $f(x)=\wh{Y}^{i,t_{i-1},x}_{t_{i-1}}$ of (\ref{eq-Yhat-tim1}) efficiently
so that the right-hand side of (\ref{eq-sparse}) is of the order of $C\Del^3$.
Consider the interpolation for the first and the remaining two terms of (\ref{eq-Yhat-tim1}),  separately.
The former has $k=3$, and the latter has only $k=1$ but it is proportional to $h_i\propto \Del^2$.
Therefore, for interpolation of $\wh{Y}^{i,t_{i-1},x}_{t_{i-1}}$, the number of nodes $N_{(q,d)}\propto \Del^\varep~(\propto n^{\varep/2}), \varep>1$ 
can maintain the same error estimate of Corollary~\ref{coro-fd-final}.

\section{Numerical examples}
In the remainder of the paper, we demonstrate our computation scheme and its empirical
convergence rate using illustrative models.
For simplicity, we use a full grid (instead of a sparse grid) at each time step
with the scaling rule $\Del=\zeta |\pi|^{1/2}$. 
As existing literature, we focus on approximating the initial value of the BSDE $(Y_0=Y_0^{0,x_0})$
and thus restrict the computations only to the relevant grid points.
More extensive tests on higher dimensional setups
with sparse grids will be left for the future works\footnote{For example, see \cite{Ma-Zabaras,GZhang}.}.

\subsection{A solvable qg-BSDE}
Let us first  consider the following model with $d=2$ similar to those studied in \cite{Richou}:
\bea
\label{eq-X-num}
X_t&=&x_0+\int_0^t \begin{pmatrix} b_1 X_s^1 \\ b_2 X_s^2 \end{pmatrix} ds+\int_0^t \begin{pmatrix}
\sigma_1 X_s^1 & 0 \\
0 & \sigma_2 X_s^2 \end{pmatrix} 
\begin{pmatrix}
1 & 0 \\
\rho & \sqrt{1-\rho^2}
\end{pmatrix} dW_s~, \\
Y_t&=&\xi(X_T)+\int_t^T \frac{a}{2}|Z_s|^2 ds-\int_t^T Z_s dW_s~, 
\label{eq-qgBSDE-num}
\eea
where $b_i,\sigma_i$, $i\in\{1,2\}$, $\rho\in[-1,1]$ and $a$ are all constants.
For this example, by using a exponential transformation  $\bigl(e^{a Y_t}, t\in[0,T]\bigr)$,
we obtain a closed form solution:
\bea
Y_t=\frac{1}{a}\log\bigl(\mbb{E}\bigl[\exp\bigl(a \xi(X_T)\bigr)\bigr|\calf_t\bigr]\bigr)~,
\label{eq-closed}
\eea
whose expectation can be obtained semi-analytically by integrating over the density of $X$.
We use
\be
 \xi(x)=3 \bigl(\sin^2(x^\bbf{1})+\sin^2(x^\bbf{2})\bigr)
\label{eq-terminal-num}
\ee
as the terminal value function, and set $x_0=(1,1)^\top$, $T=1$, $b_1=b_2=0.05$, $\rho=0.3$.
\begin{figure}[H]
\vspace{-3mm}
\begin{center}	
\includegraphics[width=95mm]{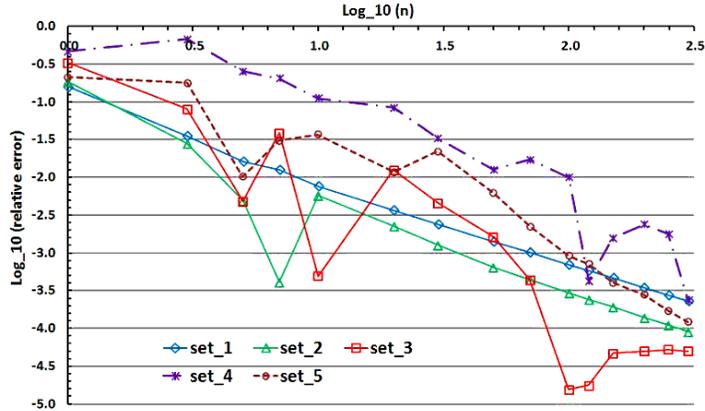}
\end{center}
\vspace{-7mm}
\caption{\small Empirical convergence of the proposed scheme for (\ref{eq-qgBSDE-num}) with ${\rm set}_i, i\in\{1,\cdots,5\}$.}
\label{Fig-qgBSDE}
\end{figure}
We have tested the following five sets of parameters $(\sigma_{i=1,2}, a)$:
\bea
&&{\rm set}_1=\bigl\{\sigma_i=0.5, a=1.0\bigr\},~{\rm set}_2=\bigl\{\sigma_i=0.5, a=2.0\bigr\},~{\rm set}_3=\bigl\{\sigma_i=0.5, a=3.0\bigr\},\nn \\
&&{\rm set}_4=\bigl\{\sigma_i=1.0, a=3.0\bigr\},~{\rm set}_5=\bigl\{\sigma_i=0.5, a=4.0\bigr\} 
\label{paramset-qg}
\eea
by changing the partition from $n=1$ to $n=300$.
In Figure~\ref{Fig-qgBSDE},  we have plotted $\log_{10}$(relative error)
against the $\log_{10}(n)$ for set$_i,i\in\{1,\cdots,5\}$, where the relative error 
is defined by
\be
\frac{\mbox{estimated $Y_0$ by the proposed scheme} -\mbox{the value obtained from (\ref{eq-closed})}
}{\mbox{the value obtained from (\ref{eq-closed})}}\nn~.
\ee
As naturally expected, the bigger ``$a$" we use, the bigger $\zeta$ is needed to keep the derivatives (calculated by finite-difference scheme) non-divergent as Assumption~\ref{assumption-posteriori} requires.

\subsubsection*{Driver truncation}
As we have emphasized, it is crucial to have stable derivatives as Assumption~\ref{assumption-posteriori}
for the proposed scheme to converge.
For the qg-BSDE (\ref{eq-qgBSDE-num}), if we increase the coefficient ``$a$" 
while keeping the factor $\zeta$ of $\Del=\zeta|\pi|^{1/2}$ constant, 
we have observed that these derivatives (and hence the estimate of $Y$) are, in fact, divergent.
In the remainder, instead of making $\zeta$ larger, let us study the truncation of the driver $f$ so that it has
a global Lipschitz constant $N$ following the scaling rule  ( see Section 2.1 of \cite{Richou} )
\be
N \propto n^{\al},\quad 0<\al<1~.
\label{eq-driver-truncation}
\ee
The error estimates for the qg-BSDEs under this truncation have been studied by 
Imkeller \& Reis (2010)~\cite{Imkeller} (Theorem 6.2) and applied to 
the backward numerical scheme by Chassagneux \& Richou (2016)~\cite{Richou}.
From Theorem 6.2~\cite{Imkeller}, one easily sees that this truncation 
does not affect the theoretical bound on the convergence rate of Theorem~\ref{th-main},
which is  also the case for the scheme studied in \cite{Richou}.

\begin{figure}[H]
\vspace{-3mm}
\begin{center}	
\includegraphics[width=100mm]{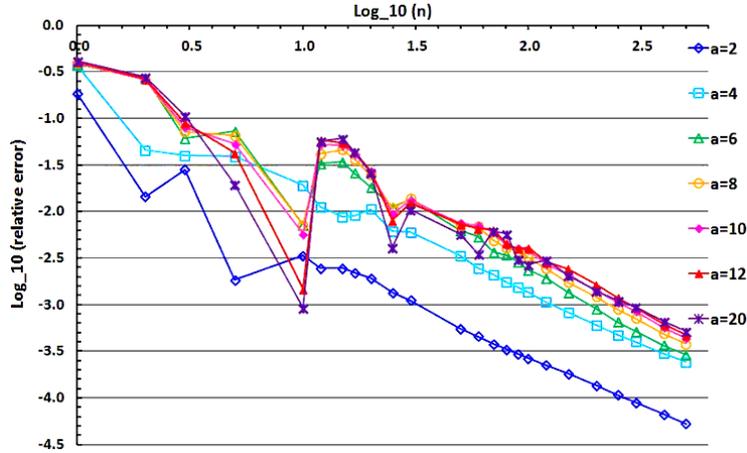}
\end{center}
\vspace{-7mm}
\caption{\small Empirical convergence of the proposed scheme for (\ref{eq-qgBSDE-num}) with a truncated driver so that the Lipschitz constant scales as $N\propto n^{1/3}$.}
\label{Fig-truncation}
\end{figure}

We have chosen the constant $\zeta$  
so that it marginally works for the $\text{set}_3$ in (\ref{paramset-qg}) without any truncation 
and adopted the scaling factor $\alpha=1/3$.
We tested the following seven cases of large quadratic coefficients; 
\be 
\label{eq-a-sets}
\{a=2,~a=4,~ a=6,~ a=8,~a=10,~a=12,~ a=20\} 
\ee
while keeping the other parameters the same, i.e.,
\be
x_0=(1,1)^\top, ~T=1, ~b_1=b_2=0.05, ~\rho=0.3, ~\sigma_1=\sigma_2=0.5~.
\label{eq-default}
\ee

In Figure~\ref{Fig-truncation}, we have plotted the $\log_{10}$(relative error) against the $\log_{10} (n)$
changing the number of partitions from $n=1$ to $n=500$.
Except for coarse partitions $n\lesssim 10$,  the truncation of the driver yields
quite stable convergence even for very large quadratic coefficients.
We find no significant change in the empirical convergence rate, and it is close to one.
The introduction of the truncation $(\ref{eq-driver-truncation})$ looks quite attractive since there is no need to adjust $\zeta$ 
according to different size of the coefficient $a$. There seems a deep relation among the stability (Assumption~\ref{assumption-posteriori}),
the scaling rule of finite difference scheme as well as the truncation of the driver $(N\propto n^\al)$.
This interesting problem requires further research.

\begin{remark}
Since the truncation introduced in Imkeller \& Reis~\cite{Imkeller} leaves the structure condition (Assumption~\ref{assumption-Y} (i)) intact,
one can still use the error estimates derived for qg-BSDEs. From Theorem 6.2~\cite{Imkeller}, one can see that the difference of the original qg-BSDE and its truncated version by the Lipschitz constant $N$ $(\del Y:=Y-Y^N, \del Z^N:=Z-Z^N)$ 
scales as $||(\del Y^N, \del Z^N)||_{\calk^p}\propto N^{-\beta}=n^{-\alpha \beta }$ with an arbitrary  $\beta>0$.
Hence, it does not affect the estimate of the convergence speed.
On the other hand,  one cannot improve the convergence analysis
by adopting the simpler proof of the Lipschitz BSDEs (see Section~\ref{sec-Lip-th}) combined with the truncation method.
This is because the coefficients of the error estimates generally depend on the Lipschitz constant exponentially 
as $C\propto e^{N T}=e^{n^\alpha T}$ through the use of the Gronwall's lemma.
\end{remark}
\subsubsection*{Non-differentiable terminal function}
We now test a case of non-differentiable terminal function 
\be
\xi(x)=\min\bigl(\max(x^{\bf{1}},1),3\bigr)+\max(2-x^{\bf{2}},0)~
\label{eq-non-d-terminal}
\ee
in (\ref{eq-qgBSDE-num}). 
If we apply the finite-difference scheme given in the last section directly, 
the 2nd and 3rd-order derivatives appearing in Assumption~\ref{assumption-posteriori} 
grow with the rate of $1/\Del$ and $1/\Del^2$, at least,  
near the boundary $t=T$.
Thus one naturally expects some instability appears if the space discretization $\Del=\zeta |\pi|^{1/2}$
becomes sufficiently small. 
Note that, one can always apply the technique of Section~\ref{sec-FD}
as long as an appropriate mollified function is chosen and kept fixed. In this case, however,
the total error of Corollary~\ref{coro-fd-final} contains, of course, an additional term arising from the mollification.
\begin{figure}[H]
\vspace{-3mm}
\begin{center}	
\includegraphics[width=75mm]{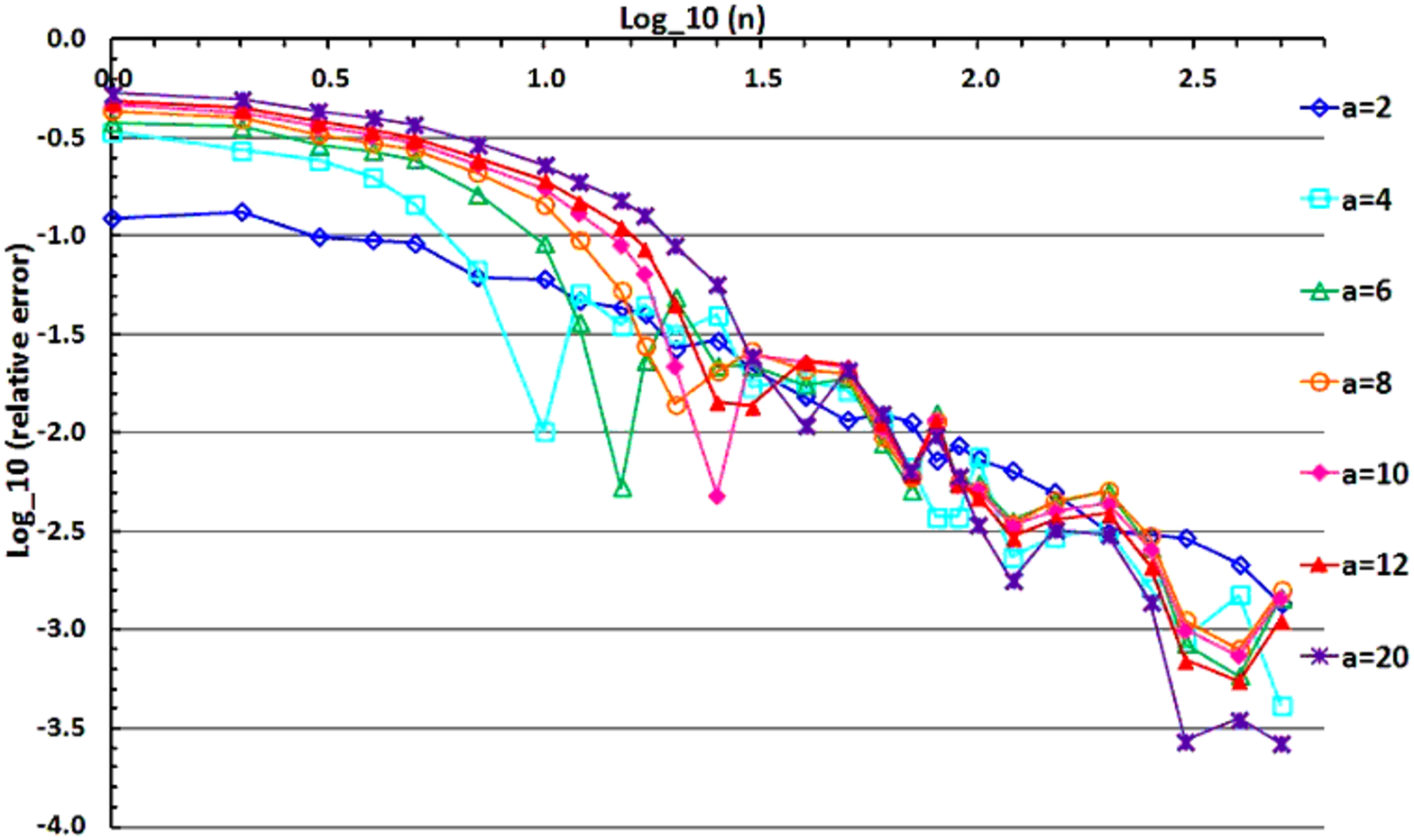}
\includegraphics[width=75mm]{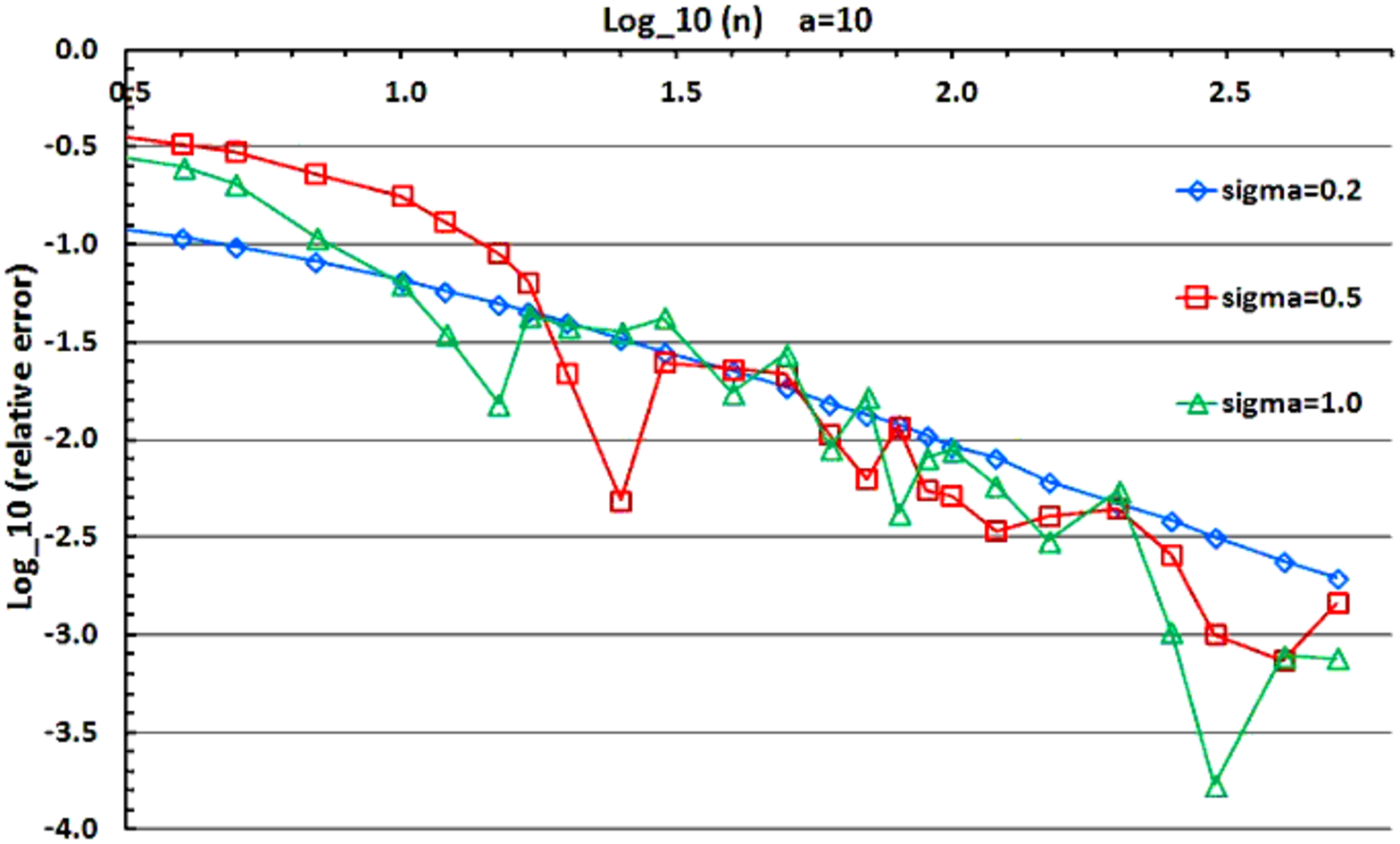}
\end{center}
\vspace{-7mm}
\caption{\small Empirical convergence of the proposed scheme for (\ref{eq-qgBSDE-num})
with a non-differentiable terminal function (\ref{eq-non-d-terminal}). 
The left one uses (\ref{eq-default}) with $a$'s in (\ref{eq-a-sets}), and the right one uses $a=10$
and (\ref{eq-default}) but with difference choices of volatilities $(\sigma_i)_{1\leq i\leq 2}=0.2,~0.5,~1.0$.} 
\label{Fig-truncation-nonD}
\end{figure}

In Figure~\ref{Fig-truncation-nonD}, we have tested the model (\ref{eq-qgBSDE-num}) 
with the  terminal function (\ref{eq-non-d-terminal}) and plotted the $\log_{10}$(relative error) 
for the number of partitions from $n=1$ to $n=500$ by directly applying the finite-difference scheme.
The same truncation of the driver $N\propto n^{1/3}$ has been used as in the previous example.
In the {\it left} figure, 
we have tested the same seven cases of ``$a$" (\ref{eq-a-sets}) with the same set of parameters (\ref{eq-default}). 
In the {\it right} one, we have kept $a=10$ but tested (\ref{eq-default}) replaced by three 
different choices of volatilities $(\sigma_i)_{1\leq i\leq 2}=0.2,~0.5,~1.0$.
The results are very encouraging. From the left one, although one actually observes some instability, the overall rate of convergence
is not much different from the previous example in Fig~\ref{Fig-truncation} of a differentiable terminal function.
From the right one, one observes that the size of volatilities does not meaningfully affect the empirical convergence rate
(but increases the instability to a certain degree).

In the computation, we observed that the maximum of the derivatives decays rather quickly when $t$ is away from the maturity.
In fact, from the expression (\ref{eq-closed}) and the {\it integration-by-parts} formula, 
one can show that the solution $Y_t^{t,\cdot}:\mbb{R}^d\rightarrow \mbb{R}$ is 
a smooth function of $x$ for $t<T$ as long as $X_T$ has a smooth density, which is the case for
the current log-normal model (\ref{eq-X-num}) for $(X_t)_{t\in [0,T]}$.
The above numerical result (and also the examples of the next subsection) suggests 
relaxing the conditions in Assumptions~\ref{assumption-Y}, \ref{assumption-barY} (and hence \ref{assumption-posteriori})
may be possible under appropriate conditions. 
Further studies on the regularity of the true solution as well as its approximation are needed.

\subsection{Lipschitz BSDEs}

\subsubsection{About the convergence estimate for Lipschitz BSDEs}
\label{sec-Lip-th}
In the reminder, let us test the proposed scheme for a Lipschitz BSDE for completeness.
The scheme given in the previous sections is equally applicable to the standard Lipschitz BSDEs
and yields the same convergence estimates.
Before providing  the numerical results, let us briefly explain the allowed setup as well as the 
associated changes in the relevant results. The major changes in the setup can be 
summarized as follows:\\
\bull $\xi(x)$ and $f(t,x,0,0)$ have at most polynomial growth in $x$.\\
\bull $\xi\in C^3(\mbb{R}^d)$ and $f$ is one-time continuously differentiable with respect to 
the spatial variables $(x,y,z)$, where $|\part_y f|, |\part_z f|$ are bounded and
$|\part_x f|$ has at most polynomial growth in $x$.\\
\bull Assumption~\ref{assumption-barY} (ii) is removed and (i) is modified as
$\max_{1\leq i\leq n}|\part_x^m \wh{u}^{i+1}(x)|\leq K^\prime(1+|x|^{(\varrho-m)\vee 0}),~\forall x\in \mbb{R}^d, m=0,\cdots,3$
with some positive integer $\varrho\geq 1$.  
The perturbation $(\del^i)_i$ are now allowed to have polynomial growth.  \\
\bull Assumption~\ref{assumption-posteriori} is modified accordingly to check the above polynomial growth condition.

Note that the boundedness of the terminal function used for qg-BSDEs is needed to guarantee
that the derivative process $(\part_x \ol{\Theta}^{i,t,x})$  is well-posed by the 
results~\cite{Briand-Confortola}.\footnote{It is also used to prove the uniqueness of the solution.}
Since the derivative process $\part_x\ol{\Theta}^{i,t,x}$ now follows a Lipschitz BSDE, the boundedness 
condition is unnecessary and similar analysis in Proposition~\ref{prop-barZ-sup} yields $|\ol{Z}^i_t|\leq C(1+|X_t|^\varrho)$.
Theorem~\ref{th-delY-delZ} still holds by the same analysis. Note that $|\gamma_t^i|$ is now bounded
by Lipschitz constant and hence Proposition~\ref{prop-barZ-bmo} is unnecessary anymore.
Since $X^{t,x}\in \mbb{S}^p$ for any $p\geq 1$, the scaling of $h_i$ (such as $Ch_i^{3p/2}$ etc.) for the 
short-term expansion in the Appendixes B and C is unchanged.
From these observations, one can show that the estimate in Theorem~\ref{th-main} still holds true.

Since $(|\part_x^m \wh{u}^i|)_{i,m}$ have polynomial growth, the constants $C$'s
appearing in Section~\ref{sec-FD} (in particular those in Lemma~\ref{lemma-uhat-approx} and 
Definition~\ref{def-connection-FD}) 
generally depend polynomially in $x$. This induces the changes in a way that
$C\Del^3\rightarrow C\Del^3 (1+|x|^\varrho)$. However, the effects can by absorbed by 
adjusting ``$\rho$" used in Section~\ref{sec-FD} appropriately in every place. It is also the case for (\ref{AE-conditional})
since the initial-value dependence of the short-term expansion is at most polynomial.
Proposition~\ref{prop-sumdel-bound} becomes unnecessary (it is only for qg-BSDEs) and hence 
the scaling factor $\del>0$ of the grid size $M$ is not restricted to  $\del< 1/\rho$.
Proposition~\ref{prop-R-fd} is proved exactly the same manner with the modified $\rho$
and Corollary~\ref{coro-fd-final}, which is the main result,  is shown to hold
by simple application of Chebyshev's inequality.

\subsubsection{Numerical examples: Option pricing with different interest rates}
We use the same scaling rule $\Del=\zeta |\pi|^{1/2}$ but, of course, no truncation of the driver.
We consider a very popular valuation problem of European options
under two different interest rates, $r$ for investing and $R~(\neq r)$ for borrowing.
Since this problem has been often used for testing the numerical schemes for  Lipschitz BSDEs,
it would be informative to compare the current scheme to the existing numerical examples based on 
Monte Carlo simulation.

Let us assume the dynamics of the security price as
\be
X_t=x_0+\int_0^t \mu X_s ds+\int_0^t \sigma X_s dW_s~,  \nn
\ee
where $d=1$ and $\mu,\sigma$ are positive constants.
For the option payoff $\Phi(X_T)$ at the expiry $T$, the option price $Y_t$ implied by the 
self-financing replication is given by
\bea
Y_t=\Phi(X_T)-\int_t^T \Bigl\{ r Y_s +\frac{\mu-r}{\sigma} Z_s-\Bigl(Y_s-\frac{Z_s}{\sigma}\Bigr)^{-}(R-r)\Bigr\}ds
-\int_t^T Z_s dW_s~.
\label{eq-BSDE-rR}
\eea
Although both the terminal and driver functions are not smooth, 
we can expect rather accurate results considering the result in Fig~\ref{Fig-truncation-nonD}
for the qg-BSDE~\footnote{Although we tested the same model with 
mollified functions,  we found no meaningful difference in the empirical convergence rate.}.

\begin{figure}[H]
\vspace{-3mm}
\begin{center}	
\includegraphics[width=95mm]{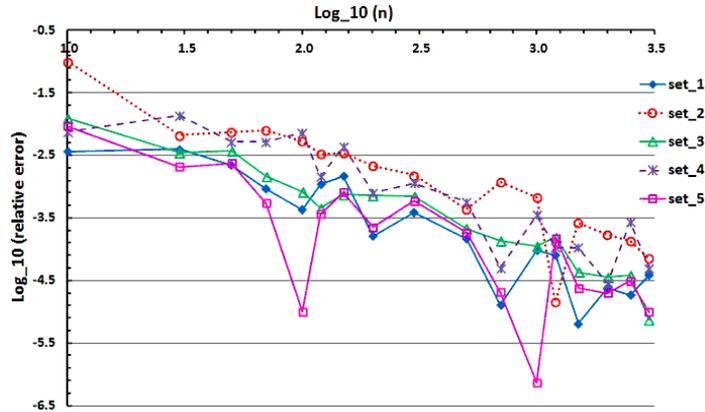}
\end{center}
\vspace{-7mm}
\caption{\small Empirical convergence of the proposed scheme for (\ref{eq-BSDE-rR}) for call options.}
\label{Fig-BS}
\end{figure}

Firstly, we study the cases where the payoff function is equal to that of a call option:
$\Phi(x)=(x-K)^+$,
where $K>0$ is the strike price. As suggested by \cite{Gobet-Warin}, this example provides
a very interesting test since the price must be exactly equal to that of Black-Scholes 
model with interest rate $R$. This is because the replicating portfolio consists
of the long-only position and hence the investor must always borrow money to fund her position.
We have chosen  the common parameters as
$\{T=1, r=0.01,~ R=0.06,~ \mu=0.06, ~X_0=100\}$
and tested the following five sets of $(K, \sigma)$~\footnote{
$K=106$ is close to {\it at the money forward} for $T=1$ with $6\%$ interest rate. The bigger strikes 
correspond to $2\sigma$ {\it out of the money}.}
with $n=10$ to $n=3000$ in Figure~\ref{Fig-BS}:
\bea
&&{\rm set}_1=\{K=106,~\sigma=0.3\}~, {\rm set}_2=\{K=166, ~\sigma=0.3\}~,{\rm set}_3=\{K=106, ~\sigma=1.0\}~,\nn \\
&& {\rm set}_4=\{K=306, ~\sigma=1.0\}~,{\rm set}_5=\{K=106, ~\sigma=2.0\}\nn.
\eea
The Black-Scholes price for each set is given by 
${\rm BS}=\{12.000,  1.117, 38.346,  11.662, 68.296 \}$ respectively.
Although the relative errors for OTM options are slightly higher, the convergence rate
to the exact BS prices is close to 1 for every case.
It is a bit striking that we do not see any deterioration in convergence rate
in spite of the non-smooth functions and rather high volatilities.
The observed irregularity of the error size is likely due to the change of the
configuration of the grids close to the terminal time relative to the non-differentiable points
of the terminal function.

\begin{figure}[H]
\vspace{-3mm}
\begin{center}	
\includegraphics[width=95mm]{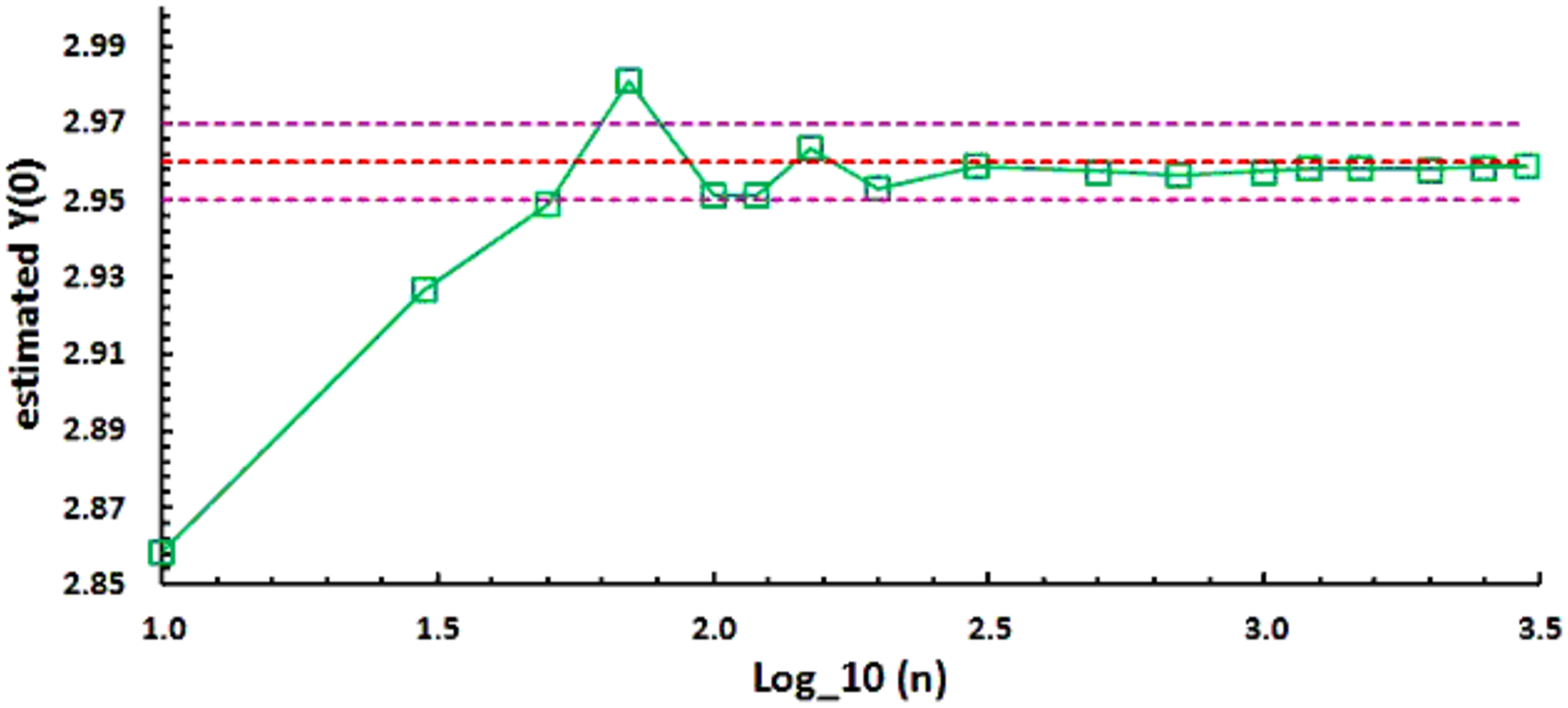}
\end{center}
\vspace{-10mm}
\caption{\small Empirical convergence of the proposed scheme for (\ref{eq-BSDE-rR}) for a call spread.}
\label{Fig-call-spread}
\begin{center}	
\includegraphics[width=95mm]{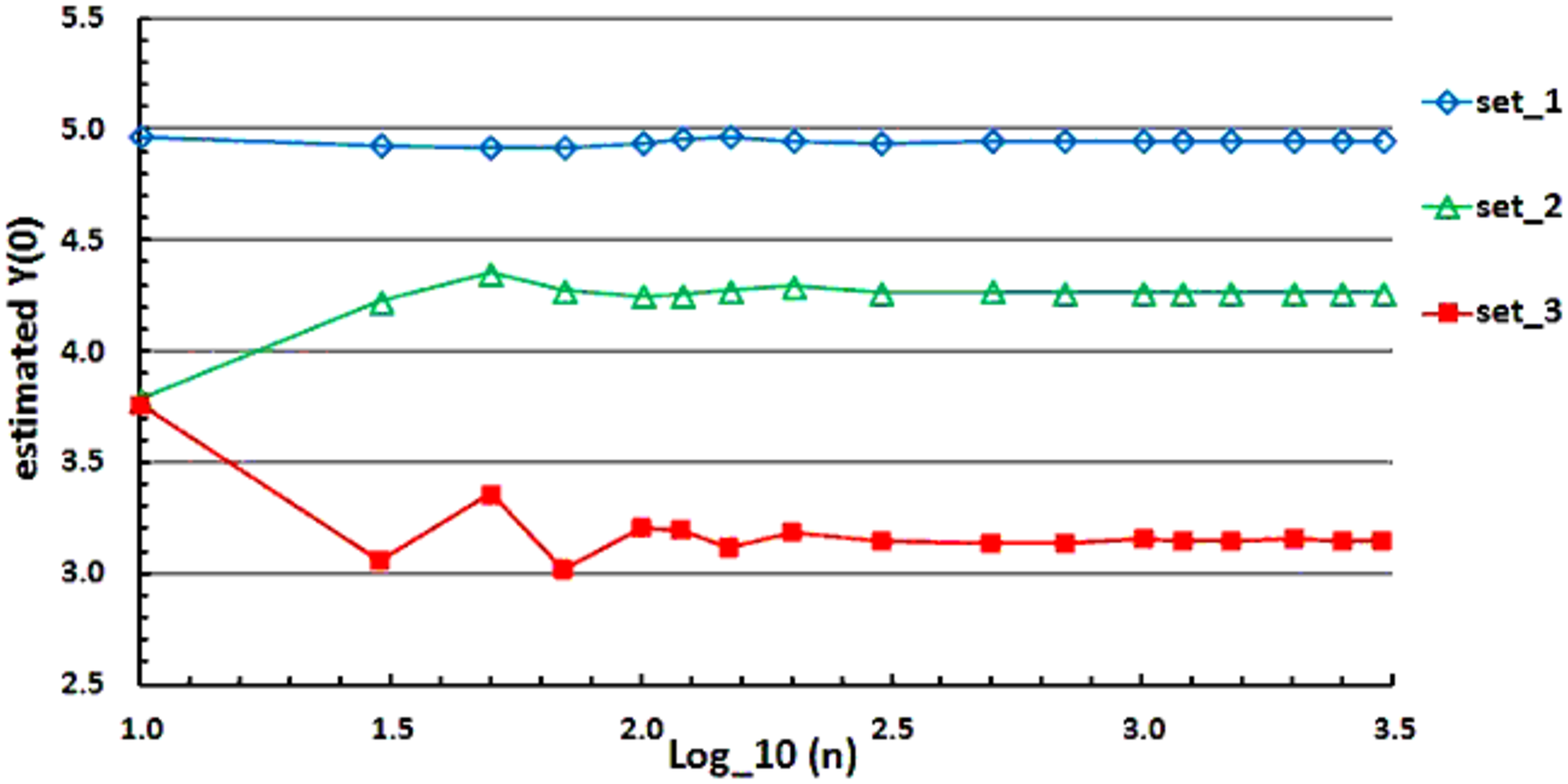}
\end{center}
\vspace{-8mm}
\caption{\small Empirical convergence of the proposed scheme for (\ref{eq-BSDE-rR}) for a call spread with 
$T=1.0$ and higher volatilities.}
\label{Fig-high-vol}
\end{figure}

Next, let us consider a call-spread case:
$\Phi(x)=(x-K_1)^+-2(x-K_2)^+$. This is exactly the same setup studied in \cite{Gobet-Warin}
and hence we can test the performance of our scheme relative to the standard 
regression-based Monte Carlo simulation.
Let us choose the same parameter sets as in \cite{Gobet-Warin}:
\be
\{r=0.01,~ R=0.06,~ \mu=0.05, ~X_0=100, ~T=0.25,~\sigma=0.2,~K_1=95, ~K_2=105\}
\label{gobet-set}
\ee
The result of \cite{Gobet-Warin} suggests that  $Y_0=2.96\pm 0.01$ or $Y_0=2.95\pm 0.01$  with one standard deviation
dependent on the choice of basis functions for the regressions.
In Figure~\ref{Fig-call-spread}, we have compared the estimated $Y_0$ from our scheme and the one in \cite{Gobet-Warin}.
The dotted lines represent $2.96\pm 0.01$ for ease of comparison.
In our scheme, $Y_0$ converges toward $2.96$.
In fact, the improvement of the regression method of \cite{Gobet-Warin} using martingale basis functions 
proposed by Bender \& Steiner (2012)~\cite{Bender-Steiner} suggests $2.96$ which is perfectly consistent with our result. 

We have also tested the convergence with a longer maturity and higher volatilities 
for the final payoff $\Phi(x)=(x-K_1)^+-(x-K_2)^+$.
We have used $\{r=0.01,~ R=0.06,~ \mu=0.05, ~X_0=100,~K_1=95, ~K_2=105\}$
as before, but with longer maturity $T=1.0$ and ${\rm set}_1:=\{\sigma=0.3\}$, ${\rm set}_2:=\{\sigma=0.5\}$
and ${\rm set}_3:=\{\sigma=1.0\}$. From Figure~\ref{Fig-high-vol}, one observes
smooth convergence for all the cases. 
The decrease in price for higher volatilities is natural from the fact that $K_2$ is closer 
to the {\it at-the-money-forward} point and hence the short position has higher sensitivity on the volatility.

\subsubsection*{An example with a large Lipschitz constant}
Bender \& Steiner~\cite{Bender-Steiner} have tested an extreme scenario with a parameter set (\ref{gobet-set}) replaced by $R=3.01$.
In this case, the non-linearity of the driver has a Lipschitz constant $(R-r)/\sigma=15$. 
Their experiments suggest that the standard method of \cite{Gobet-Warin} fails to converge for this example
under the simulation settings they tried.
Their improved method with martingale basis functions  (see Table 3 in \cite{Bender-Steiner}) gives $Y_0\simeq 6.47$ with $n=128$ and 
$Y_0\simeq 6.44$ with the finest partition $n=181$.
\begin{figure}[H]
\vspace{-3mm}
\begin{center}	
\includegraphics[width=90mm]{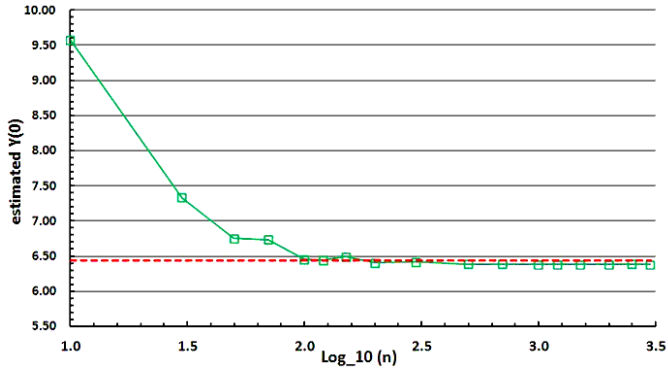}
\end{center}
\vspace{-8mm}
\caption{\small Empirical convergence of the proposed scheme for (\ref{eq-BSDE-rR}) with $R=3.01$.}
\label{Fig-R301}
\end{figure}
In Figure~\ref{Fig-R301}, we have plotted estimated $Y_0$ from our scheme with $n=10$ to $n=3000$.
The dotted line corresponds to the value $6.44$ given in \cite{Bender-Steiner}.
In our scheme, $Y_0$ seems to converge to $6.38$. In particular, with the same discretization $n=181$, 
our scheme yields $Y_0\simeq 6.43$ showing a nice consistency.
Note that the method \cite{Bender-Steiner} requires to change the basis functions
based on the law of $X$. 

\begin{remark}
Our numerical computation is implemented in Microsoft Excel VBA with Xeon X5570 cpu @ 2.93GHz. It takes
around $8$ seconds for $2$-dimensional examples of qg-BSDE with $n=100$ time steps,
and about $0.8$ seconds for $1$-dimensional examples of Lipschitz BSDE with $n=1000$ time steps.
\end{remark}

\section{Conclusion}
In this paper, we have developed a semi-analytic computation scheme for Markovian qg-BSDEs
by connecting the short-term expansions that are given in explicit form by the asymptotic expansion
technique. The method can  be applied also to the standard Lipschitz BSDEs in almost the same way.
At least for low dimensional setups, the scheme is quite easy to implement 
as a slightly elaborated tree method and has high accuracy even for very large quadratic coefficients
and Lipschitz constants. For high-dimensional setups, we have suggested a sparse grid scheme
as a promising candidate to overcome the curse-of-dimensionality at least to a certain degree.
Testing this interesting idea is left for future work.

From a theoretical viewpoint, the main difficulty has arisen from controlling the bound 
of the derivatives $|\part_x^m \wh{u}^i|$ uniformly, which 
remains as an open question. This forces us to require {\it a posteriori} checks for these bounds 
as in Assumption~\ref{assumption-posteriori} to guarantee the convergence for a given range.
Under appropriate assumptions, the properties of the corresponding semilinear PDEs may provide 
the necessary regularities. Pursuing this issue is left for future work.

\appendix
\small
\setstretch{1}
\section{BMO-martingale and its properties}
\label{app-BMO}
In this section, let us summarize the properties of BMO-martingales, the associated $\calh^2_{BMO}$-space and their
properties which play an important role in the discussions.
\begin{definition}
A BMO-martingale $M$ is a square integrable martingale satisfying $M_0=0$
and 
\bea
||M||^2_{BMO}:=\sup_{\tau\in\calt^T_0}\Bigl|\Bigl|\mbb{E}\Bigl[\langle M \rangle_T-
\langle M \rangle_\tau|\calf_\tau \Bigr]\Bigr|\Bigr|_{\infty}<\infty~, \nn
\eea
where the supremum is taken over all stopping times $\tau\in\calt^T_0$.
\end{definition}

\begin{definition}
\label{def-h2bmo}
$\calh^2_{BMO}(\mbb{R}^k)$ is the set of $\mbb{R}^k$-valued progressively measurable processes $Z$
satisfying
\be
||Z||^2_{\calh^2_{BMO}}:=\sup_{\tau\in\calt^T_0}\Bigl|\Bigl|\mbb{E}\Bigl[\int_\tau^T |Z_s|^2 ds\Bigr|\calf_\tau\Bigr]
\Bigr|\Bigr|_{\infty}<\infty~.\nn
\ee
\end{definition}

Note that if $Z\in \calh^2_{BMO}(\mbb{R}^{1\times  d})$, we have
\bea
\Bigl|\Bigl|\int_0^\cdot Z_s dW_s\Bigr|\Bigr|_{BMO}^2=\sup_{\tau\in\calt^T_0}\Bigl|\Bigl|\mbb{E}\Bigl[\int_\tau^T |Z_s|^2 ds\Bigr|\calf_\tau\Bigr]
\Bigr|\Bigr|_{\infty}=||Z||^2_{\calh^2_{BMO}}<\infty~,\nn
\eea
and hence $Z*W:=\int_0^\cdot Z_s dW_s$ is a BMO-martingale.
The next result is well-known as {\it energy inequality}.
\begin{lemma}
\label{lemma-energy}
Let $Z$ be in $\calh^2_{BMO}$. Then, for any $n\in\mbb{N}$,
\bea
\mbb{E}\Bigl[\Bigl(\int_0^T  |Z_s|^2 ds\Bigr)^n\Bigr]\leq n! \Bigl(||Z||^2_{\calh^2_{BMO}}\Bigr)^n~. \nn
\eea
\begin{proof}
See proof of Lemma 9.6.5 in \cite{Cvitanic}.
\end{proof}
\end{lemma}

Let $\cale(M)$ be a Dol\'eans-Dade exponential of $M$.
\begin{lemma}{(Reverse H\"older inequality)}
\label{lemma-reverse}
Let $M$ be a BMO-martingale. Then, $\bigl(\cale_t(M), t\in[0,T]\bigr)$
is a uniformly integrable martingale, and for every stopping time $\tau\in\calt^T_0$,
there exists some positive constant $r^*>1$ such that the inequality
\bea
\mbb{E}\Bigl[\cale_T(M)^{r}|\calf_\tau\Bigr]\leq C_{r,M}\cale_\tau(M)^r \nn~,
\eea
holds for every $1<r\leq r^*$ with some positive constant $C_{r,M}$ depending 
only on $r$ and $||M||_{BMO}$.
\begin{proof}
See Theorem 3.1 of Kazamaki (1994)~\cite{Kazamaki}.
\end{proof}
\end{lemma}

\begin{lemma}
\label{lemma-BMO-PQ}
Let $M$ be a square integrable martingale and $\wh{M}:=\langle M \rangle-M$.
Then, $M\in BMO(\mbb{P})$ if and only if $\wh{M}\in BMO(\mbb{Q})$ with 
$d\mbb{Q}/d\mbb{P}=\cale_T(M)$. Furthermore, $||\wh{M}||_{BMO(\mbb{Q})}$
is determined by some function of $||M||_{BMO(\mbb{P})}$.
\begin{proof}
See Theorem 2.4 and 3.3 in \cite{Kazamaki}.
\end{proof}
\end{lemma}

\begin{remark}
Theorem 3.1~\cite{Kazamaki} also tells that
there exists some decreasing function $\Phi(r)$ with $\Phi(1+)=\infty$
and $\Phi(\infty)=0$ such that if $||M||_{BMO(\mbb{P})}$ satisfies
$||M||_{BMO(\mbb{P})}<\Phi(r)$
then $\cale(M)$ allows the reverse H\"older inequality with power $r$.
This implies together with Lemma~\ref{lemma-BMO-PQ}, one can take 
a common positive constant $\bar{r}$ satisfying $1<\bar{r}\leq r^*$ such that both of 
the $\cale(M)$ and $\cale(\hat{M})$ satisfy the reverse H\"older inequality
with power $\bar{r}$ under the respective probability measure $\mbb{P}$ and $\mbb{Q}$.
Furthermore,  the upper bound $r^*$ is determined only by $||M||_{BMO(\mbb{P})}$ (or equivalently by $||M||_{BMO(\mbb{Q})}$).
\end{remark}

\section{Short-term expansion: Step 1}
\label{sec-short1}
In the following two sections, we approximate the solution  $(\ol{Y}^i,\ol{Z}^i)$ 
of the BSDE (\ref{eq-BSDE-barY}) semi-analytically and also obtain its error estimate.
We need two steps for achieving this goal, which involve the linearization method  and the small-variance
expansion method for BSDEs proposed in Fujii \& Takahashi (2012)~\cite{FT-analytic} and (2015)~\cite{FT-AE},
respectively~\footnote{
Note that the small-variance asymptotic expansion has been widely applied for the pricing of European contingent claims
since the initial attempts by Takahashi (1999)~\cite{T-APFM} and Kunitomo \& Takahashi (2003)~\cite{Kunitomo-Takahashi}.}.

When there is no confusion, we adopt the so-called Einstein convention 
assuming the obvious summation of duplicate indexes (such as $i\in\{1,\cdots, d\}$ of $x^i$)
without explicitly using the summation symbol $\sum$. For example, 
$\part_{x^i,x^j}\xi(X_T)\part_x X_T^{i} \part_x X_T^{j}$
assumes the summation about indexes $i$ and $j$ so that it denotes
$\sum_{i,j=1}^d \part_{x^i,x^j}\xi(X_T)\part_x X_T^{i} \part_x X_T^{j}$. 
\\\\
{\it(Standing Assumptions for Section~\ref{sec-short1})
We make Assumptions~\ref{assumption-X}, \ref{assumption-Y} and
Assumption~\ref{assumption-barY} (i) the standing assumptions for this section. \\
}

Let us introduce the next decomposition of the BSDE (\ref{eq-BSDE-barY})
for each interval $t\in I_i$, $i\in\{1,\cdots,n\}$:
\bea
\label{eq-BSDE-al-0}
&&\ol{Y}_t^{i,[0]}=\wh{u}^{i+1}(X_{t_i})-\int_t^{t_i} \ol{Z}_r^{i,[0]} dW_r~, \\
&&\ol{Y}_t^{i,[1]}=\int_t^{t_i}f\bigl(r,X_r,\ol{Y}^{i,[0]}_r,\ol{Z}_r^{i,[0]}\bigr)dr-\int_t^{t_i}\ol{Z}_r^{i,[1]}dW_r~.
\label{eq-BSDE-al-1}
\eea
They are the leading 
contributions in the linearization method \cite{FT-analytic, Takahashi-Yamada}.

\begin{lemma}
\label{lemma-barY-al-0}
For every interval $I_i, i\in\{1,\cdots,n\}$, there exists a unique solution $(\ol{Y}^{i,[0]},\ol{Z}^{i,[0]})$
to the BSDE (\ref{eq-BSDE-al-0}) satisfying,  with some $(i,n)$-independent positive constants $C$ and $C_p$, that
$\displaystyle ||\ol{Y}^{i,[0]}||_{\cals^\infty[t_{i-1},t_i]}+||\ol{Z}^{i,[0]}||_{\calh^2_{BMO}[t_{i-1},t_i]}\leq C$,
and also $||\ol{Z}^{i,[0]}||_{\cals^p[t_{i-1},t_i]}\leq C_p$
for any $p\geq 2$.
\begin{proof}
The boundedness $||\ol{Y}^{i,[0]}||_{\cals^\infty}\leq C$ follows easily from Assumption~\ref{assumption-barY} (i),
which then implies $||\ol{Z}^{i,[0]}||_{\calh^2_{BMO}}\leq C$. The second claim follows from the similar arguments used in Proposition~\ref{prop-barZ-sup}~.
\end{proof}
\end{lemma}

\begin{lemma}
For every interval $I_i$, $i\in\{1,\cdots,n\}$, there exists a unique solution 
$(\ol{Y}^{i,[1]},\ol{Z}^{i,[1]})$ to the BSDE (\ref{eq-BSDE-al-1}) satisfying, with some $(i,n)$-independent positive constant $C_p$, 
that
\be
||\ol{Y}^{i,[1]}||_{\cals^p[t_{i-1},t_i]}+||\ol{Z}^{i,[1]}||_{\calh^p[t_{i-1},t_i]}\leq C_p\nn
\ee
for any $p\geq 2$.
\begin{proof}
Since it is a Lipschitz BSDE (with zero Lipschitz constant), the existence of a unique solution easily follows.
The standard estimate (see, for example, \cite{BDH03})
and Assumption~\ref{assumption-Y} (i) implies
\bea
&&\Bigl|\Bigl|(\ol{Y}^{i,[1]},\ol{Z}^{i,[1]})\Bigr|\Bigr|^p_{\calk^p[t_{i-1},t_i]} \leq C_p \mbb{E}\Bigl[\Bigl(\int_{t_{i-1}}^{t_i}|f(r,X_r,\ol{Y}_r^{i,[0]},\ol{Z}_r^{i,[0]})|dr\Bigr)^p\Bigr]\nn \\
&&\qquad \leq C_p \mbb{E}\Bigl[\Bigl(\int_{t_{i-1}}^{t_i}\bigl[l_r+\beta |\ol{Y}^{i,[0]}_r|+
\frac{\gamma}{2}|\ol{Z}_r^{i,[0]}|^2 \bigr]dr\Bigr)^p\Bigr] \nn \\
&&\qquad \leq C_p \Bigl( ||l||^p_T+||\ol{Y}^{i,[0]}||^p_{\cals^p[t_{i-1},t_i]}+||\ol{Z}^{i,[0]}||^{2p}_{\cals^{2p}[t_{i-1},t_i]}\Bigr)
~.\nn
\eea
Thus one obtains the desired result by Lemma~\ref{lemma-barY-al-0}.
\end{proof}
\end{lemma}

We now define the process $(\wt{\ol{Y}}^i,\wt{\ol{Z}}^i)$
for each interval $t\in I_i$, $i\in\{1,\cdots,n\}$ by
\be
\wt{\ol{Y}}^i_t:=\ol{Y}^{i,[0]}_t+\ol{Y}_t^{i,[1]}~, \qquad \wt{\ol{Z}}^i_t:=\ol{Z}^{i,[0]}_t+\ol{Z}_t^{i,[1]}~.\nn
\ee
\begin{proposition}
\label{prop-AE-alpha}
There exists some $(i,n)$-independent positive constant $C_p$ such that the inequality
\bea
\mbb{E}\Bigl[\bigl|\bigl|\ol{Y}^i-\wt{\ol{Y}}^i\bigr|\bigr|^p_{[t_{i-1},t_i]}
+\Bigl(\int_{t_{i-1}}^{t_i}\bigl|\ol{Z}_r^i-\wt{\ol{Z}}^i_r\bigr|^2 dr\Bigr)^\frac{p}{2}\Bigr]\leq C_p h_i^{3p/2}\nn
\eea
holds for every interval $I_i$, $i\in\{1,\cdots,n\}$ with any $p\geq 2$.
\begin{proof}
For notational simplicity, let us put
\bea
&&\del Y^{i,[0]}_t:=\ol{Y}^i_t-\ol{Y}^{i,[0]}_t, \quad \del Z^{i,[0]}_t:=\ol{Z}^i_t-\ol{Z}_t^{i,[0]}\nn \\
&&\del Y^{i,[1]}_t:=\ol{Y}^i_t-\wt{\ol{Y}}^i_t, \qquad \del Z^{i,[1]}_t:=\ol{Z}^i_t-\wt{\ol{Z}}^i_t~ \nn
\eea
for each interval $t\in I_i, i\in\{1,\cdots,n\}$. Then,  they are given by the solutions to the 
following BSDEs respectively:
\bea
&&\del Y^{i,[0]}_t=\int_t^{t_i}f(r,X_r,\ol{Y}^i_r,\ol{Z}_r^i)dr-\int_t^{t_i}\del Z^{i,[0]}_r dW_r\nn~, \\
&&\del Y^{i,[1]}_t=\int_t^{t_i}\Bigl(f(r,X_r,\ol{Y}^i_r,\ol{Z}_r^i)-f(r,X_r,\ol{Y}^{i,[0]}_r,\ol{Z}_r^{i,[0]})\Bigr)dr
-\int_t^{t_i} \del Z_r^{i,[1]}dW_r~.\nn
\eea
By the stability result for the Lipschitz BSDEs (see, for example, \cite{BDH03}),
Assumption~\ref{assumption-Y} (i), (\ref{eq-barY-bound}) and Proposition~\ref{prop-barZ-sup}, one obtains
\bea
&&\Bigl|\Bigl|(\del Y^{i,[0]},\del{Z}^{i,[0]})\Bigr|\Bigr|^p_{\calk^p[t_{i-1},t_i]} \leq C_p \mbb{E}\Bigl[\Bigl(\int_{t_{i-1}}^{t_i}\bigl[l_r+\beta |\ol{Y}^i_r|+\frac{\gamma}{2}|\ol{Z}_r^i|^2 \bigr]
dr\Bigr)^p\Bigr] \nn \\
&&\qquad \leq C_p h_i^p \left( ||l||_T^p+||\ol{Y}^i||_{\cals^\infty[t_{i-1},t_i]}^p+
||\ol{Z}^i||^{2p}_{\cals^{2p}[t_{i-1},t_i]}\right)\leq C_p h_i^p ,
\label{eq-pre-0}
\eea
with some $(i,n)$-independent positive constant $C_p$ for $\forall p\geq 2$.
Similar analysis for $(\del Y^{i,[1]}, \del Z^{i,[1]})$ using  Assumption~\ref{assumption-Y} (ii) 
yields
\bea
&& \Bigl|\Bigl|(\del Y^{i,[1]},\del{Z}^{i,[1]})\Bigr|\Bigr|^p_{\calk^p[t_{i-1},t_i]}\leq C_p\mbb{E}\Bigl[\Bigl(\int_{t_{i-1}}^{t_i}\Bigl[ |\del Y^{i,[0]}_r|+(1+|\ol{Z}_r^{i}|+|\ol{Z}_r^{i,[0]}|)
|\del Z_r^{i,[0]}|\Bigr]dr\Bigr)^p\Bigr]\nn \\
&&\quad \leq C_p\Bigl(
h_i^p ||\del Y^{i,[0]}||^p_{\cals^p[I_i]}+\mbb{E}\Bigl[1+||\ol{Z}^{i}||^{2p}_{I_i}
+||\ol{Z}^{i,[0]}||^{2p}_{I_i}\Bigr]^\frac{1}{2}\mbb{E}\Bigl[\Bigl(h_i\int_{t_{i-1}}^{t_i} |\del Z_r^{i,[0]}|^2 dr\Bigr)^p
\Bigr]^\frac{1}{2}
\Bigr). \nn
\eea
By applying Proposition~\ref{prop-barZ-sup}, Lemma~\ref{lemma-barY-al-0} and the previous estimate (\ref{eq-pre-0}),
one obtains the desired result.
\end{proof}
\end{proposition}

\section{Short-term expansion: Step 2}
\label{sec-short2}
In the second step, we obtain 
simple analytic approximation for the BSDEs (\ref{eq-BSDE-al-0}) and (\ref{eq-BSDE-al-1})
while keeping the same order of accuracy given in Proposition~\ref{prop-AE-alpha}.
We use the small-variance expansion method for BSDEs proposed in \cite{FT-AE}
which renders all the problems into a set of simple ODEs.
Furthermore, we shall see that these ODEs can be approximated by a {\it single-step} Euler method
for each interval $I_i$. 
\\ \\
{\it(Standing Assumptions for Section~\ref{sec-short2})
Similarly to the last section, 
we make Assumptions~\ref{assumption-X}, \ref{assumption-Y} and
Assumption~\ref{assumption-barY} (i)  the standing assumptions for this section. 
}

\subsection{Approximation for $(\ol{Y}^{i,[0]}, \ol{Z}^{i,[0]})$}
For each interval, we introduce a new parameter $\ep$ satisfying $\ep\in (-c,c)$ with some constant $c>1$ to perturb (\ref{eq-SDE-org}) and
(\ref{eq-BSDE-al-0}):
\bea
\label{eq-X-ep}
&&X_t^\ep=X_{t_{i-1}}+\int_{t_{i-1}}^t b(r,X_r^{\ep}) dr+\int_{t_{i-1}}^t \ep \sigma(r,X_r^{\ep})dW_r~. \\
\label{eq-barY-0-ep}
&&\ol{Y}_t^{i,[0],\ep}=\wh{u}^{i+1}(X_{t_i}^\ep)-\int_t^{t_i}\ol{Z}_r^{i,[0],\ep}dW_r~.
\eea
for $t\in I_i=[t_{i-1},t_i]$, $i\in\{1,\cdots,n\}$.
Notice the way $\ep$ is introduced to $X^\ep$, by
which we have a different process for each interval $I_i$.\footnote{
It would be more appropriate to write $X_t^{i,\ep}$ to emphasize the dependence on the interval $t\in I_i$,
but we have omitted ``$i$" to lighten the notation.}
In the following, in order to avoid confusion between the index specifying the interval 
and the one for the component of $x\in\mbb{R}^d$, we use
the bold Gothic symbols such as $\{\bold{i}, \bold{j},\cdots\}$ for the latter,
each of which runs through $1$ to $d$.
\begin{lemma}
\label{lemma-delep}
The classical derivatives of $(X^\ep,\ol{Y}^{i,[0],\ep}, \ol{Z}^{i,[0],\ep})$
with respect to $\ep$
\bea
\part_\ep^k X^{\ep}_t:=\frac{\part^k}{\part \ep^k}X_t^\ep,\quad
\part_\ep^k \ol{Y}^{i,[0],\ep}:=\frac{\part^k}{\part \ep^k}\ol{Y}^{i,[0],\ep}_t,
\quad \part_\ep^k \ol{Z}^{i,[0],\ep}:=\frac{\part^k}{\part \ep^k}\ol{Z}^{i,[0],\ep}_t \nn
\eea
for $k=\{1,2,3\}$ are given by the solutions to the following forward- and backward-SDEs:
\bea
&&\part_\ep X_t^{\ep,\bbf{i}}=\int_{t_{i-1}}^t \part_{x^\bbf{j}}b^{\bbf{i}}(r,X_r^\ep)\part_\ep X_r^{\ep,\bbf{j}}dr
+\int_{t_{i-1}}^t \bigl[ \sigma^{\bbf{i}}(r,X_r^\ep)+\ep (\part_\ep X_r^{\ep,\bbf{j}})\part_{x^\bbf{j}}
\sigma^{\bbf{i}}(r,X_r^\ep)\bigr]dW_r ~\nn, \\
&&\part_\ep^2 X_t^{\ep,\bbf{i}}=\int_{t_{i-1}}^t
\bigl[ \part_{x^\bbf{j}} b^{\bbf{i}}(r,X_r^\ep)\part_\ep^2 X_r^{\ep,\bbf{j}}+
\part^2_{x^\bbf{j},x^\bbf{k}}b^\bbf{i}(r,X_r^\ep)\part_\ep X_r^{\ep,\bbf{j}}
\part_\ep X_r^{\ep,\bbf{k}}\bigr]dr\nn \\
&&\hspace{5mm}+\int_{t_{i-1}}^t\bigl[2(\part_\ep X_r^{\ep,\bbf{j}})\part_{x^\bbf{j}}\sigma^{\bbf{i}}(r,X_r^\ep)
+\ep (\part_\ep^2 X_r^{\ep,\bbf{j}}) \part_{x^\bbf{j}}\sigma^\bbf{i}(r,X_r^\ep)+\ep(\part_\ep X_r^{\ep,\bbf{j}})(\part_\ep X_r^{\ep,\bbf{k}})\part^2_{x^\bbf{j},x^\bbf{k}}\sigma^\bbf{i}(r,X_r^\ep)\bigr]
dW_r~ \nn, \\
&&\part_\ep^3 X_t^{\ep,\bbf{i}}=\int_{t_{i-1}}^t \bigl[
\part_{x^\bbf{j}}b^\bbf{i}(r,X_r^\ep)\part_\ep^3 X_r^{\ep,\bbf{j}}
+3\part^2_{x^\bbf{j},x^\bbf{k}}b^{\bbf{i}}(r,X_r^\ep)\part^2_\ep X_r^{\ep,\bbf{j}}\part_\ep X_r^{\ep,\bbf{k}} \nn\\
&&\hspace{8mm}+\part^3_{x^\bbf{j},x^\bbf{k},x^\bbf{m}}b^\bbf{i}(r,X_r^\ep)\part_\ep X_r^{\ep,\bbf{j}}\part_\ep X_r^{\ep,\bbf{k}}
\part_\ep X_r^{\ep,\bbf{m}}\bigr]dr+\int_{t_{i-1}}^t \bigl[
3(\part_\ep^2X_r^{\ep,\bbf{j}})\part_{x^\bbf{j}}\sigma^\bbf{i}(r,X_r^\ep)\nn \\
&&\hspace{8mm}+3(\part_\ep X_r^{\ep,\bbf{j}})(\part_\ep X_r^{\ep,\bbf{k}})\part^2_{x^\bbf{j},x^\bbf{k}}\sigma^\bbf{i}
(r,X_r^\ep)+\ep (\part_\ep^3 X_r^{\ep,\bbf{j}})\part_{x^\bbf{j}}\sigma^\bbf{i}(r,X_r^\ep) \nn \\
&&\hspace{8mm}+3\ep (\part_\ep^2 X_r^{\ep,\bbf{j}})(\part_\ep X_r^{\ep,\bbf{k}})
\part^2_{x^\bbf{j},x^\bbf{k}}\sigma^\bbf{i}(r,X_r^\ep)
+\ep (\part_\ep X_r^{\ep,\bbf{j}})(\part_\ep X_r^{\ep,\bbf{k}})(\part_\ep X_r^{\ep,\bbf{m}})
\part^3_{x^\bbf{j},x^{\bbf{k}},x^\bbf{m}}\sigma^\bbf{i}(r,X_r^\ep)\bigr]dW_r\nn~, 
\eea
\bea
&&\part_\ep \ol{Y}_t^{i,[0],\ep}=\part_{x^\bbf{j}}\wh{u}^{i+1}(X_{t_i}^\ep)\part_\ep X_{t_i}^{\ep,\bbf{j}}
-\int_t^{t_i}\part_\ep \ol{Z}_r^{i,[0],\ep}dW_r \nn~, \\
&&\part_\ep^2 \ol{Y}_t^{i,[0],\ep}=\part_{x^\bbf{j}}\wh{u}^{i+1}(X_{t_i}^\ep)\part_\ep^2 X_{t_i}^{\ep,\bbf{j}}
+\part^2_{x^\bbf{j},x^\bbf{k}}\wh{u}^{i+1}(X_{t_i}^\ep)\part_\ep X_{t_i}^{\ep,\bbf{j}}
\part_\ep X_{t_i}^{\ep,\bbf{k}}-\int_t^{t_i}\part^2_\ep\ol{Z}_r^{i,[0],\ep}dW_r\nn~, \\
&&\part_\ep^3 \ol{Y}_t^{i,[0],\ep}=\part_{x^\bbf{j}}\wh{u}^{i+1}(X_{t_i}^\ep)\part_\ep^3 X_{t_i}^{\ep,\bbf{j}}
+3\part^2_{x^\bbf{j},x^\bbf{k}}\wh{u}^{i+1}(X_{t_i}^\ep)(\part_\ep^2 X_{t_i}^{\ep,\bbf{j}})
(\part_\ep X_{t_i}^{\ep,\bbf{k}})\nn \\
&&\hspace{15mm}+\part^3_{x^\bbf{j},x^\bbf{k},x^\bbf{m}}\wh{u}^{i+1}(X_{t_i}^\ep)
(\part_\ep X_{t_i}^{\ep,\bbf{j}})(\part_\ep X_{t_i}^{\ep,\bbf{k}})
(\part_\ep X_{t_i}^{\ep,\bbf{m}})-\int_t^{t_i}\part_\ep^3 \ol{Z}_r^{i,[0],\ep}dW_r\nn~,
\eea
for $t\in I_i=[t_{i-1},t_i]$. Einstein convention is used with $\{\bbf{i},\bbf{j},\cdots\}$ running through $1$ to $d$.
\begin{proof}
The classical differentiability can be shown by
following the arguments of Theorem 3.1 in \cite{Ma-Zhang}.
See Section 6 of \cite{FT-AE} for more details.
\end{proof}
\end{lemma}

\begin{lemma}
\label{lemma-delX-norm}
For $k=\{1,2,3\}$, there exists some $(i,n)$-independent positive constant $C_{p,k}$ 
such that the inequality
\bea
\mbb{E}\Bigl[\bigl|\bigl|\part_\ep^k X^{\ep}\bigr|\bigr|_{[t_{i-1},t_i]}^p\Bigr]\leq C_{p,k}h_i^{kp/2} \nn
\eea
holds for every interval $I_i,~i\in\{1,\cdots,n\}$ with any $p\geq 2$.
\begin{proof}
This can be shown by applying the standard estimates for the Lipschitz SDEs given, for example, in Appendix A of \cite{FT-AE}.
For $k=1$,
\bea
\mbb{E}\Bigl[||\part_\ep X^\ep||_{[t_{i-1},t_i]}^p\Bigr] 
&\leq& C_p\mbb{E}\Bigl[\Bigl(\int_{t_{i-1}}^{t_i}|\sigma(r,X_r^\ep)|^2dr\Bigr)^{p/2}\Bigr]\leq C_p h_i^{p/2}~.\nn
\eea
For $k=2$, one obtains
\bea
&&\mbb{E}\Bigl[||\part_\ep^2 X^\ep||_{[t_{i-1},t_i]}^p\Bigr]\leq
C_p\mbb{E}\Bigl[\Bigl(\int_{t_{i-1}}^{t_i}|\part_\ep X_r^\ep|^2dr\Bigr)^p
+\Bigl(\int_{t_{i-1}}^{t_i}\bigl[|\part_\ep X_r^\ep|^2+|\part_\ep X_r^\ep|^4\bigr]dr\Bigr)^\frac{p}{2}
\Bigr]\nn \\
&&\quad\leq C_p\left( h_i^p\mbb{E}\Bigl[||\part_\ep X^\ep||^{2p}_{I_i}\Bigr]+h_i^{p/2}\mbb{E}\Bigl[||\part_\ep X^\ep||^p_{I_i}+
||\part_\ep X^\ep||^{2p}_{I_i}\Bigr]\right)\leq C_p h_i^p ~,\nn 
\eea
as desired. One can show the last case $k=3$ in a similar manner.
\end{proof}
\end{lemma}

Let introduce the following processes, with $k\in\{0,1,2\}$, 
\bea
&& X_t^{[k]}:=\frac{\part^k}{\part \ep^k} X_t^\ep\Bigr|_{\ep=0},~\ol{Y}_t^{i,[0],[k]}:=\frac{\part^k}{\part \ep^k} \ol{Y}_t^{i,[0],\ep}\Bigr|_{\ep=0}, \quad
\ol{Z}_t^{i,[0],[k]}:=\frac{\part^k}{\part \ep^k} \ol{Z}_t^{i,[0],\ep}\Bigr|_{\ep=0}\nn
\eea
and also
\bea
\wt{\ol{Y}}^{i,[0]}_t:=\sum_{k=0}^2\frac{1}{k!}\ol{Y}_t^{i,[0],[k]},\quad \wt{\ol{Z}}^{i,[0]}_t:=\sum_{k=0}^2\frac{1}{k!}\ol{Z}_t^{i,[0],[k]}
\label{eq-tilde-bar-Y-0}
\eea
for each interval $t\in I_i,~i\in\{1,\cdots,n\}$.

\begin{lemma}
\label{lemma-barY-0-ep}
There exists some $(i,n)$-independent positive constant $C_p$ such that
the inequality 
\bea
\mbb{E}\Bigl[\bigl|\bigl|\ol{Y}^{i,[0]}-\wt{\ol{Y}}^{i,[0]}\bigr|\bigr|_{[t_{i-1},t_i]}^p+
\Bigl(\int_{t_{i-1}}^{t_i}|\ol{Z}_r^{i,[0]}-\wt{\ol{Z}}_r^{i,[0]}|^2 dr\Bigr)^{p/2}\Bigr]
\leq C_p h_i^{3p/2}\nn
\eea
holds for every interval $I_i,i\in\{1,\cdots,n\}$ with any $p\geq 2$.
\begin{proof}
We can use the residual formula of Taylor expansion
thanks to the classical differentiability of $\ol{\Theta}^{i,[0],\ep}$ with respect to $\ep$;
\bea
&&\mbb{E}\Bigl[\bigl|\bigl|\ol{Y}^{i,[0]}-\wt{\ol{Y}}^{i,[0]}\bigr|\bigr|_{[t_{i-1},t_i]}^p+
\Bigl(\int_{t_{i-1}}^{t_i}|\ol{Z}_r^{i,[0]}-\wt{\ol{Z}}_r^{i,[0]}|^2 dr\Bigr)^{p/2}\Bigr]\nn \\
&&\quad \leq C_p\mbb{E}\Bigl[\sup_{r\in I_i}\Bigl|\frac{1}{2}\int_0^1(1-\ep)^2 \part_\ep^3 \ol{Y}^{i,[0],\ep}_r d\ep \Bigl|^p
+\Bigl(
\int_{t_{i-1}}^{t_i}\Bigl|\frac{1}{2}\int_0^1 (1-\ep)^2\part_\ep^3 \ol{Z}_r^{i,[0],\ep}d\ep \Bigr|^2 dr\Bigr)^{p/2}\Bigr] \nn \\
&&\quad \leq C_p\int_0^1\Bigr( \mbb{E}\Bigr[ \bigl|\bigl|\part_\ep^3 \ol{Y}^{i,[0],\ep}\bigr|\bigr|^p_{[t_{i-1},t_i]}
+\Bigl(\int_{t_{i-1}}^{t_i}\bigl|\part_\ep^3 \ol{Z}_r^{i,[0],\ep}\bigr|^2 dr\Bigr)^{p/2}\Bigr]
\Bigr)d\ep~.\nn
\eea
Applying the standard estimates of the Lipschitz BSDEs (see, for example, \cite{BDH03}), 
the boundedness of $\part_x^k \wh{u}^{i+1}$ as well as Lemma~\ref{lemma-delX-norm}, one obtains
\bea
&&\mbb{E}\Bigl[\bigl|\bigl|\ol{Y}^{i,[0]}-\wt{\ol{Y}}^{i,[0]}\bigr|\bigr|_{[t_{i-1},t_i]}^p+
\Bigl(\int_{t_{i-1}}^{t_i}\bigl|\ol{Z}_r^{i,[0]}-\wt{\ol{Z}}_r^{i,[0]}\bigr|^2 dr\Bigr)^{p/2}\Bigr]\nn \\
&&\quad \leq C_p\int_0^1\left(\mbb{E}\Bigl[||\part_\ep^3 X^\ep||^p_{I_i}+||\part_\ep^2 X^\ep||^p_{I_i}||\part_\ep X^\ep||^p_{I_i}
+||\part_\ep X^\ep||^{3p}_{I_i}
\Bigr]\right)d\ep \leq C_p h_i^{3p/2}\nn
\eea
as desired.
\end{proof}
\end{lemma}

The last lemma implies that it suffices to obtain $(\wt{\ol{Y}}^{i,[0]}, \wt{\ol{Z}}^{i,[0]})$  
for our purpose\footnote{See Remark~\ref{remark-del-order}.}, which is the second order approximation of $(\ol{Y}^{i,[0]}, \ol{Z}^{i,[0]})$.
Furthermore, as we shall see next, the solution of these BSDEs can be obtained explicitly by simple ODEs
thanks to the grading structure introduced by the asymptotic expansion.
The relevant system of FBSDEs is summarized below:
\bea
&&X_t^{[0]}=X_{t_{i-1}}+\int_{t_{i-1}}^t b(r,X_r^{[0]})dr\nn~, \\
&&X_t^{[1],\bbf{i}}=\int_{t_{i-1}}^t \part_{x^\bbf{j}}b^{\bbf{i}}(r,X_r^{[0]})X_r^{[1],\bbf{j}}dr
+\int_{t_{i-1}}^t \sigma^\bbf{i}(r,X_r^{[0]})dW_r\nn~, \\
&&X_t^{[2],\bbf{i}}=\int_{t_{i-1}}^t \Bigl(\part_{x^\bbf{j}}b^\bbf{i}(r,X_r^{[0]})X_r^{[2],\bbf{j}}
+\part^2_{x^\bbf{j},x^\bbf{k}}b^\bbf{i}(r,X_r^{[0]})X_r^{[1],\bbf{j}}X_r^{[1],\bbf{k}}\Bigr)dr+\int_{t_{i-1}}^t 2X_r^{[1],\bbf{j}}\part_{x^\bbf{j}}\sigma^\bbf{i}(r,X_r^{[0]})dW_r\nn~, \\
\label{eq-barY-0-0}
&&\ol{Y}_t^{i,[0],[0]}=\wh{u}^{i+1}(X_{t_i}^{[0]})-\int_t^{t_i}\ol{Z}_r^{i,[0],[0]}dW_r ~,\\
\label{eq-barY-0-1}
&&\ol{Y}_t^{i,[0],[1]}=\part_{x^\bbf{j}}\wh{u}^{i+1}(X_{t_i}^{[0]})X_{t_{i}}^{[1],\bbf{j}}
-\int_t^{t_i}\ol{Z}_r^{i,[0],[1]}dW_r~,  \\
&&\ol{Y}_t^{i,[0],[2]}=\part_{x^\bbf{j}}\wh{u}^{i+1}(X_{t_i}^{[0]})X_{t_i}^{[2],\bbf{j}}
+\part^2_{x^\bbf{j},x^\bbf{k}}\wh{u}^{i+1}(X_{t_i}^{[0]})X_{t_i}^{[1],\bbf{j}}X_{t_i}^{[1],\bbf{k}}
-\int_t^{t_i}\ol{Z}_r^{i,[0],[2]}dW_r~, 
\label{eq-barY-0-2}
\eea
for $t\in I_i, i\in\{1,\cdots,n\}$ with Einstein convention for $\{\bbf{i},\bbf{j},\cdots\}$.

\begin{definition}(Coefficient functions)\\
\label{def-coefficients}
We define the set of functions
$\chi:I_i \times \mbb{R}^d\rightarrow \mbb{R}^d$,~
$y:I_i\times \mbb{R}^d \rightarrow \mbb{R}$,~
 $\bbf{y}^{[1]}:I_i\times \mbb{R}^d\rightarrow \mbb{R}^d$,
~$\bbf{y}^{[2]}:I_i\times \mbb{R}^d\rightarrow \mbb{R}^d$, ~
$G^{[2]}:I_i\times \mbb{R}^d\rightarrow \mbb{R}^{d\times d}$,~
$y^{[2]}_0: I_i\times \mbb{R}^d\rightarrow \mbb{R}$ by
\bea
&&\chi(t,x):=x+\int_{t_{i-1}}^t b(r,\chi(r,x))dr\nn ~,\\
&&y(t,x):= \wh{u}^{i+1}(\chi(t_i,x)) \nn~, 
\eea
\bea
&&\bbf{y}^{[1]}_{\bbf{j}}(t,x):=\part_{x^\bbf{j}}\wh{u}^{i+1}(\chi(t_i,x))+\int_t^{t_i}
\part_{x^\bbf{j}}b^\bbf{k}(r,\chi(r,x))\bbf{y}^{[1]}_{\bbf{k}}(r,x)dr \nn~,\\
&&G^{[2]}_{\bbf{j},\bbf{k}}(t,x):=\part^2_{x^\bbf{j},x^\bbf{k}}\wh{u}^{i+1}(\chi(t_i,x))+\int_t^{t_i}\Bigl\{ \Bigl(\bigl[\part_x b(r,\chi(r,x))\bigr]G^{[2]}(r,x)\Bigr)^\leftrightarrow_{\bbf{j},\bbf{k}}\nn \\
&&\hspace{20mm}+\part^2_{x^\bbf{j},x^\bbf{k}}b^\bbf{m}(r,\chi(r,x))\bbf{y}^{[2]}_\bbf{m}(r,x)
\Bigr\}dr\nn ~,\\
&&y^{[2]}_0(t,x):=\int_t^{t_i}{\rm Tr}\Bigl(G^{[2]}(r,x)[\sigma\sigma^\top](r,\chi(r,x))\Bigr)dr\nn~,
\eea
and $\bbf{y}^{[2]}=\bbf{y}^{[1]}$ for $(t,x)\in I_i\times \mbb{R}^d$, $i\in\{1,\cdots, n\}$. We have used Einstein convention 
and the notation $\bigl([\part_x b(r,x)]_{\bbf{i},\bbf{j}}=\part_{x^\bbf{i}}b^\bbf{j}(r,x), \bbf{i},\bbf{j}\in\{1,\cdots,d\}\bigr)$. 
We denote the symmetrization by $A^{\leftrightarrow}:=A+A^\top$ for a $d\times d$-matrix $A$~\footnote{Hence, $G^{[2]}$
is symmetric matrix valued.}.
\end{definition}
Note that the above coefficient functions are given by the ODEs for a given $x\in\mbb{R}^d$ in each period.
The solution of the BSDEs are expressed by these functions in the following way:
\begin{lemma}
\label{lemma-al0-AE}
For each period $t\in I_i, i\in\{1,\cdots,n\}$, the solutions of the BSDEs (\ref{eq-barY-0-0}), (\ref{eq-barY-0-1}) 
and (\ref{eq-barY-0-2}) are given by, with Einstein convention,
\bea
&&\ol{Y}^{i,[0],[0]}_t=y(t,X_{t_{i-1}}), \quad \ol{Z}^{i,[0],[0]}_t\equiv 0
\qquad \text{(0-th order)}  \nn \\
&&\ol{Y}^{i,[0],[1]}_t=\bbf{y}^{[1]}_{\bbf{j}}(t,X_{t_{i-1}})X_t^{[1],\bbf{j}}, \quad 
\ol{Z}^{i,[0],[1]}_t=\bbf{y}^{[1]}_{\bbf{j}}(t,X_{t_{i-1}})\sigma^\bbf{j}(t,\chi(t,X_{t_{i-1}})),
\quad \text{(1st order)} \nn
\eea
and lastly, for the 2nd order
\bea
&&\ol{Y}^{i,[0],[2]}_t=\bbf{y}^{[2]}_{\bbf{j}}(t,X_{t_{i-1}})X_t^{[2],\bbf{j}}+
G^{[2]}_{\bbf{j},\bbf{k}}(t,X_{t_{i-1}})X_t^{[1],\bbf{j}}X_t^{[1],\bbf{k}}+y^{[2]}_0(t,X_{t_{i-1}})~,\nn \\
&&\ol{Z}^{i,[0],[2]}_t=2\Bigl(\bbf{y}^{[2]}_\bbf{j}(t,X_{t_{i-1}})X_t^{[1],\bbf{k}}
\part_{x^\bbf{k}}\sigma^\bbf{j}(t,\chi(t,X_{t_{i-1}}))+G^{[2]}_{\bbf{j},\bbf{k}}(t,X_{t_{i-1}})
X_t^{[1],\bbf{j}}\sigma^\bbf{k}(t,\chi(t,X_{t_{i-1}}))\Bigr)\nn ~.
\eea
\begin{proof}
This is a special case of the results of Section 8 of \cite{FT-AE}.
The existence of the unique solution to the BSDEs (\ref{eq-barY-0-0}), (\ref{eq-barY-0-1}) and (\ref{eq-barY-0-2})
is obvious. The expression can be directly checked by applying It\^o formula to the
suggested forms using the ODEs given in Definition~\ref{def-coefficients},  and compare the results with the BSDEs.
\end{proof}
\end{lemma}

Since each interval $I_i$ has a very short span $h_i$, we expect that
we can approximate the above ODEs by just a single-step of Euler method 
without affecting the order of error given in Lemma~\ref{lemma-barY-0-ep}.
\begin{definition}(Approximated coefficient functions)\\
\label{def-Euler}
We define the set of functions;
$\ol{\chi}:I_i\times \mbb{R}^d\rightarrow\mbb{R}^d$, ~$\ol{y}:I_i\times \mbb{R}^d\rightarrow \mbb{R}$,
~$\ol{\bbf{y}}^{[1]}:I_i\times \mbb{R}^d\rightarrow \mbb{R}^d$,~$\ol{\bbf{y}}^{[2]}:I_i\times \mbb{R}^d\rightarrow \mbb{R}^d$,~
$\ol{G}^{[2]}:I_i\times \mbb{R}^d\rightarrow \mbb{R}^{d\times d}$,~
$\ol{y}^{[2]}_0:I_i\times \mbb{R}^d\rightarrow \mbb{R}$ by
\bea
\ol{\chi}(t,x)&:=&x+\Del(t)b(t_{i-1},x)\nn ~,\\
\ol{y}(t,x)&:=& \wh{u}^{i+1}(\ol{\chi}(t_i,x))\nn ~,\\
\ol{\bbf{y}}^{[1]}_\bbf{j}(t,x)&:=&\part_{x^\bbf{j}}\wh{u}^{i+1}(\ol{\chi}(t_i,x))+
\del(t)\part_{x^\bbf{j}}b^\bbf{k}(t_i,\ol{\chi}(t_i,x))\part_{x^\bbf{k}}\wh{u}^{i+1}(\ol{\chi}(t_i,x))\nn ~, \\
\ol{G}^{[2]}_{\bbf{j},\bbf{k}}(t,x)&:=&\part^2_{x^\bbf{j},x^\bbf{k}}\wh{u}^{i+1}(\ol{\chi}(t_i,x))
+\del(t)\Bigl\{\Bigl([\part_x b(t_i,\ol{\chi}(t_i,x))]\part^2_{x,x}\wh{u}^{i+1}(\ol{\chi}(t_i,x))
\Bigr)^{\leftrightarrow}_{\bbf{j},\bbf{k}} \nn \\
&&+ \part^2_{x^\bbf{j},x^\bbf{k}}b^\bbf{m}(t_i,\ol{\chi}(t_i,x))\part_{x^\bbf{m}}\wh{u}^{i+1}(\ol{\chi}(t_i,x))
\Bigr\}\nn ~, \\
\ol{y}^{[2]}_0(t,x)&:=&\del(t){\rm Tr}\Bigl(\ol{G}^{[2]}(t_i,x)[\sigma\sigma^\top](t_i,\ol{\chi}(t_i,x))\Bigr)\nn~,
\eea
and $\ol{\bbf{y}}^{[2]}=\ol{\bbf{y}}^{[1]}$
for $(t,x)\in I_i\times \mbb{R}^d$, $i\in\{1,\cdots,n\}$. We have used Einstein convention and 
the notations $\Del(t):=t-t_{i-1},~\del(t):=t_i-t$.
\end{definition}
The functions in Definition~\ref{def-Euler} provide 
good approximations for the coefficient functions in Definition~\ref{def-coefficients}
in the following sense:
\begin{lemma}
\label{lemma-Euler}
There exists some $(i,n)$-independent positive constant $C_p$ satisfying
\bea
&&\mbb{E}\left\{ \sup_{t\in I_i}\Bigl|\chi(t,X_{t_{i-1}})-\ol{\chi}(t,X_{t_{i-1}})\Bigr|^p+
\sup_{t\in I_i}\Bigl|y(t,X_{t_{i-1}})-\ol{y}(t,X_{t_{i-1}})\Bigr|^p \right.  \nn \\
&&\quad +\sum_{k=1}^2\sup_{t\in I_i}\Bigl|\bbf{y}^{[k]}(t,X_{t_{i-1}})-\ol{\bbf{y}}^{[k]}(t,X_{t_{i-1}})\Bigr|^p 
+\sup_{t\in I_i}\Bigl|G^{[2]}(t,X_{t_{i-1}})-\ol{G}^{[2]}(t,X_{t_{i-1}})\Bigr|^p\nn \\
&&\quad \left.+\sup_{t\in I_i}\Bigl|y^{[2]}_0(t,X_{t_{i-1}})-\ol{y}^{[2]}_0(t,X_{t_{i-1}})\Bigr|^p
\right\}\leq C_p h_i^{3p/2}~ ,\nn 
\eea
for every interval $I_i, i\in\{1,\cdots,n\}$
with any $p\geq 2$.
\begin{proof}
Firstly, let us consider $(\chi,\ol{\chi})$.
Using $1/2$-H\"older continuity in $t$, the global Lipschitz and linear growth properties of $b$ in $x$, we have
\bea
|\chi(t,x)-\ol{\chi}(t,x)|&\leq& \int_{t_{i-1}}^t |b(r,\chi(r,x))-b(t_{i-1},x)|dr\nn \\
&\leq & K\int_{t_{i-1}}^t\Bigl[\Del(r)^{1/2}+|\chi(r,x)-\ol{\chi}(r,x)|+\Del(r)|b(t_{i-1},x)|\Bigr]dr\nn \\
&\leq & C(1+|x|h_i^{1/2})h_i^{3/2}+K\int_{t_{i-1}}^t |\chi(r,x)-\ol{\chi}(r,x)|dr \nn
\eea
and hence by Gronwall inequality, $
\sup_{t\in I_i}|\chi(t,x)-\ol{\chi}(t,x)|\leq e^{K h_i}C(1+|x|\sqrt{h_i})h_i^{3/2}$.
Thus
\bea
\mbb{E}\Bigl[\sup_{t\in I_i}|\chi(t,X_{t_{i-1}})-\ol{\chi}(t,X_{t_{i-1}})|^p\Bigr]
\leq C_p h_i^{3p/2}\left(1+h_i^{p/2}\mbb{E}\Bigl[|X_{t_{i-1}}|^p\Bigr]\right) \leq C_p h_i^{3p/2}~ 
\label{app-chi}
\eea
with some $(i,n)$-independent positive constant $C_p$.
Since $|\part_x \wh{u}^{i+1}|\leq K^\prime$ by Assumption~\ref{assumption-barY} (i), 
\bea
|y(t,x)-\ol{y}(t,x)|&=&|\wh{u}^{i+1}(\chi(t_i,x))-\wh{u}^{i+1}(\ol{\chi}(t_i,x))|\leq 
K^\prime|\chi(t_i,x)-\ol{\chi}(t_i,x)|~. \nn
\eea
Thus from (\ref{app-chi}),
\bea
\label{eq-bar-y-error}
\mbb{E}\Bigl[\sup_{t\in I_i}|y(t,X_{t_{i-1}})-\ol{y}(t,X_{t_{i-1}})|^p\Bigr]\leq C_p h_i^{3p/2} 
\eea
with some $(i,n)$-independent positive constant $C_p$.

Let us now consider
\be
\ol{\bold{y}}^{[1]}(t,x)=\part_x \wh{u}^{i+1}(\ol{\chi}(t_i,x))+\del(t)
\bigl[\part_x b(t_i,\ol{\chi}(t_i,x))\bigr]\part_x\wh{u}^{i+1}(\ol{\chi}(t_i,x))~.\nn
\ee
Since both $|\part_x \wh{u}^{i+1}|$ and $|\part_x b|$ are bounded, it is easy to see 
\be
\sup_{(t,x)\in I_i\times \mbb{R}^d}|\ol{\bold{y}}^{[1]}(t,x)|\leq C
\label{eq-bar-y1-bound}
\ee
with some positive constant $C$.
For $t\in I_i$ with a given $x\in\mbb{R}^d$, we have
\bea
&&\bold{y}^{[1]}(t,x)-\ol{\bold{y}}^{[1]}(t,x)=\part_x \wh{u}^{i+1}(\chi(t_i,x))-\part_x \wh{u}^{i+1}(\ol{\chi}(t_i,x))\nn \\
&&\hspace{15mm}+\int_t^{t_i}\Bigl(
\part_x b(r,\chi(r,x))\bold{y}^{[1]}(r,x)-\part_x b(t_i,\ol{\chi}(t_i,x))\ol{\bold{y}}^{[1]}(t_i,x)\Bigr)dr ~.\nn 
\eea
From (\ref{eq-bar-y1-bound}), $1/2$-H\"older continuity and global Lipschitz property of $\part_x b$, we obtain
\bea
&&\hspace{-8mm}|\bold{y}^{[1]}(t,x)-\ol{\bold{y}}^{[1]}(t,x)| \leq |\part_x\wh{u}^{i+1}(\chi(t_i,x))-\part_x\wh{u}^{i+1}(\ol{\chi}(t_i,x))|
+\int_t^{t_i}\left\{|\part_x b(r,\chi(r,x))||\bold{y}^{[1]}(r,x)-\ol{\bold{y}}^{[1]}(r,x)| \right. \nn \\
&&~\qquad \left. +|\part_x b(r,\chi(r,x))||\ol{\bold{y}}^{[1]}(r,x)-\ol{\bold{y}}^{[1]}(t_i,x)|+|\part_x b(r,\chi(r,x))-\part_x b(t_i,\ol{\chi}(t_i,x))||\ol{\bold{y}}^{[1]}(t_i,x)|
\right\}dr\nn\\
&&\quad\leq K^\prime|\chi(t_i,x)-\ol{\chi}(t_i,x)|+K\int_t^{t_i}|\bold{y}^{[1]}(r,x)-\ol{\bold{y}}^{[1]}(r,x)|dr\nn \\
&&~\qquad+Ch_i^2+C\int_t^{t_i}\Bigl(\del(r)^{1/2}+|\chi(r,x)-\ol{\chi}(r,x)|+|\ol{\chi}(r,x)-\ol{\chi}(t_i,x)|\Bigr)dr\nn  \\
&&\quad \leq K\int_t^{t_i}|\bold{y}^{[1]}(r,x)-\ol{\bold{y}}^{[1]}(r,x)|dr+C h_i^{3/2}\bigl(1+|x|\sqrt{h_i})~.\nn
\eea
Thus the backward Gronwall inequality (see, for example, Corollary 6.62 in \cite{Pardoux-Rascanu}) gives 
\bea
\sup_{t\in I_i}|\bold{y}^{[1]}(t,x)-\ol{\bold{y}}^{[1]}(t,x)|\leq  C h_i^{3/2}(1+|x|\sqrt{h_i})e^{K h_i}\nn~,
\eea
and hence
\bea
\label{eq-bar-y1-error}
\mbb{E}\Bigl[\sup_{t\in I_i}|\bold{y}^{[1]}(t,X_{t_{i-1}})-\ol{\bold{y}}^{[1]}(t,X_{t_{i-1}})|^p\Bigr]
\leq C_p h_i^{3p/2}~,
\eea
with some $(i,n)$-independent constant $C_p$ as desired.

By the boundedness of $|\part_x^m \wh{u}^{i+1}(x)|$ and $|\part_x^m b|$ with $m\in\{1,2\}$, it is easy to see
that $|\ol{G}^{[2]}|$ is also bounded
\bea
\sup_{(t,x)\in I_i\times \mbb{R}^d}|\ol{G}^{[2]}(t,x)|\leq C
\label{eq-bar-G2-bound}
\eea
with some positive constant $C$.
Similar analysis done for $\ol{\bold{y}}^{[1]}$ using (\ref{eq-bar-G2-bound}),
$1/2$-H\"older and Lipschitz continuities for $\part_x b, \part_x^2 b$, the backward Gronwall inequality yields
\bea
\sup_{t\in I_i}|G^{[2]}(t,x)-\ol{G}^{[2]}(t,x)|\leq C h_i^{3/2}(1+|x|\sqrt{h_i}) \nn~,
\eea
and hence
\bea
\label{eq-bar-G2-error}
\mbb{E}\Bigl[\sup_{t\in I_i}|G^{[2]}(t,X_{t_{i-1}})-\ol{G}^{[2]}(t,X_{t_{i-1}})|^p\Bigr]\leq C_p h_i^{3p/2}
\eea
with some $(i,n)$-independent positive constant $C_p$ as desired.

Finally, we consider
\bea
\ol{y}^{[2]}_0(t,x)=\del(t){\rm Tr}\Bigl(\ol{G}^{[2]}(t_i,x)[\sigma\sigma^\top]\bigl(t_i,\ol{\chi}(t_i,x)\bigr)\Bigr)~.\nn
\eea
From (\ref{eq-bar-G2-bound}) and the linear-growth property of $\sigma$, 
\be
|\ol{y}^{[2]}_0(t,x)|\leq C \del(t)\bigl(1+|x|^2\bigr)~, 
\label{eq-y20-bound}
\ee
is satisfied for every $(t,x)\in I_i\times \mbb{R}^d$ with some positive constant $C$.
We have
\bea
y^{[2]}_0(t,x)-\ol{y}^{[2]}_0(t,x)=\int_t^{t_i}{\rm Tr}\Bigl(
G^{[2]}(r,x)[\sigma\sigma^\top](r,\chi(r,x))-\ol{G}^{[2]}(t_i,x)[\sigma\sigma^\top](t_i,\ol{\chi}(t_i,x))\Bigr)dr\nn
\eea
and thus
\bea
&&|y^{[2]}_0(t,x)-\ol{y}^{[2]}_0(t,x)| \leq \int_t^{t_i}{\rm Tr}\Bigl\{
\Bigl(|G^{[2]}(r,x)-\ol{G}^{[2]}(r,x)|+|\ol{G}^{[2]}(r,x)-\ol{G}^{[2]}(t_i,x)|\Bigr)\bigl|[\sigma\sigma^\top](r,\chi(r,x))\bigr|
 \nn \\
&&\hspace{37mm}  +\bigl|\ol{G}^{[2]}(t_i,x)\bigr|\Bigl|[\sigma\sigma^\top](r,\chi(r,x))-
[\sigma\sigma^\top](t_i,\ol{\chi}(t_i,x))\Bigr|\Bigr\}dr\nn \\
&&\quad \leq Ch_i^{3/2}(1+|x|)+C h_i^2 |x|^2(1+h_i|x|) \nn
\eea
with some $(i,n)$-independent constant $C$. Thus we obtain, for any $p\geq 2$,
\bea
&&\mbb{E}\Bigl[\sup_{t\in I_i}|y^{[2]}_0(t,X_{t_{i-1}})-\ol{y}^{[2]}_0(t,X_{t_{i-1}})|^p\Bigr] \leq C_p h_i^{3p/2}
\label{eq-bar-y20-error}
\eea
as desired.
From (\ref{app-chi}), (\ref{eq-bar-y-error}), (\ref{eq-bar-y1-error}), (\ref{eq-bar-G2-error}),
(\ref{eq-bar-y20-error}) and $\bold{y}^{[2]}=\bold{y}^{[1]}$, the claim is proved.
\end{proof}
\end{lemma}

We now introduce the processes $(\wh{Y}_t^{i,[0]}, \wh{Z}_t^{i,[0]})$  for each period $t\in I_i$. They are defined by
$(\wt{\ol{Y}}_t^{i,[0]}, \wt{\ol{Z}}_t^{i,[0]})$ of (\ref{eq-tilde-bar-Y-0}) with the coefficient functions
in Definition~\ref{def-coefficients} replaced 
by the approximations in Definition~\ref{def-Euler}, i.e.;
\bea
\label{eq-Yhat-i0}
\wh{Y}^{i,[0]}_t&:=&\ol{y}(t,X_{t_{i-1}})+(X_t^{[1]})^\top\ol{\bold{y}}^{[1]}(t,X_{t_{i-1}}) \nn \\
&+&\frac{1}{2}\Bigl( (X_t^{[2]})^\top \ol{\bold{y}}^{[2]}(t,X_{t_{i-1}})+(X_t^{[1]})^\top \ol{G}^{[2]}(t,X_{t_{i-1}})X_t^{[1]}
+\ol{y}^{[2]}_0(t,X_{t_{i-1}})\Bigr), \\
\wh{Z}^{i,[0]}_t&:=&\ol{\bold{y}}^{[1]\top}(t,X_{t_{i-1}}) \sigma(t,\ol{\chi}(t,X_{t_{i-i}}))\nn \\
&&\hspace{-12mm}+\Bigl((X_t^{[1]})^\top\part_x \sigma(t,\ol{\chi}(t,X_{t_{i-1}}))\ol{\bold{y}}^{[2]}(t,X_{t_{i-1}})
+(X_t^{[1]})^\top\ol{G}^{[2]}(t,X_{t_{i-1}})\sigma(t,\ol{\chi}(t,X_{t_{i-1}}))
\Bigr),
\label{eq-Zhat-i0}
\eea
where we have used Matrix notation for simplicity. The details of indexing can be checked from those given in
Lemma~\ref{lemma-al0-AE}.

\begin{lemma}
\label{lemma-Yhat-i0-error}
There exists some $(i,n)$-independent positive constant $C_p$ such that the inequality
\bea
\mbb{E}\Bigl[\bigl|\bigl|\wt{\ol{Y}}^{i,[0]}-\wh{Y}^{i,[0]}\bigr|\bigr|_{[t_{i-1},t_i]}^p\Bigr]
+\mbb{E}\Bigl[\bigl|\bigl|\wt{\ol{Z}}^{i,[0]}-\wh{Z}^{i,[0]}\bigr|\bigr|^p_{[t_{i-1},t_i]}\Bigr]\leq C_p h_i^{3p/2}~ \nn
\eea
holds for every interval $I_i,i\in \{1,\cdots,n\}$ for any $p\geq 2$.
\begin{proof}
It can be shown easily from Lemmas~\ref{lemma-delX-norm} and \ref{lemma-Euler}.
\end{proof}
\end{lemma}
\begin{corollary}
\label{lemma-Yhat-i0-error-final}
There exists some $(i,n)$-independent positive constant $C_p$ such that 
\bea
\mbb{E}\Bigl[||\ol{Y}^{i,[0]}-\wh{Y}^{i,[0]}||^p_{[t_{i-1},t_i]}+\Bigl(\int_{t_{i-1}}^{t_i}
|\ol{Z}_r^{i,[0]}-\wh{Z}_r^{i,[0]}|^2 dr\Bigr)^{p/2}\Bigr]\leq C_p h_i^{3p/2} \nn
\eea
holds for every interval $I_i,i\in\{1,\cdots,n\}$ with any $p\geq 2$.
\begin{proof}
It follows directly from Lemmas~\ref{lemma-barY-0-ep} and \ref{lemma-Yhat-i0-error}.
\end{proof}
\end{corollary}

Since $X_{t_{i-1}}^{[1]}=X_{t_{i-1}}^{[2]}=0$, we have a very simple expression 
at the starting time $t_{i-1}$ of each period $I_i=[t_{i-1},t_i]$:
\bea
\wh{Y}^{i,[0]}_{t_{i-1}}&=&\ol{y}(t_{i-1},X_{t_{i-1}})+\frac{1}{2}\ol{y}^{[2]}_0(t_{i-1},X_{t_{i-1}}),\nn \\
\wh{Z}^{i,[0]}_{t_{i-1}}&=& \ol{\bold{y}}^{[1] \top}(t_{i-1},X_{t_{i-1}})\sigma(t_{i-1},X_{t_{i-1}}). \nn
\eea
We have the following continuity property of the approximated solution $(\wh{Y}^{i,[0]}, \wh{Z}^{i,[0]})$:
\begin{lemma}
\label{lemma-continuity-Yhat}
There exists some $(i,n)$-independent positive constant $C_p$ such that the inequality
\bea
\mbb{E}\Bigl[\sup_{t\in I_i}\Bigl|\wh{Y}^{i,[0]}_t-\wh{Y}^{i,[0]}_{t_{i-1}}\Bigr|^p\Bigr]
+\mbb{E}\Bigl[\sup_{t\in I_i}\Bigl|\wh{Z}^{i,[0]}_t-\wh{Z}^{i,[0]}_{t_{i-1}}\Bigr|^p\Bigr]\leq C_p h_i^{p/2} ~,\nn
\eea
holds for every interval $I_i,i\in\{1,\cdots,n\}$ with any $p\geq 2$.
\begin{proof}
Since $\ol{y}(t,x)=\ol{y}(t_{i-1},x)$ for $(t,x)\in I_i\times \mbb{R}^d$, we have
\bea
\wh{Y}^{i,[0]}_t-\wh{Y}^{i,[0]}_{t_{i-1}}&=&X_t^{[1]\top} \ol{\bold{y}}^{[1]}(t,X_{t_{i-1}})
+\frac{1}{2}\left( X_t^{[2]\top} \ol{\bold{y}}^{[2]}(t,X_{t_{i-1}})+
X_t^{[1]\top}\ol{G}^{[2]}(t,X_{t_{i-1}})X_t^{[1]}
\right)~\nn \\
&+&\frac{1}{2}\Bigl( \ol{y}^{[2]}_0(t,X_{t_{i-1}})-\ol{y}^{[2]}_0(t_{i-1},X_{t_{i-1}})\Bigr)~.\nn
\eea
The bounds in (\ref{eq-bar-y1-bound}) (remember that $\ol{\bold{y}}^{[1]}=\ol{\bold{y}}^{[2]}$), (\ref{eq-bar-G2-bound}) and (\ref{eq-y20-bound}) 
as well as the estimates in Lemma~\ref{lemma-delX-norm} imply
\bea
\mbb{E}\Bigl[\sup_{t\in I_i}\Bigl|\wh{Y}^{i,[0]}_t-\wh{Y}^{i,[0]}_{t_{i-1}}\Bigr|^p\Bigr]\leq C_p\mbb{E}\Bigl[
||X^{[1]}||^p_{I_i}+||X^{[2]}||^p_{I_i}+||X^{[1]}||^{2p}_{I_i}+h_i^p\Bigl(1+|X_{t_{i-1}}|^{2p}\Bigr)
\Bigr]\leq C_p h_i^{p/2}\nn
\eea
as desired.
Similarly we have
\bea
&&|\wh{Z}^{i,[0]}_t-\wh{Z}^{i,[0]}_{t_{i-1}}|\leq|\ol{\bold{y}}^{[1]}(t,X_{t_{i-1}})-\ol{\bold{y}}^{[1]}(t_{i-1},X_{t_{i-1}})|
|\sigma(t,\ol{\chi}(t,X_{t_{i-1}}))|\nn \\
&&\hspace{25mm}+|\ol{\bold{y}}^{[1]}(t_{i-1},X_{t_{i-1}})||\sigma(t,\ol{\chi}(t,X_{t_{i-1}}))-\sigma(t_{i-1},X_{t_{i-1}})|\nn \\
&&\hspace{25mm}+|X_t^{[1]}|\Bigl|\part_x\sigma(t,\ol{\chi}(t,X_{t_{i-1}}))\ol{\bold{y}}^{[2]}(t,X_{t_{i-1}})
+\ol{G}^{[2]}(t,X_{t_{i-1}})\sigma\bigr(t,\ol{\chi}(t,X_{t_{i-1}})\bigl)\Bigr|\nn\\
&&\qquad \leq C\Del(t)\Bigl(1+|X_{t_{i-1}}|\Bigr)+C\Bigl(\Del(t)^{1/2}+\Del(t)(1+|X_{t_{i-1}}|)\Bigr)+C|X_t^{[1]}|\Bigl(1+|X_{t_{i-1}}|\Bigr)\nn~,
\eea
with some positive  constant $C$. Thus, we obtain $
\mbb{E}\left[\sup_{t\in I_i}\Bigl|\wh{Z}^{i,[0]}_t-\wh{Z}^{i,[0]}_{t_{i-1}}\Bigr|^p\right]\leq C_p h_i^{p/2}$ 
as desired.
\end{proof}
\end{lemma}

\subsection{Approximation for $(\ol{Y}^{i,[1]}, \ol{Z}^{i,[1]})$} 
We now want to approximate the remaining BSDE (\ref{eq-BSDE-al-1}) appeared in the decomposition of $(\ol{Y}^i,\ol{Z}^i)$.
We shall see below that this can be achieved in a very simple fashion.
We define the process $(\wh{Y}^{i,[1]},\wh{Z}^{i,[1]})$ by
\bea
\label{eq-Yhat-i1}
\wh{Y}^{i,[1]}_t&:=&\del(t)f\Bigl(t_{i-1}, X_{t_{i-1}},\wh{Y}^{i,[0]}_{t_{i-1}},\wh{Z}^{i,[0]}_{t_{i-1}}\Bigr)~,  \\
\wh{Z}^{i,[1]}_t&:=&0~, 
\label{eq-Zhat-i1}
\eea
for each period $t\in I_i, i\in\{1,\cdots,n\}$. Here, $\del(t)=t_i-t$ as before.
\begin{lemma}
\label{lemma-Yhat-i1-error}
There exists some $(i,n)$-independent positive constant $C_p$ such that the inequality
\bea
\mbb{E}\Bigl[\bigl|\bigl| \ol{Y}^{i,[1]}-\wh{Y}^{i,[1]}\bigr|\bigr|^p_{[t_{i-1},t_i]}
+\Bigl(\int_{t_{i-1}}^{t_i}|\ol{Z}_r^{i,[1]}|^2dr\Bigr)^{p/2}\Bigr]\leq C_p h_i^{3p/2}\nn
\eea
holds for every interval $I_i, i\in\{1,\cdots,n\}$ with any $p\geq 2$.
\begin{proof}
Let us put $~\del Y^{i,[1]}_t:=\ol{Y}^{i,[1]}_t-\wh{Y}^{i,[1]}_t$, $~~\del Z^{i,[1]}_t:=\ol{Z}^{i,[1]}_t$ 
for $t\in I_i$. Then, $(\del Y^{i,[1]}, \del Z^{i,[1]})$ is the solution of the following Lipschitz BSDE:
\bea
\del Y^{i,[1]}_t=\int_t^{t_i}\del f(r)dr-\int_t^{t_i}\del Z_r^{i,[1]}dW_r~,\nn
\eea
where $\displaystyle \del f(r):=f(r,X_r,\ol{Y}^{i,[0]}_r,\ol{Z}_r^{i,[0]})-f(t_{i-1},X_{t_{i-1}},\wh{Y}^{i,[0]}_{t_{i-1}},
\wh{Z}^{i,[0]}_{t_{i-1}})$.
From Assumption~\ref{assumption-Y} (ii), it satisfies with the positive constant $K$ that
\bea
|\del f(r)| &\leq& |f(r,X_r,\ol{Y}^{i,[0]}_r,\ol{Z}^{i,[0]}_r)-f(t_{i-1},X_{t_{i-1}},\ol{Y}^{i,[0]}_r,\ol{Z}_r^{i,[0]})|\nn \\
&&+|f(t_{i-1},X_{t_{i-1}},\ol{Y}^{i,[0]}_r,\ol{Z}^{i,[0]}_r)-f(t_{i-1},X_{t_{i-1}},\wh{Y}^{i,[0]}_r,\wh{Z}_r^{i,[0]})|\nn \\ 
&&+|f(t_{i-1},X_{t_{i-1}},\wh{Y}^{i,[0]}_r,\wh{Z}_r^{i,[0]})-f(t_{i-1},X_{t_{i-1}},\wh{Y}^{i,[0]}_{t_{i-1}},
\wh{Z}_{t_{i-1}}^{i,[0]})|\nn \\
&\leq &K\Bigl(\Del(r)^{1/2}+(1+|\ol{Y}^{i,[0]}_r|+|\ol{Z}_r^{i,[0]}|^2)|X_r-X_{t_{i-1}}|\Bigr) \nn \\
&&+K|\ol{Y}^{i,[0]}_r-\wh{Y}^{i,[0]}_r|+K(1+|\ol{Z}_r^{i,[0]}|+|\wh{Z}^{i,[0]}_r|)|\ol{Z}_r^{i,[0]}
-\wh{Z}^{i,[0]}_r|\nn \\
&&+K|\wh{Y}^{i,[0]}_r-\wh{Y}^{i,[0]}_{t_{i-1}}|+
K(1+|\wh{Z}_r^{i,[0]}|+|\wh{Z}^{i,[0]}_{t_{i-1}}|)|\wh{Z}_r^{i,[0]}-\wh{Z}^{i,[0]}_{t_{i-1}}|\nn ~.
\eea

From Lemma~\ref{lemma-barY-al-0}, we know that 
$||\ol{Y}^{i,[0]}||_{\cals^\infty[t_{i-1},t_i]}+||\ol{Z}^{i,[0]}||_{\cals^p[t_{i-1},t_i]}\leq C_p$
for any $p\geq 2$. From (\ref{eq-Yhat-i0}), (\ref{eq-Zhat-i0}), Lemma~\ref{lemma-delX-norm}, and the boundedness properties 
of $(\ol{y},\ol{\bold{y}}^{[i]},\ol{G}^{[2]})$ shown in the proof for Lemma~\ref{lemma-Euler},  a similar inequality
$||\wh{Y}^{i,[0]}||_{\cals^p[t_{i-1},t_i]}+||\wh{Z}^{i,[0]}||_{\cals^p[t_{i-1},t_i]}\leq C_p$
holds. The continuity property of the Lipschitz SDE $\mbb{E}\Bigl[\sup_{t\in I_i}|X_t-X_{t_{i-1}}|^p\Bigr]\leq C_p h_i^{p/2}$ 
is also well-known to hold for any $p\geq 2$. Then,
\bea
&&\mbb{E}\Bigl[||\del Y^{i,[1]}||^p_{I_i}+\Bigl(\int_{t_{i-1}}^{t_i}|\del Z_r^{i,[1]}|^2 dr\Bigr)^{p/2}\Bigr]
\leq C_p\mbb{E}\Bigl[\Bigl(\int_{t_{i-1}}^{t_i}|\del f(r)|dr\Bigr)^p\Bigr]\nn \\
&&\quad \leq C_p\left\{\LDis h_i^{3p/2}+h_i^p\mbb{E}\Bigl[1+||\ol{Y}^{i,[0]}||^{2p}_{I_i}+
||\ol{Z}^{i,[0]}||_{I_i}^{4p}\Bigr]^\frac{1}{2}\mbb{E}\Bigl[\sup_{t\in I_i}|X_t-X_{t_{i-1}}|^{2p}\Bigr]^\frac{1}{2}
\right. \nn \\
&&\hspace{15mm}+h_i^p\left( \mbb{E}\Bigl[||\ol{Y}^{i,[0]}-\wh{Y}^{i,[0]}||^p_{I_i}\Bigr]+
\mbb{E}\Bigl[\sup_{t\in I_i}\bigl|\wh{Y}^{i,[0]}_t-\wh{Y}^{i,[0]}_{t_{i-1}}\bigr|^p\Bigr]\right)\nn \\
&&\hspace{12mm}+\mbb{E}\Bigl[1+||\ol{Z}^{i,[0]}||^{2p}_{I_i}+||\wh{Z}^{i,[0]}||^{2p}_{I_i}\Bigr]^\frac{1}{2}
\mbb{E}\Bigl[\Bigl(h_i\int_{t_{i-1}}^{t_i}|\ol{Z}_r^{i,[0]}-\wh{Z}_r^{i,[0]}|^2 dr\Bigr)^p\Bigr]^\frac{1}{2}\nn \\
&&\hspace{11mm}\left.\LDis+h_i^p\mbb{E}\Bigl[1+||\wh{Z}^{i,[0]}||^{2p}_{I_i}+|\wh{Z}^{i,[0]}_{t_{i-1}}|^{2p}\Bigr]^\frac{1}{2}
\mbb{E}\Bigl[\sup_{t\in I_i}\bigl|\wh{Z}^{i,[0]}_t-\wh{Z}^{i,[0]}_{t_{i-1}}\bigr|^{2p}\Bigr]^\frac{1}{2}
\right\} \leq C_p h_i^{3p/2}\nn 
\eea
follows from Corollary~\ref{lemma-Yhat-i0-error-final} and Lemma~\ref{lemma-continuity-Yhat}.
\end{proof}
\end{lemma}

The main result regarding the short-term approximation can be summarized as the next theorem.
\begin{theorem}
\label{th-barY-hatY}
Under Assumptions~\ref{assumption-X}, \ref{assumption-Y}
and Assumption~\ref{assumption-barY}(i), the process $(\wh{Y}^i,\wh{Z}^i)$ defined by 
$(\wh{Y}^i_t:=\wh{Y}^{i,[0]}_t+\wh{Y}^{i,[1]}_t,\quad \wh{Z}^i_t:=\wh{Z}^{i,[0]}_t, t\in I_i)$
is the short-term approximation of  the solution $(\ol{Y}^i,\ol{Z}^i)$ of the qg-BSDE (\ref{eq-BSDE-barY})
and satisfies, with some $(i,n)$-independent positive constant $C_p$,  that
\bea
\mbb{E}\Bigl[||\ol{Y}^i-\wh{Y}^i||^p_{[t_{i-1},t_i]}+\Bigl(\int_{t_{i-1}}^{t_i}|\ol{Z}_r^i-\wh{Z}_r^i|^2 dr\Bigr)^{p/2}\Bigr]
\leq C_p h_i^{3p/2}\nn~,
\eea
for every period $I_i, i\in\{1,\cdots,n\}$ and $\forall p\geq 2$.
\begin{proof}
It follows directly from Proposition~\ref{prop-AE-alpha}, Corollary~\ref{lemma-Yhat-i0-error-final} and 
Lemma~\ref{lemma-Yhat-i1-error}.
\end{proof}
\end{theorem}
\subsubsection*{Acknowledgement}
The research is partially supported by Center for Advanced Research in Finance (CARF).



\end{document}